\documentclass[acmsmall,screen]{acmart}

\usepackage{amsmath}

\usepackage{amssymb}
\usepackage{amsfonts}
\usepackage{bm}
\usepackage{siunitx}
\usepackage{booktabs}
\usepackage{multirow}
\usepackage{array}
\usepackage{enumitem}
\usepackage{algorithm}
\usepackage{algorithmic}
\usepackage{xcolor}
\usepackage{graphicx}
\usepackage{subcaption}
\usepackage{tabularx}
\usepackage{makecell}
\usepackage{booktabs}
\usepackage{multirow}
\usepackage{graphicx}
\usepackage{placeins}

\renewcommand\footnotetextcopyrightpermission[1]{}
\newcommand{\NA}{N/A}

\title{PROPEL: A Memory-Driven, Adaptive Vector-Flow-Field Router with Process-Aware Waveguide Generation for Large-Scale Photonic Integrated Circuits}

\author{A Abdur Rahman Akib}
\author{Eunso Shin}
\author{Jonathan Rogers}
\author{Jonathan Wierer}
\author{Stanley Cheung}
\affiliation{
  \institution{North Carolina State University}
  \city{Raleigh}
  \state{North Carolina}
  \country{USA}
}
\email{aakib@ncsu.edu}

\begin{document}

\begin{abstract}
Photonic integrated circuits (PICs) are rapidly increasing in scale and routing complexity, driven by photonic computing, optical switching, programmable interferometer meshes, wavelength-routed optical networks-on-chip, and chip-to-chip optical interconnects. Manual waveguide planning becomes increasingly impractical as designs approach hundreds to thousands of photonic components, because routing must satisfy geometric, optical, electrical, and manufacturing constraints.

This paper presents \textbf{Photonic Routing Optimization and Placement Engine (PROPEL)}, a memory-driven, native-accelerated adaptive vector-flow-field routing framework with process-aware waveguide generation for large-scale PIC physical design. PROPEL builds a routing kernel based on a bounded vector-flow-field algorithm and DRC validation. The same kernel supports passive optical routing, active optical--electrical routing, and topology-driven net assignment. C++ accelerates computationally intensive field and routing calculations, while Python performs detailed DRC replay, legality checking, route commitment, and GDSFactory-based geometry generation.

Compared with state-of-the-art routing solutions, PROPEL is faster on 17 of 18 shared passive-routing cases, with a median speedup of $2.6\times$ and a geometric-mean speedup of $2.5\times$, while maintaining zero final design-rule violations. PROPEL also supports active optical--electrical routing, topology-driven net assignment, length, delay, and optical-path-length matching, process-map-aware waveguide generation, route-memory reuse, selective rip-up/reroute, and GDS-level legality replay. These capabilities address heterogeneous PIC layouts where optical, electrical, process, topology, matching, and manufacturability constraints must be handled together. In incremental-layout experiments, PROPEL restores more than 95\% of previously verified routes after small local component movements and reroutes only the affected subset, reducing modified-design routing from tens or hundreds of seconds to a few seconds while maintaining zero final design-rule violations.
\end{abstract}


\maketitle
\section{Introduction}
\label{sec:introduction}

\begin{figure}[!t]
    \centering
    \includegraphics[width=0.93\linewidth]{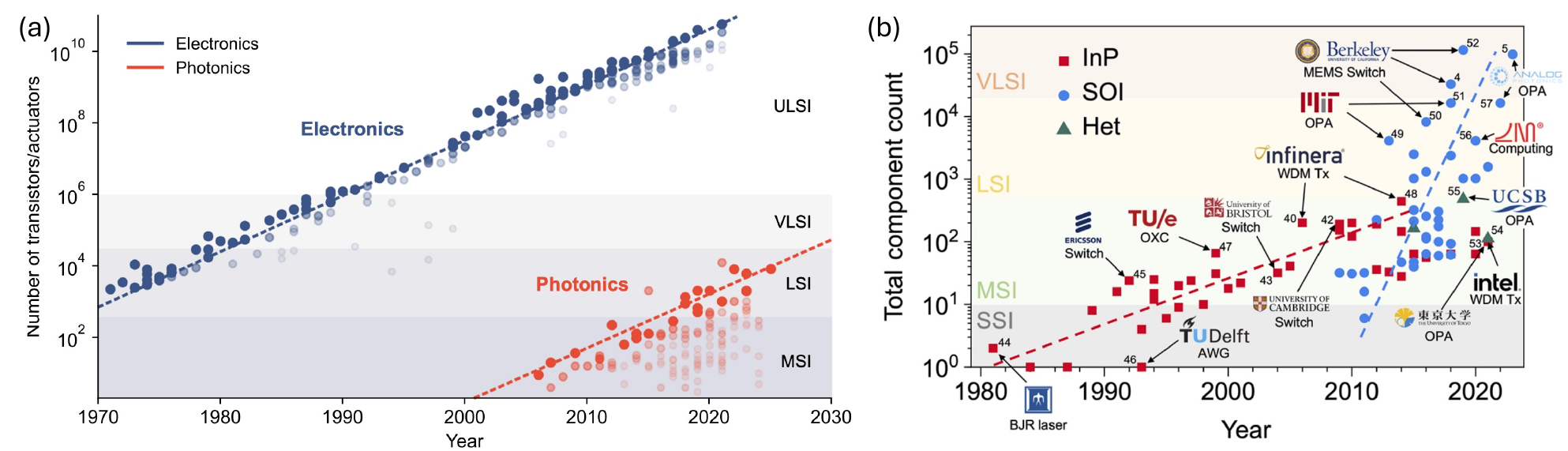}
    \caption{(a) Time evolution of photonic component integration vs. electronics \cite{Prez-Lopez2025_NP}. (b) Breakdown of competing photonic integrated circuit platforms \cite{Wang2025_APL}.}
    \label{Fig_NumPhotonicComp}
\end{figure}

Photonic integrated circuits (PICs) are rapidly increasing in scale and routing complexity, driven by photonic computing, optical switching, programmable interferometer meshes, wavelength-routed optical networks-on-chip (WRONoCs), field-programmable photonic arrays, and chip-to-chip optical interconnects~\cite{zhou2025lidar,zhou2025lidar2,zhou2025apollo,zhou2026lidar3,tseng2019wronoc,tan2011gwor,cheng2020siliconswitch,bogaerts2020programmable,lopez2020autorouting,Tossoun2025_JSTQE, Cheung2022_JLT,Soares2011_PJ,Liang2022_JSTQE}. As these systems move from small manually routable layouts toward large and heterogeneous designs, physical routing becomes a central bottleneck. A modern PIC router must connect hundreds to thousands of optical and electrical terminals while respecting geometric design rules, photonic loss constraints, port orientations, crossing restrictions, bend-radius limits, and layout manufacturability. Looking into the future, Figure \ref{Fig_NumPhotonicComp} illustrates the evolution of photonic components per year with an estimated doubling every 24 months. Ultra-large-scale integration (ULSI) systems comprising > 1,000,000 components are projected to emerge by 2030 \cite{Prez-Lopez2025_NP,Wang2025_APL}.

Modern PIC layout remains largely schematic-driven. Designers manually place devices, draw or script waveguide paths, insert crossings, tune local port access, and repeatedly revise the layout when routing conflicts appear. This flow is manageable for small and highly regular circuits, but it becomes increasingly impractical for dense photonic tensor cores, multiport interferometer fabrics, large switch networks, programmable meshes, active photonic systems, and future ULSI photonics \cite{Prez-Lopez2025_NP,Wang2025_APL}. In these layouts, a locally short route may block another port, consume the only feasible crossing region, violate bend clearance, or force long detours elsewhere. Therefore, large-scale PIC layout requires routing methods that are not only fast, but also geometry-aware, loss-aware, congestion-aware, and design-rule-aware~\cite{boos2013proton,beuningen2016protonplus,beuningen2016platon,chuang2018planaronoc,zhou2025lidar2,deng2025automatic}.

PIC routing differs fundamentally from electronic routing. In VLSI and PCB routing, wires can often use multiple metal layers, vias, sharp corners, and layer-direction conventions. In contrast, optical waveguides require smooth curvilinear geometry with a finite bend footprint and technology dependent minimum bend radius. Photonic ports also require a directional and precise alignment. Same-layer waveguide crossings are allowed, but they are physical optical structures that introduce insertion loss, crosstalk, and increase device footprint and density. Thus, photonic routing depends not only on the route centerline, but also on heading, curvature, crossing compatibility, local clearance, terminal approach direction, and final GDS geometry~\cite{condrat2013crossing,condrat2014tcad,zhou2025lidar2,deng2025automatic}.

Prior automated photonic physical-design work has addressed important parts of this problem. PROTON, PROTON+, and PLATON introduced automatic placement and routing for optical networks-on-chip, showing that physical implementation directly affects insertion loss and optical power budget~\cite{boos2013proton,beuningen2016protonplus,beuningen2016platon}. PlanarONoC and crossing-aware optical channel-routing approaches emphasized the importance of reducing unnecessary crossings and preserving planarity when possible~\cite{chuang2018planaronoc,condrat2013crossing,condrat2014tcad}. LiDAR and LiDAR~2.0 moved the field toward GDS-level detailed waveguide routing by incorporating curvy waveguide geometry, port-access planning, crossing insertion, congestion-aware ordering, crossing-space preservation, and hierarchical reuse~\cite{zhou2025lidar,zhou2025lidar2}. These works established that PIC routing must be solved as a physically constrained detailed-routing problem rather than as abstract logical path planning.

However, passive optical waveguide routing alone is not sufficient for a complete PIC physical-design flow. Active PICs also require electrical routing for modulators, heaters, phase shifters, photodetectors, microring resonators, monitor circuits, ground lines, and I/O pads. Electrical routing in active PICs cannot be treated as ordinary VLSI routing because PICs often have few usable metal layers, long millimeter-scale pin-to-pad wires, weak layer-direction preference, packaging-driven pad constraints, and electro-optical interaction constraints. Metal routes must satisfy electrical design rules while avoiding or controlling overlap with sensitive photonic components and routed waveguides~\cite{zhou2026lidar3,zhou2025apollo}. This creates the need for an integrated routing framework that can handle both passive optical routing and active optical--electrical layout closure. Programmable photonics further expands the role of routing. In field-programmable photonic gate arrays and programmable photonic meshes, routing is both a layout problem and a configuration problem. Optical paths must be selected through a reconfigurable mesh while considering loss, power, damaged elements, resource occupation, wavelength-dependent behavior, and valid light propagation~\cite{bogaerts2020programmable,lopez2020autorouting,chen2019graphbased,kerchove2023resourceadaptation,wang2024multicasting}. Recent work on routing and placing filters in programmable photonics also shows that placement, routing, and functional configuration must be coordinated when multiple functions share the same programmable mesh~\cite{kerchove2025routingfilters}. These works reinforce the same conclusion: photonic routing must be aware of physical resources, optical loss, topology, and reconfigurable constraints.

Topology is another important dimension. In WRONoCs and optical switch fabrics, the topology already contains useful physical structure through wavelength assignment, microring placement, switching stages, and internal waveguide organization. Generic optical router and GWOR designs show that scalable non-blocking optical routers can be constructed from structured wavelength-routed waveguide groups~\cite{tseng2019wronoc,tan2011gwor}. WRONoC design-automation studies further show that topology design, wavelength assignment, component placement, and waveguide routing are strongly coupled~\cite{tseng2019wronoc}. ToPro demonstrates that treating an optimized WRONoC topology as a pure point-to-point netlist can destroy useful internal structure and introduce unnecessary crossings or long detours; instead, topology-aware projection and routing-order optimization can preserve physical quality~\cite{zheng2021topro}. These observations motivated us to develop a topology-driven routing mode in which physical connectivity can be assigned or refined before detailed routing.

Despite these advances, several gaps remain. Existing methods solve important subproblems, but the overall PIC physical-design flow is still fragmented across passive waveguide routing, active electrical routing, and topology-aware net assignment. Dense routing also suffers from redundant exploration, where the router may examine many locally feasible but globally unproductive routing choices. Furthermore, routing guidance, crossing awareness, bend feasibility, port access, and congestion negotiation are often treated as separate heuristics rather than as components of a unified directional model. A scalable router needs a common physical kernel that can guide routing, enforce photonic design rules, preserve crossing and bend feasibility, and support multiple design stages using the same geometric abstraction.

This paper presents \textbf{PROPEL Router}, a memory-driven, native-accelerated, adaptive vector-flow-field routing framework with process-aware waveguide generation for large-scale PIC physical design. The central idea is to use an adaptive vector flow field (VFF) as the primary routing guidance model. Instead of relying only on local distance-to-target behavior, PROPEL constructs bounded directional guidance over the routing region and uses this field to bias routing toward physically promising directions while respecting bend, crossing, port-access, obstacle, and congestion constraints. The VFF does not replace design-rule checking; rather, it provides a directional guidance layer that reduces unnecessary exploration and improves routing consistency in dense layouts.

PROPEL is built around a common geometry-aware routing kernel. This kernel includes bounded VFF construction, group-aware VFF guidance, adaptive routing bounds, orientation-aware routing states, bend-aware motion primitives, crossing-aware legality checks, congestion and history costs, and detailed design-rule checking. For acceleration, computationally intensive field construction and routing calculations can be executed in C++, while Python remains responsible for detailed DRC checking and enforcement, port-access validation, crossing legality, route commitment, and GDSFactory-based \cite{gdsfactory} geometry generation. This separation preserves correctness while accelerating repeated calculations that dominate runtime in large layouts.

\begin{figure}[!t]
    \centering
    \includegraphics[width=0.92\linewidth]{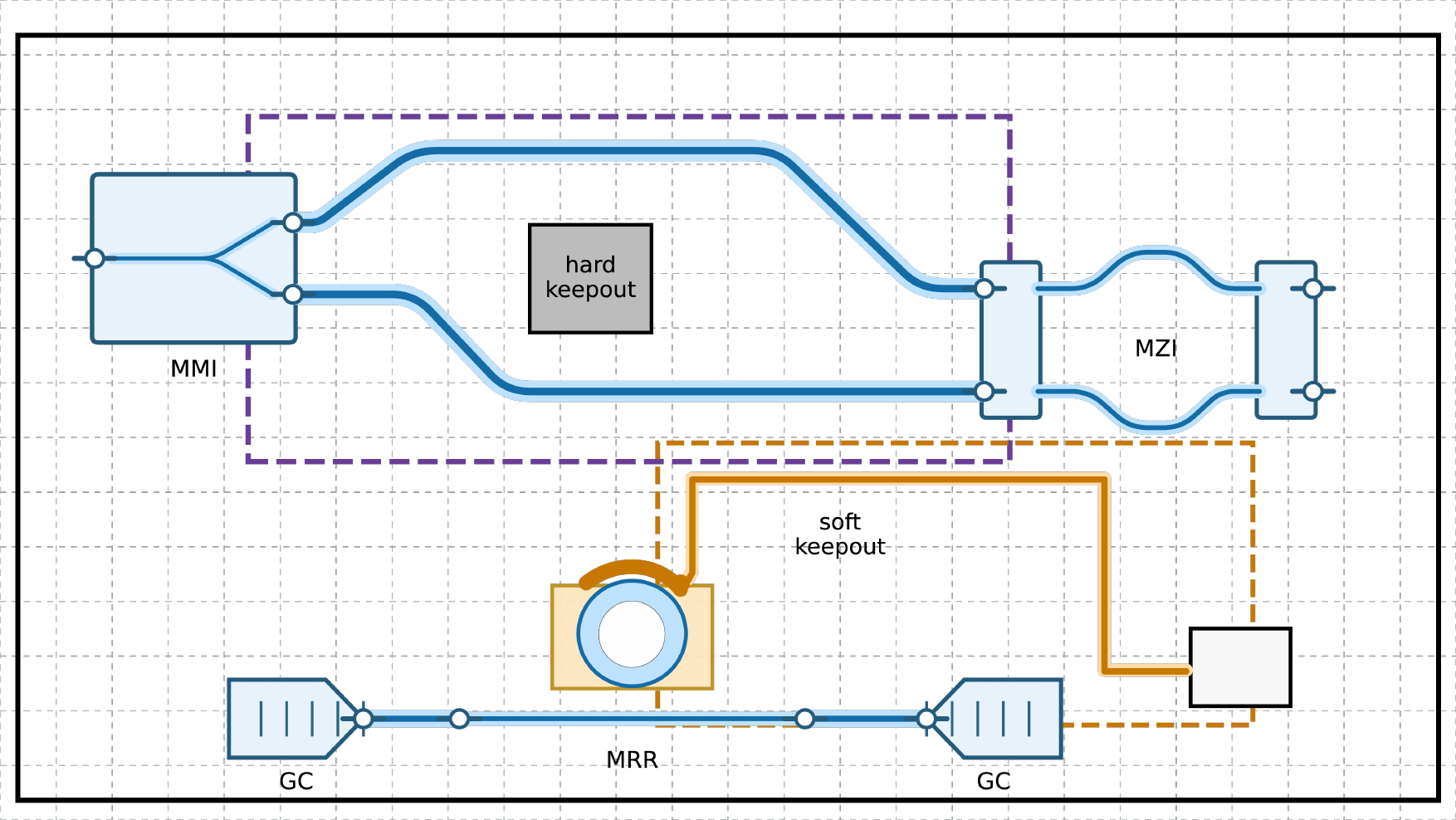}
    \caption{Overview of the PROPEL routing framework. PROPEL uses a common geometry-aware routing model to coordinate optical waveguide routing, electrical metal routing, hard and soft keep-out handling, port-access constraints, and bounded routing regions. The figure illustrates how optical and electrical routes are generated in the same physical layout while preserving route legality around photonic devices and keep-out regions.}
    \label{fig:PROPEL_overview}
\end{figure}

On top of this common kernel, PROPEL supports three connected physical-design modes, as summarized in Figure~\ref{fig:PROPEL_overview}. First, for \textbf{Passive PIC Optical Routing}, PROPEL generates optical waveguide routes using bounded VFF guidance, adaptive port-access reservation, group-aware routing, crossing-aware legality, nested failed-net rip-up and reroute, optional length and group-delay matching, and process-aware routing. Second, for \textbf{Active PIC Electro-Optical Routing}, PROPEL extends the flow to include electrical nets while checking metal spacing, legal layers, device overlap, waveguide interaction, and user-defined electro-optical isolation constraints. Third, for \textbf{Topological Routing}, PROPEL can synthesize or refine physically meaningful optical connections before detailed routing using port side, orientation, length, obstacle, and crossing-aware costs.

The main contributions of this work are as follows:
\begin{itemize}
    \item We propose PROPEL, a bounded vector-flow-field routing framework for large-scale PIC physical design that combines directional guidance with strict final DRC replay.
    \item We introduce a common routing kernel that supports passive optical routing, active optical--electrical routing, topology-driven assignment, matching refinement, route-memory reuse, and process-aware geometry generation.
    \item We demonstrate DRC-clean passive routing across spacious and compact PIC benchmarks.
    \item We evaluate length-, delay-, and optical-path-length-matching refinement with zero final mismatch and zero final DRC violations across representative matched-routing benchmarks.
    \item We evaluate active electrical routing and topology-driven routing as extensions of the same DRC-aware routing kernel.
\end{itemize}

\begin{figure}[!t]
    \centering
    \includegraphics[width=0.92\linewidth]{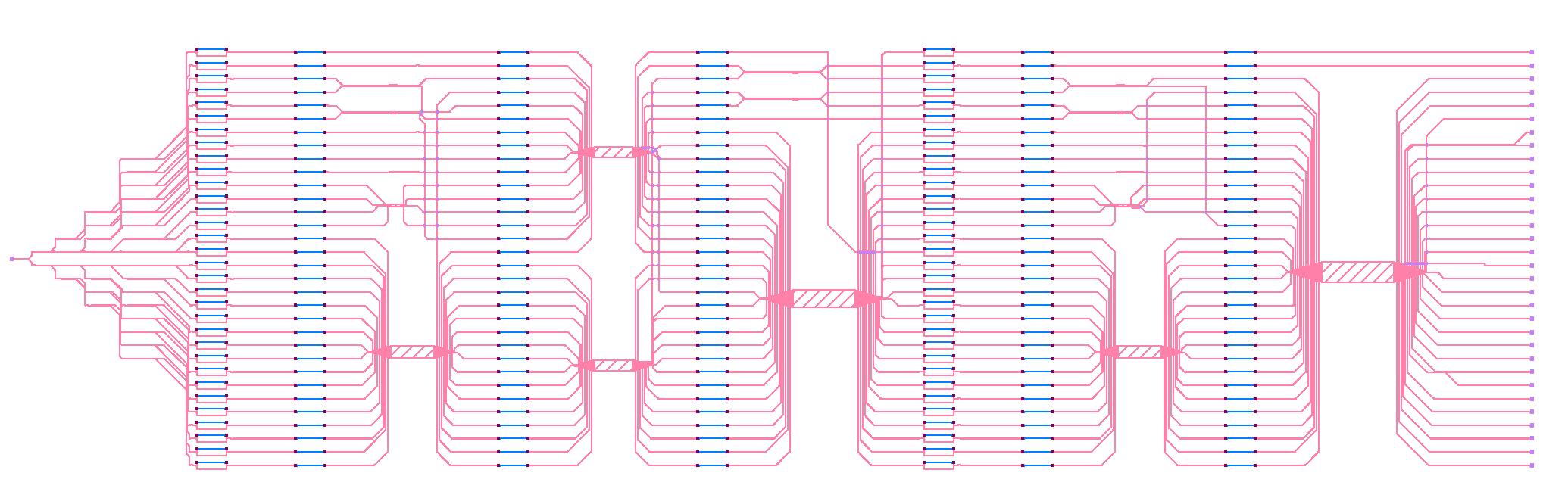}
    \caption{Large-scale fully routed photonic chip generated in under 8 minutes.}
    \label{fig:gds_mmi32}
\end{figure}


\section{Background and Motivation}
\label{sec:background}
\subsection{Routing Across Domains}

Routing is the problem of finding feasible paths between endpoints in a constrained space. The abstract formulation appears in robotics, PCB design, VLSI routing, and PIC layout. However, the physical meaning of a route differs sharply across domains. In robotics, a planner finds a trajectory for a moving agent through obstacles. The result is fed to a robot to find the most optimal path between two waypoints.

PCB routing introduces persistent physical traces, multiple nets, trace width, spacing rules, layer changes, and vias. Multiple paths compete for the same substrate, so ordering and congestion matter. However, PCB routing normally has several metal layers and can use vias as a congestion escape mechanism. VLSI routing is even more mature, with many metal layers, preferred routing directions, via stacks, and strong design-rule engines. Most routes are Manhattan or near-Manhattan, and a 90-degree bend is essentially a zero-footprint corner compared with a photonic bend ~\cite{mcmurchie1995pathfinder}.

PIC routing is usually restricted to one optical waveguide layer. Same-layer crossings are allowed but lossy. Bends are real curvilinear structures with finite footprint and minimum radius. Ports are directional and require exact approach orientation. Therefore, a PIC detailed router must reason about continuous geometry even when the search itself is performed on a discrete grid.

\begin{table}[t]
\centering
\caption{Comparison of routing constraints across domains.}
\label{tab:routing_comparison}
\small
\resizebox{\columnwidth}{!}{
\begin{tabular}{@{}lcccc@{}}
\toprule
\textbf{Constraint} & \textbf{Robotics} & \textbf{PCB} & \textbf{VLSI} & \textbf{PIC} \\
\midrule
Multiple simultaneous routes & N/A & \checkmark & \checkmark & \checkmark \\
Persistent physical routes & \checkmark & \checkmark & \checkmark & \checkmark \\
Path Constraints / DRC & \checkmark & \checkmark & \checkmark & \checkmark \\
Multiple routing layers & N/A & \checkmark & \checkmark & Limited \\

Curvature-aware geometry & Optional & Limited & No & Required \\
Direction dependence & Limited & Moderate & Pin-access dependent & Strict \\
Phase / length matching & No & Moderate & Moderate & Stringent \\
\bottomrule
\end{tabular}
}
\end{table}

\subsection{Photonic Design Rules}

PIC routing must satisfy several photonic-specific constraints. Waveguide connections must preserve compatible optical mode, layer, width, and cross-section. Incompatible waveguides require transitions or tapers. A crossing is a physical photonic structure with its own footprint, alignment requirement, and optical loss.

Unlike VLSI wires, photonic waveguides cannot turn with zero-radius corners. Bends must be implemented using circular, Euler, sine, or related smooth curves. Typical silicon waveguides can support bend radii of several micrometers, while lower-index-contrast platforms such as silicon nitride often require much larger bend radii. Crossings are also physical devices. Each crossing requires a footprint, local alignment, compatible waveguide type, and sufficient straight run-up. Crossings introduce insertion loss and crosstalk, so they must be inserted intentionally rather than treated as accidental line intersections.

Photonic ports have positions and orientations. A routed waveguide must approach a port with the correct tangent direction. Finally, the primary quality metric is insertion loss. Long waveguides increase propagation loss; bends add bend loss; crossings add crossing loss. The critical path loss sets the optical power budget and directly affects the required power.

\section{Core Routing Model and PROPEL Kernel}
\label{sec:PROPEL_kernel}

\subsection{Layout, Ports, Nets, and Routing Domains}
\label{subsec:layout_domains}

PROPEL uses a common physical model for all routing modes. A photonic layout is represented as a bounded two-dimensional design region containing placed component instances, oriented ports, routing obstacles, already committed routes, and optional technology-dependent keep-out regions. This shared representation allows passive optical routing, active optical-electrical routing, and netlist-free topological routing to use the same geometric database instead of maintaining separate models for each task.

Each placed component has a physical footprint and a set of ports. A port is treated as an oriented terminal. For optical routing, this distinction is essential because a waveguide must approach the port with a compatible tangent direction. A path that reaches the correct point but arrives from the wrong direction is not a valid photonic connection. Electrical pins and pads are represented similarly, but their legality depends on metal-layer, pin-access, pad-access, and electrical design-rule constraints. Let $\Omega\subset\mathbb{R}^{2}$ denote the bounded chip routing domain.

A routed connection is modeled as a two-terminal net,
\begin{equation}
    n_i = (p_s,p_t,\eta_i),
\end{equation}
where $p_s$ and $p_t$ are the source and target ports, and $\eta_i$ stores the routing-domain information for the net. This metadata includes the net type, routing layer, waveguide cross-section, bend-radius rule, crossing policy, electrical rule class, matching group, or topology-assignment information. Multi-output photonic behavior is represented through physical devices such as splitters, MMIs, resonators, or topology blocks. After these devices are instantiated, each detailed routed segment is still treated as a two-terminal connection.

For each net, PROPEL generates a physical route geometry within a definite chip layout region. During the search, the waveguide routing path is represented as an orientation-aware path or centerline. After routing, the path is converted into physical layout geometry, such as an optical waveguide, metal trace, crossing structure, or taper. A route is only committed when it finds a path and satisfies the design rules of its routing domain.

The layout can contain both fixed and dynamic obstacles. Fixed obstacles come from placed devices, locked geometry, chip boundaries, keep-out regions, and routing blockages. Dynamic obstacles come from routes that have already been committed. After each successful route, PROPEL updates the routed-geometry database so that later nets see the new route as an obstacle, crossing candidate, congestion source, or soft penalty region depending on the domain and interaction type.

Table~\ref{tab:PROPEL_domain_model} summarizes how each routing domain uses the shared model. The important point is that the geometry engine is common, while the legality rules and cost terms are domain-specific.

\begin{table}[!t]
\centering
\caption{Routing domains in the PROPEL physical model.}
\label{tab:PROPEL_domain_model}
\small
\begin{tabularx}{\columnwidth}{@{}lXX@{}}
\toprule
\textbf{Domain} & \textbf{Input representation} & \textbf{Main physical constraints} \\
\midrule
Optical routing &
Fixed optical port-to-port nets &
Port orientation, bend radius, waveguide spacing, crossing legality, optical loss \\

Electrical routing &
Electrical pin-to-pad or pin-to-pin nets &
Metal width/spacing, layer legality, via policy, pad access, photonic keep-out \\

Topological routing &
Input/output port sets without fixed pairing &
Assignment distance, side compatibility, orientation compatibility, crossing pressure \\

\bottomrule
\end{tabularx}
\end{table}

This domain separation keeps the main routing formulation compact. PROPEL does not need a different physical model for each operating mode. 

\subsection{Routing Objectives and Reported Metrics}
\label{subsec:routing_objectives_metrics_compact}

PROPEL treats routing as a constrained physical-design problem. PROPEL's first objective is legality rather than length, crossing count, or runtime. A route is useful only if it satisfies the design rules of its routing domain. Therefore, PROPEL follows a feasibility-first objective: complete all required connections, commit only DRC-clean geometry, and then evaluate physical quality metrics on the verified layout. For a routed solution, the primary feasibility condition is
\begin{equation}
    \mathrm{DRV}(\Gamma)=0,
\end{equation}
where $\mathrm{DRV}$ is the number of design-rule violations in the committed layout. PROPEL also reports the number of successfully routed nets, because compact benchmarks may expose partial-routing behavior before a fully clean solution is obtained. A complete routing solution requires all nets in the target netlist to be routed and verified.

For optical routing, PROPEL reports route length, crossing count, bend contribution, design-rule violations, runtime, and insertion loss. The net-level optical loss is evaluated as

\begin{equation}
    IL(n_i)
    =
    \alpha_w L(n_i)
    +
    \alpha_b B(n_i)
    +
    \alpha_c X(n_i),
\end{equation}
where $L(n_i)$ is routed waveguide length, $B(n_i)$ is bend count or accumulated bend contribution, $X(n_i)$ is crossing count, and $\alpha_w$, $\alpha_b$, and $\alpha_c$ are technology-dependent propagation, bend, and crossing loss coefficients.

For path-level evaluation, PROPEL adds the routed-net losses and device losses along each evaluated optical path:
\begin{equation}
    IL(p)
    =
    \sum_{m_i\in p} IL(m_i)
    +
    \sum_{n_j\in p} IL(n_j).
\end{equation}
The reported critical-path insertion loss is
\begin{equation}
    IL_{\max}
    =
    \max_{p\in\mathcal{P}_{\mathrm{eval}}} IL(p),
\end{equation}
where $\mathcal{P}_{\mathrm{eval}}$ is the set of evaluated optical paths. If a benchmark provides only net-level routing data, PROPEL reports the maximum configured netlist-path loss using the available route geometry and loss parameters.

The routing cost used during search does not need to be identical to the final physical loss model. Search-time weights are used to guide exploration, avoid congestion, discourage unnecessary bends and crossings, and bias motion toward the vector-flow field. Final reported metrics are computed only after the route is generated, verified, and committed. This separation prevents the search heuristic from being confused with the physical evaluation model.

For active electro-optical routing, PROPEL reports both optical and electrical metrics. Optical metrics are evaluated using the same DRC quantities described above. Electrical metrics include total metal length, via count, number of used metal layers, electrical DRC violations, and electrical-routing runtime.

For netlist-free topological routing, PROPEL additionally reports assignment quality before detailed routing. This includes the number of generated connections, estimated assignment cost, crossing-pressure estimate, and final routability of the generated netlist. These values are useful because a poor input--output assignment can make a layout difficult to route even when a legal routing solution exists for a better pairing.

\begin{table}[!t]
\centering
\caption{Metrics reported by PROPEL across routing modes.}
\label{tab:PROPEL_reported_metrics}
\small
\begin{tabularx}{\columnwidth}{@{}lX@{}}
\toprule
\textbf{Metric} & \textbf{Meaning} \\
\midrule
$N_{\mathrm{routed}}$ &
Number of successfully routed and committed nets \\

$\mathrm{DRV}$ &
Number of hard design-rule violations in the committed layout \\

$L_{\mathrm{tot}}$ &
Total routed optical waveguide length or electrical metal length \\

$X_{\mathrm{tot}}$ &
Total number of committed optical crossings \\

$IL_{\max}$ &
Maximum evaluated optical path insertion loss \\

$N_{\mathrm{via}}$ &
Number of vias used during electrical routing \\

$N_{\mathrm{layer}}$ &
Number of metal layers used by electrical routes \\

$\mathrm{USV}_{\mathrm{eo}}$ &
Strict user-specified electrical--optical interaction violations \\

$C_{\mathrm{eo}}$ &
Soft electrical--optical interaction penalty \\

$T_{\mathrm{runtime}}$ &
Runtime for the evaluated routing stage \\
\bottomrule
\end{tabularx}
\end{table}

Table~\ref{tab:PROPEL_reported_metrics} summarizes all the reported metrics by the PROPEL router.
\section{PROPEL Routing Kernel}
\label{sec:vff}

The three PROPEL operating modes share a common routing kernel. This kernel is responsible for representing the layout, building routing guidance, evaluating local route feasibility, tracking congestion, and committing legal geometry. The central idea is the adaptive vector flow field: a bounded directional guidance model that biases routing toward physically promising regions while leaving final legality to detailed design-rule checking. This section describes the common kernel before specializing it to passive optical routing, active electro-optical routing, and topology-driven routing.

\subsection{Bounded Vector-Flow-Field Search}
\label{subsec:bounded_vff_search}

\begin{figure}[!t]
    \centering
    \includegraphics[width=\linewidth]{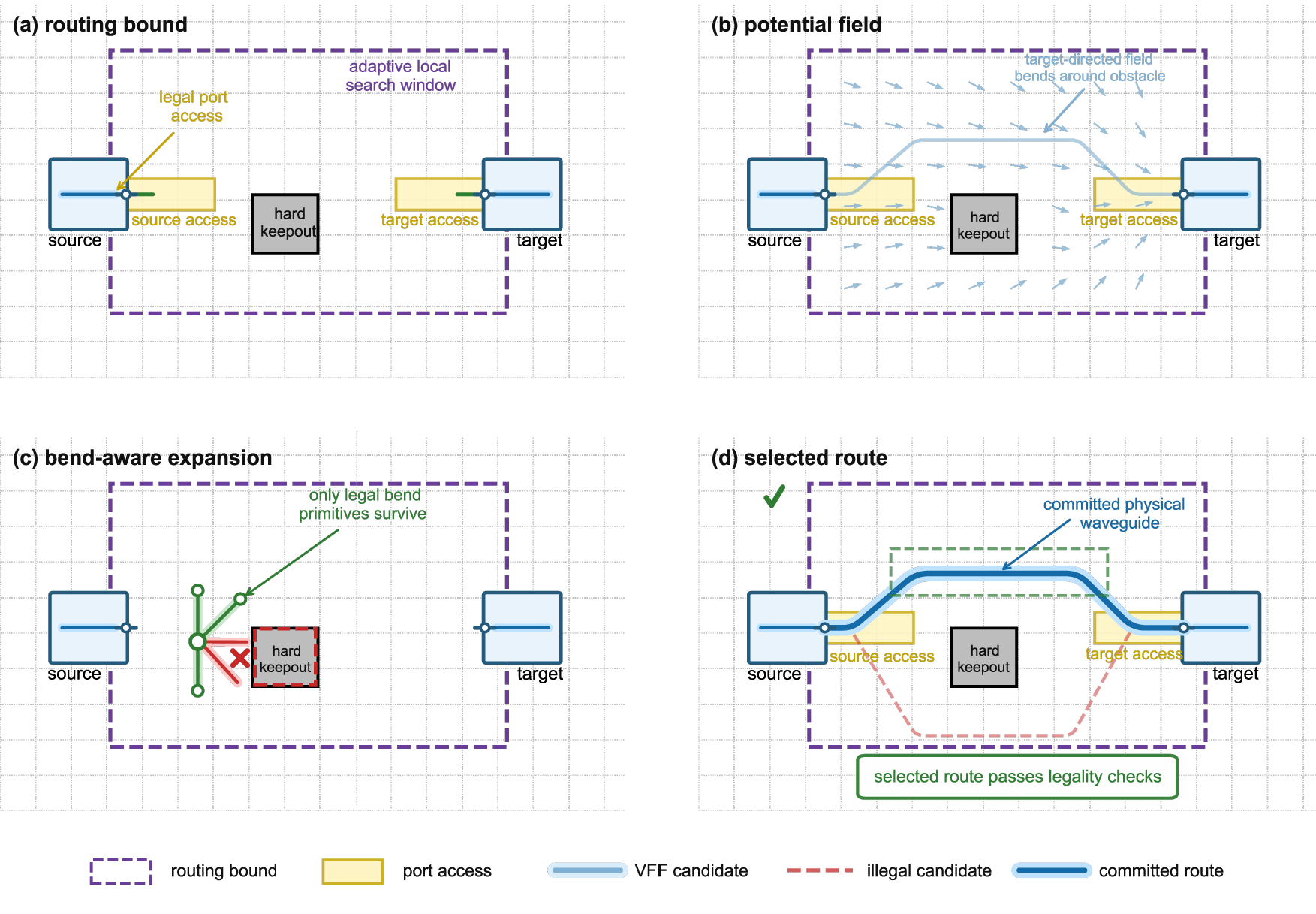}
    \caption{Bounded vector-flow-field search for one optical net. PROPEL first constructs legal source and target access regions, then forms a target-directed field inside the local routing window. Bend-aware candidate expansion is evaluated against hard keep-outs and port-access legality, and only the selected legal route is converted into a committed physical waveguide.}
    \label{fig:bounded_vff_search}
\end{figure}

The core routing operation in PROPEL is bounded vector-flow-field search. For each net, PROPEL constructs a local routing window, builds directional guidance toward the target-access region, and expands a bend-aware route through legal candidate states. Figure~\ref{fig:bounded_vff_search} illustrates this process from port access to final route commitment. PROPEL first identifies legal source and target access regions, then builds a target-directed vector field inside the local routing window, expands bend-aware candidate routes, rejects illegal primitives, and finally commits only the selected DRC-clean physical waveguide. The purpose of the vector-flow field is to guide the search so that the physical legality can be enforced by the design-rule checker before any route is committed.
A routing state stores both position and direction:
\begin{equation}
    q=(i,j,\theta),
\end{equation}
where $(i,j)$ is the grid location and $\theta$ is the current propagation direction. This orientation-aware representation is necessary for photonic routing because the legality of the next movement depends on how the route arrives at the current location. A waveguide reaching a point from the east has different bend, crossing, and port-access options than a waveguide reaching the same point from the north. For electrical routing, the state can be extended with a metal layer, but the same principle remains: the router tracks enough local state to evaluate physically meaningful movements.

For a net $n_i=(p_s,p_t,\eta_i)$, PROPEL first defines a bounded routing region,
\begin{equation}
    \mathcal{B}_i \subseteq \Omega ,
\end{equation}
that contains the source port, target port, required access regions, and an adaptive margin around the expected routing corridor. The bound prevents the router from exploring unrelated parts of the chip and makes vector-field construction local to the current routing task. If routing fails, the bound can be expanded during later retries, but the first attempt remains focused on the most relevant physical region.

Inside the routing bound, PROPEL constructs a scalar potential field $\Phi_i$ directed toward the target-access region. The vector-flow field is defined as
\begin{equation}
    \mathbf{F}_i(x,y)=-\nabla \Phi_i(x,y).
\end{equation}
In the discrete implementation, each candidate next state is scored using the local potential value and the route cost.

The local routing cost combines physical movement cost, congestion cost, and VFF guidance:
\begin{equation}
    C(q,q')
    =
    w_l C_{\mathrm{len}}(q,q')
    +
    w_b C_{\mathrm{bend}}(q,q')
    +
    w_x C_{\mathrm{cross}}(q,q')
    +
    w_g C_{\mathrm{cong}}(q')
    +
    w_h C_{\mathrm{hist}}(q')
    +
    w_{\Phi} C_{\mathrm{VFF}}(q').
\end{equation}
Here, $q'$ is a candidate next state. The length and bend terms discourage unnecessarily long or highly curved paths. The crossing term discourages unnecessary optical crossings while still allowing legal crossings when needed. The congestion and history terms discourage repeatedly contested regions. The VFF term biases the route toward the target-access region.

\begin{algorithm}[!t]
\small
\caption{PROPEL-Guided Routing for One Net}
\label{alg:PROPEL_single_net}
\begin{algorithmic}[1]
\STATE \textbf{Input:} net $n_i=(p_s,p_t,\eta_i)$, layout region $\Omega$, fixed obstacles, committed routes $\Gamma_{\mathrm{routed}}$, design rules
\STATE Build source and target access states from port positions and orientations
\STATE Construct an adaptive routing bound $\mathcal{B}_i$ around the source, target, access regions, and expected routing corridor
\STATE Build obstacle, congestion, history, and soft-penalty maps inside $\mathcal{B}_i$
\STATE Compute target-directed potential field $\Phi_i$ inside $\mathcal{B}_i$
\STATE Initialize the priority queue with the legal source-access state
\WHILE{the priority queue is not empty}
    \STATE Pop the lowest-cost state $q=(i,j,\theta)$
    \IF{$q$ is a legal target-access state}
        \STATE Reconstruct the candidate path $\hat{\gamma}_i$
        \STATE \textbf{return} $\hat{\gamma}_i$
    \ENDIF
    \FOR{each bend-aware candidate transition $q\rightarrow q'$}
        \STATE Evaluate transition cost:
        \[
        C(q,q') =
        w_l C_{\mathrm{len}}
        + w_b C_{\mathrm{bend}}
        + w_x C_{\mathrm{cross}}
        + w_g C_{\mathrm{cong}}
        + w_h C_{\mathrm{hist}}
        + w_{\Phi} C_{\mathrm{VFF}}
        \]
        \IF{candidate transition satisfies local DRC}
            \STATE Insert or update $q'$ in the priority queue
            \STATE Store parent pointer for path reconstruction
        \ENDIF
    \ENDFOR
\ENDWHILE
\STATE Expand the routing bound or mark the net as failed
\STATE \textbf{return} failure
\end{algorithmic}
\end{algorithm}

From each state, the router generates movement primitives with realizable bend geometry.

The target is also orientation-aware. A route is complete only when it reaches a legal target-access state. The route must approach the target port from the correct direction so that the final waveguide can connect to the port without an invalid snap, sharp turn, or post-processing correction. This is especially important near dense devices, where the terminal access region can be the limiting factor rather than the global source-to-target distance.

\begin{figure}[!t]
    \centering
    \includegraphics[width=\linewidth]{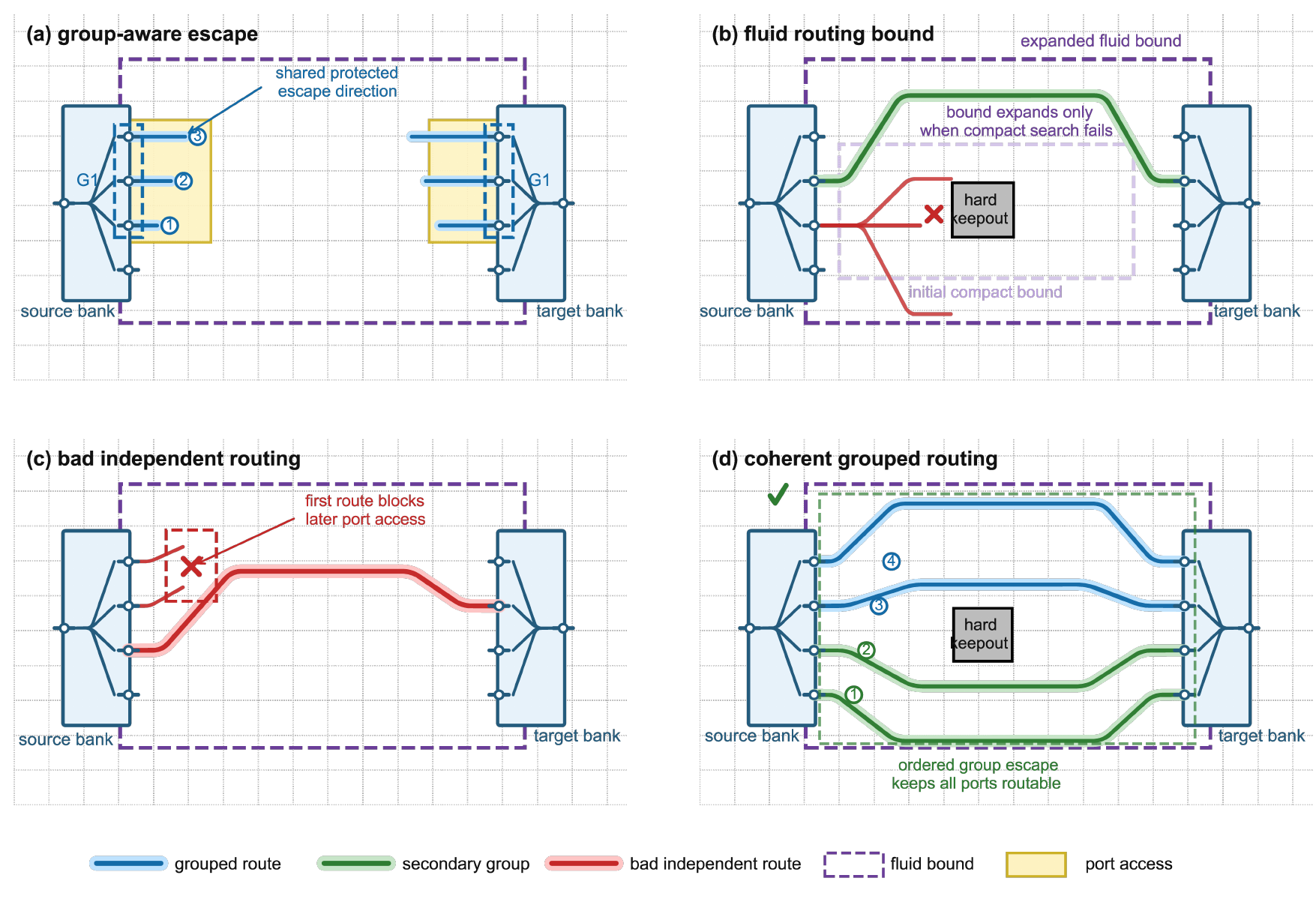}
    \caption{Group-aware guidance and fluid routing-bound control in PROPEL. In dense port banks, related nets first share an ordered escape direction before separating toward their individual targets. The routing bound is initially compact and expands only when compact routing fails, which avoids unnecessary global exploration while still allowing legal detours. The figure also contrasts an independent route that blocks later port access with coherent grouped routing that keeps all ports routable.}
    \label{fig:group_fluid_bounds}
\end{figure}

PROPEL also supports group-aware guidance for dense port banks, as illustrated in Figure~\ref{fig:group_fluid_bounds}. Nets that originate from nearby ports, terminate at nearby ports, or share a common device side can be grouped so that their initial routing directions remain coherent. This prevents early routes from consuming the natural escape corridor of related later routes. Group guidance biases related nets toward preserving shared routing resources before they separate toward individual targets.

The routing bound can also be fluid rather than fixed. A fluid bound is initialized around the source-target region and then adapted based on port direction, routing distance, group size, obstacle context, and retry history. Short local nets receive compact bounds, while long or congested nets can receive larger bounds. If a net fails because the compact bound does not contain a legal path, the bound can expand gradually instead of immediately allowing global exploration. This behavior is shown in Figure~\ref{fig:group_fluid_bounds}(b), where the routing region expands only when additional detour space is needed. This improves runtime while still allowing controlled detours when local routing fails.

\subsection{Legality Checking and Route Commitment}
\label{subsec:legality_commitment}

The bounded vector-flow field determines where the router searches; design-rule checking determines what it may commit. PROPEL therefore separates directional guidance from design-rule legality. 
For a candidate transition from a current state $q$ to a next state $q'$, PROPEL evaluates the physical primitive associated with that transition. This primitive may be a straight waveguide segment, a bend, a crossing approach, a terminal access segment, a metal segment, or a via-connected electrical step. If the candidate violates a hard rule, it is rejected regardless of its VFF score. Thus, VFF guidance affects the order in which candidates are explored, while design-rule checking determines whether they can actually be used.

For optical nets, PROPEL checks the candidate geometry against photonic design rules. The main optical checks are obstacle clearance, waveguide-to-waveguide spacing, bend-radius compliance, terminal orientation, crossing validity, and waveguide cross-section compatibility. Obstacle clearance prevents the route from intersecting fixed devices, locked geometries, chip boundaries, and hard keep-out regions. Spacing checks prevent the new waveguide from being placed too close to previously routed waveguides or other photonic structures. Bend-radius checks ensure that each turn can be realized as a manufacturable optical bend rather than as an unrealistically sharp grid corner.

\begin{figure}[!t]
    \centering
    \includegraphics[width=\linewidth]{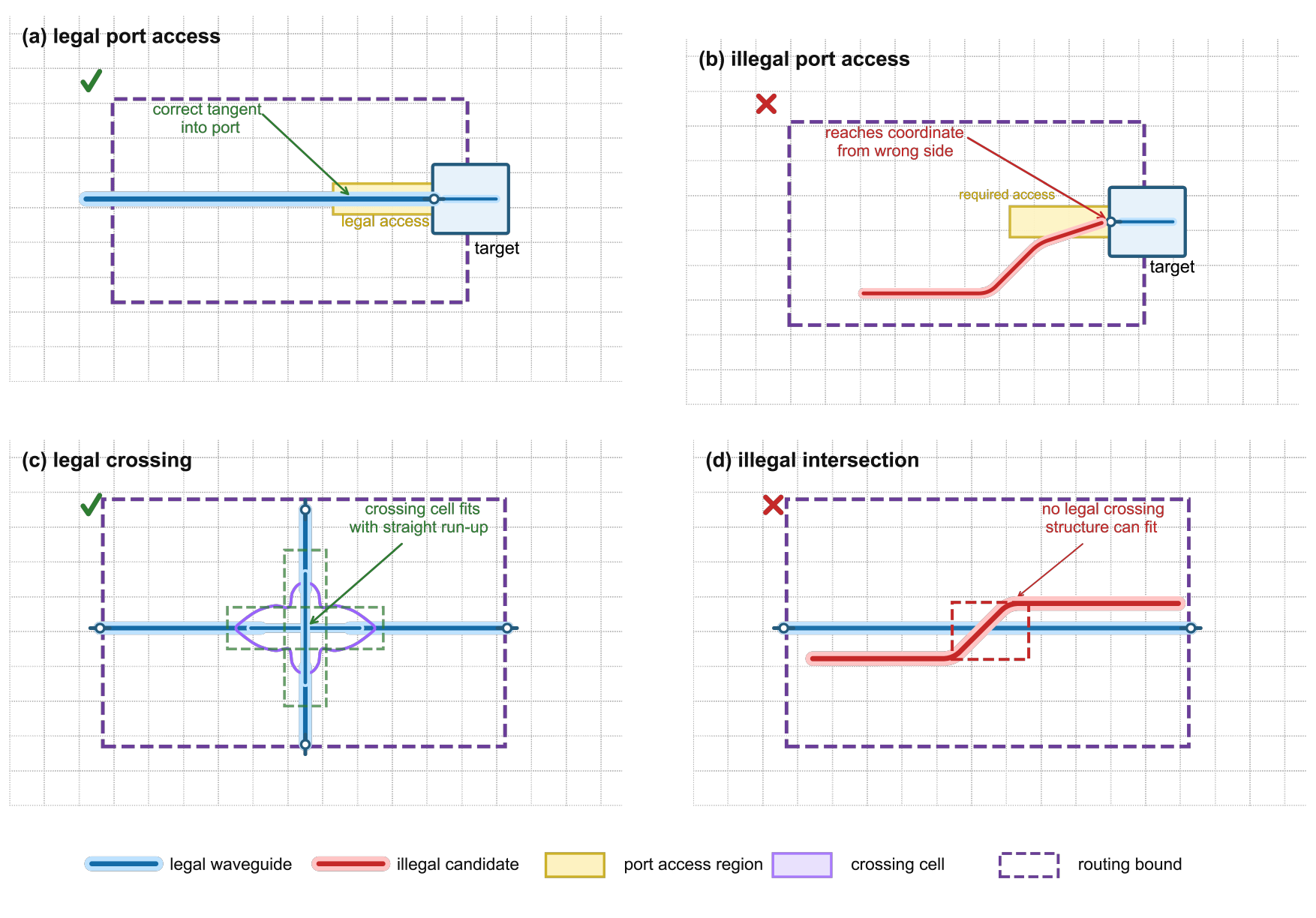}
    \caption{Optical legality checks used before route commitment. A route is legal only when it reaches the port with the correct tangent direction and when any intersection can be implemented as a valid crossing structure. Candidates that reach a port from the wrong side or intersect another waveguide without enough crossing-cell space are rejected even if their centerlines appear connected.}
    \label{fig:optical_legality_examples}
\end{figure}

Port legality is handled explicitly, as shown in Figure~\ref{fig:optical_legality_examples}(a)--(b). A photonic port is an oriented optical interface, not only a target coordinate. Therefore, PROPEL does not accept a route simply because it reaches the port location. The route must leave the source port in a compatible direction and must enter the target port from the correct opposing direction. If the final route tangent does not match the port orientation, the path is rejected even if the coordinate is correct. This avoids invalid post-routing snapping, uncontrolled terminal correction, and artificial routes that cannot be converted into a real waveguide.

Crossings are also handled explicitly, as illustrated in Figure~\ref{fig:optical_legality_examples}(c)--(d). In PIC routing, a crossing is not an ordinary geometric intersection. It is a physical photonic structure with its own footprint, alignment requirement, and optical loss. When a candidate optical route intersects an already committed waveguide, PROPEL first determines whether the interaction can be converted into a legal crossing. A legal crossing must satisfy angle compatibility, sufficient local clearance, waveguide cross-section compatibility, and enough straight run-up before and after the crossing. If the crossing cell cannot fit, or if the intersection occurs too close to a bend, port, obstacle, or another crossing, the interaction is treated as an illegal overlap.

For electrical nets, PROPEL applies metal-specific legality checks. A candidate metal route must satisfy metal width, metal spacing, allowed layer usage, via policy, pin access, pad access, and short/open rules. Electrical legality is not limited to ordinary metal DRC because active PICs contain nearby optical waveguides, resonators, modulators, heaters, photodetectors, and optical I/O regions. A metal route may be electrically legal but still undesirable or forbidden from the photonic perspective. PROPEL therefore supports photonic-aware hard keep-out regions and soft interaction regions for electrical routing.

Table~\ref{tab:legality_checks} summarizes the main legality checks used by PROPEL before route commitment. The exact rule values are technology- and configuration-dependent, but the same checking structure is used across routing modes.

\begin{table}[!t]
\centering
\caption{Main legality checks before route commitment.}
\label{tab:legality_checks}
\small
\begin{tabularx}{\columnwidth}{@{}lX@{}}
\toprule
\textbf{Check} & \textbf{Purpose} \\
\midrule
Obstacle clearance &
Prevents routes from intersecting fixed devices, chip boundary, locked geometry, and hard keep-out regions \\

Port orientation &
Ensures that optical waveguides enter and leave ports with physically valid tangent directions \\

Bend radius &
Prevents sharp turns that cannot be realized as manufacturable waveguide bends \\

Waveguide spacing &
Maintains required clearance between optical routes and neighboring photonic geometry \\

Crossing legality &
Allows crossings only when angle, footprint, straight run-up, and cross-section compatibility are valid \\

Metal DRC &
Checks electrical width, spacing, layer usage, via rules, shorts, opens, and pad/pin access \\

Electrical--optical interaction &
Rejects or penalizes metal routes near sensitive optical structures depending on hard or soft rules \\
\bottomrule
\end{tabularx}
\end{table}
\begin{algorithm}[!t]
\small
\caption{Route Verification and Commitment}
\label{alg:route_verification_commitment}
\begin{algorithmic}[1]
\STATE \textbf{Input:} candidate path $\hat{\gamma}_i$, net metadata $\eta_i$, committed routes $\Gamma_{\mathrm{routed}}$, design rules
\STATE Convert the candidate state sequence into physical route primitives
\IF{net is optical}
    \STATE Build terminal-access segments, straight waveguides, bends, and candidate crossing structures
    \STATE Check port orientation, bend radius, waveguide spacing, obstacle clearance, crossing legality, and cross-section compatibility
\ELSIF{net is electrical}
    \STATE Build metal segments, layer assignments, vias, pin connections, and pad-access geometry
    \STATE Check metal width, spacing, layer legality, via policy, pin/pad access, shorts, opens, and photonic keep-outs
\ENDIF
\STATE Replay the full route geometry in Python-side DRC
\IF{full replay fails}
    \STATE Record the failure cause for diagnostic rerouting
    \STATE \textbf{return} rejection
\ENDIF
\STATE Generate final layout geometry using the configured route builder
\STATE Recheck the final generated geometry against fixed obstacles and committed routes
\IF{final geometry is not DRC-clean}
    \STATE Record conflict information
    \STATE \textbf{return} rejection
\ENDIF
\STATE Commit the route:
\[
\Gamma_{\mathrm{routed}}
\leftarrow
\Gamma_{\mathrm{routed}}
\cup
\{\gamma_i\}
\]
\STATE Update dynamic obstacles, crossing records, congestion maps, route metrics, and route memory
\STATE \textbf{return} committed route $\gamma_i$
\end{algorithmic}
\end{algorithm}
Candidate-level legality is not the final step. After a complete candidate path is found, PROPEL performs route-level replay before commitment. During replay, the discrete route is converted into the physical representation that will be inserted into the layout. For optical nets, this includes terminal access, bends, straight segments, crossing structures, smoothing, and GDSFactory-based waveguide construction. For electrical nets, this includes metal segments, layer assignments, vias, pad access, and pin connection geometry. The route is committed only if the replayed physical geometry remains valid.

This replay step is necessary because a path can be valid at the search-state level but invalid after final geometry construction. For example, a centerline may pass through a narrow gap, but the actual waveguide width and bend footprint may not fit. A route may appear to intersect another waveguide at a point, but the crossing cell may not have enough straight approach length. A metal route may satisfy grid-level spacing, but the final metal polygon may violate a keep-out region or pad-access rule. Route-level replay prevents these mismatches from entering the committed layout.

Only after this step are the dynamic obstacle map, congestion map, crossing records, route-length statistics, loss statistics, and route-memory records updated. Later nets then see the committed route as a physical object. Depending on the interaction type, it may act as a blockage, a legal crossing candidate, a soft penalty region, or a congestion source.

PROPEL uses two checking modes: strict checking and diagnostic checking. Strict checking is used for route commitment. In strict mode, any hard design-rule violation rejects the candidate. Diagnostic checking is used only to understand failures. In diagnostic mode, PROPEL may record which previously routed net, port region, obstacle, crossing condition, or keep-out region caused a failure. These diagnostic conflicts are not considered valid geometry. They are passed to the rip-up and reroute stage so that PROPEL can decide which failed or blocking nets should be reconsidered.

The legality flow is also connected to acceleration. PROPEL can accelerate repeated numerical calculations such as bounded VFF construction, grid-cost updates, local field lookup, neighbor ordering, and route-search expansion. These operations are suitable for C++ execution because they operate on compact arrays and routing-window data.

C++ accelerates numerical search and candidate generation, while Python remains the final authority for physical-geometry replay, legality verification, and route commitment.

\begin{figure*}[!t]
    \centering
    \includegraphics[width=\textwidth]{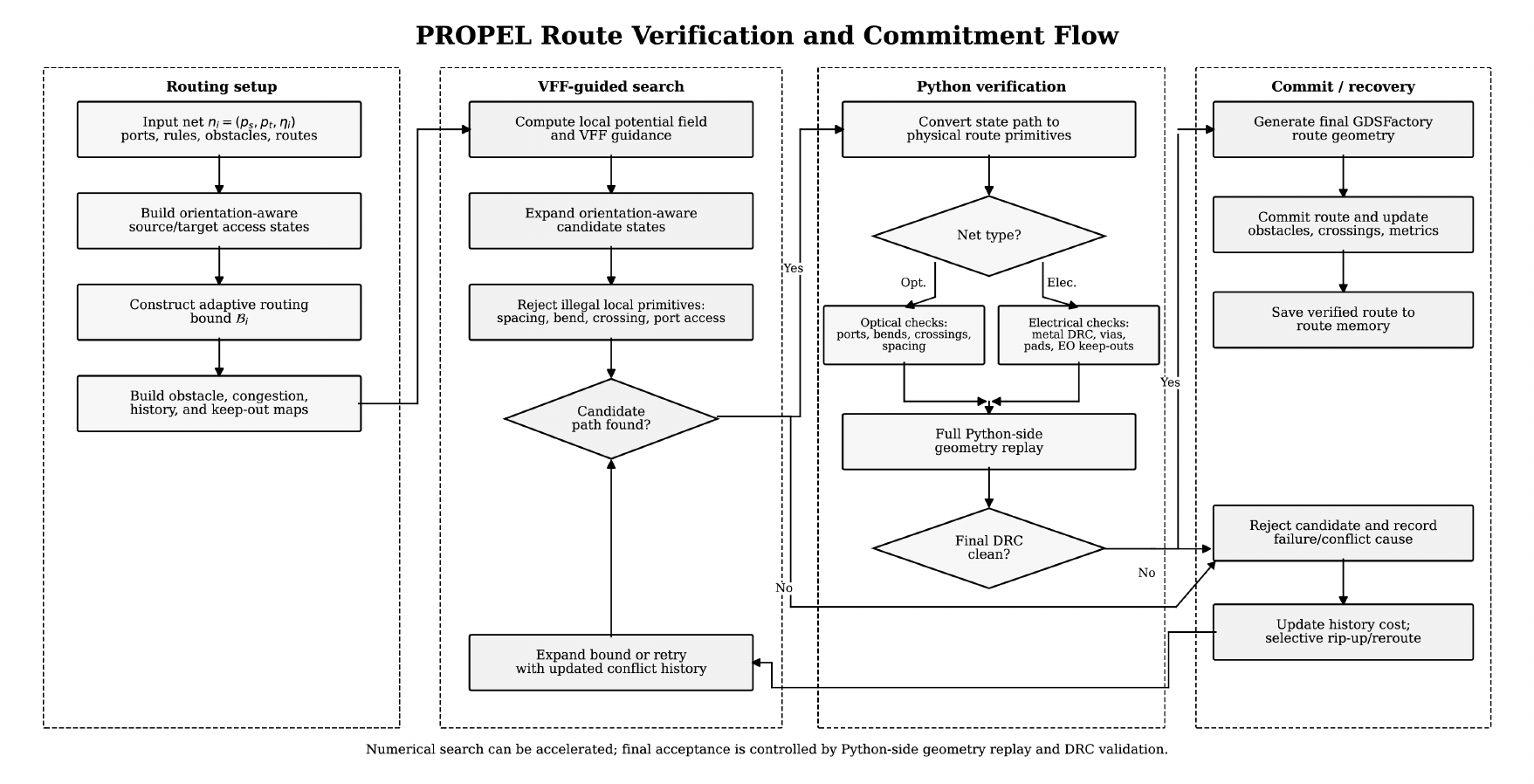}
    \caption{Route verification and commitment flow in PROPEL. The router builds orientation-aware access states, constructs a bounded VFF-guided search region, proposes a candidate route, replays the resulting physical geometry through Python-side optical or electrical DRC, and commits only verified geometry. Failed candidates update conflict history and trigger bound expansion or selective rip-up/reroute.}
    \label{fig:route_commitment_flow}
\end{figure*}

\subsection{Congestion, Rip-Up, and Incremental Routing Memory}
\label{subsec:congestion_memory_ripup}

Sequential routing is sensitive to the order in which nets are committed. A route that is legal by itself may still consume a port-access region, bend area, crossing location, or narrow corridor that is needed by a later net. This behavior is especially common in dense PIC layouts, where many waveguides compete for limited escape space near MMIs, grating-coupler arrays, switch fabrics, tensor-core blocks, and compact interferometer meshes. PROPEL therefore uses congestion history, selective rip-up, and route memory to improve convergence without repeatedly rerouting the entire layout.

PROPEL maintains separate current-congestion and historical-congestion maps. Current congestion describes routed geometry already present in the layout, while historical congestion records regions that repeatedly caused routing failures or conflicts across previous attempts. These maps are defined as
\begin{equation}
    C_{\mathrm{cong}}(q)=G(q),
    \qquad
    C_{\mathrm{hist}}(q)=H(q),
\end{equation}
where $G(q)$ is the current congestion near state $q$ and $H(q)$ is the accumulated history cost. The weights $w_g$ and $w_h$ are applied only in the transition-cost equation and are not included again in the map definitions.

When a route fails, PROPEL records the failed net and the cause of failure. Let the failed-net set after routing iteration $k$ be
\begin{equation}
    \mathcal{F}^{(k)}
    =
    \{n_i \mid n_i \text{ fails during iteration } k\}.
\end{equation}
For each failed net, PROPEL may also record a conflict set
\begin{equation}
    \mathcal{C}(n_i)
    =
    \{n_j \mid n_j \text{ blocks or conflicts with } n_i\}.
\end{equation}
The conflict set is generated from diagnostic DRC, port-access scans, crossing failures, and route-overlap analysis. These records help distinguish a net that failed because no legal path exists from a net that failed because one or two earlier routes consumed a critical region.

\begin{figure}[!t]
    \centering
    \includegraphics[width=\linewidth]{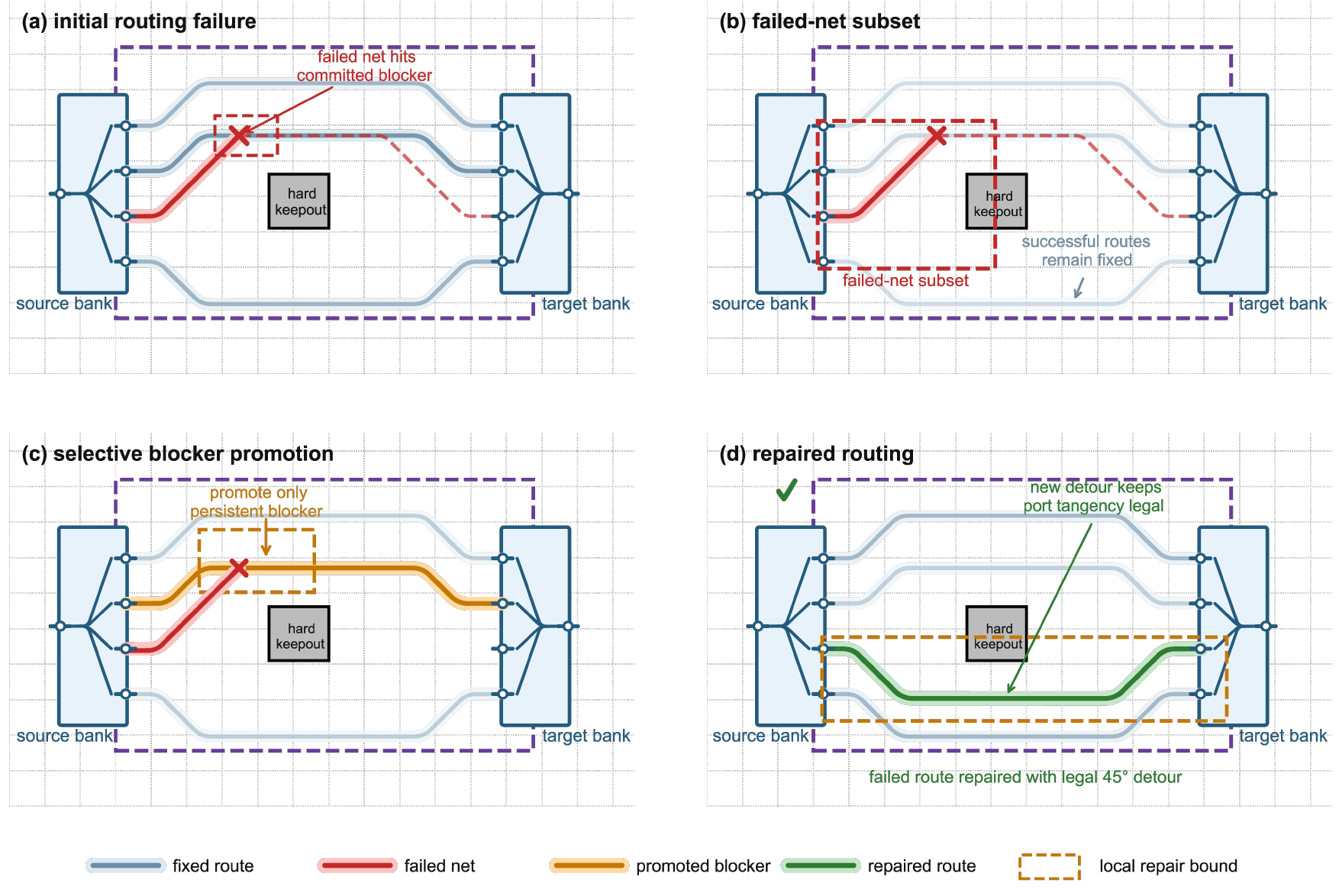}
    \caption{Selective rip-up and reroute in PROPEL. An initially failed net is first treated as the local retry target rather than forcing a global rip-up. If the failed route is repeatedly blocked by an already committed route, that route is promoted as a persistent blocker. PROPEL then reroutes only the failed and promoted subset, preserving legal routes that do not participate in the conflict.}
    \label{fig:selective_ripup_reroute}
\end{figure}

The rip-up strategy is selective, as illustrated in Figure~\ref{fig:selective_ripup_reroute}. PROPEL does not remove all routes whenever a failure occurs. Instead, it begins with the failed subset and reroutes only those nets first. If the failed route is blocked by an already committed route, PROPEL first treats the failed net as the retry target and promotes the blocking route only when diagnostic conflict records repeatedly identify it as a persistent blocker. At retry step $r$, only nets in $\mathcal{R}^{(r)}$ are removed from the committed route database:
\begin{equation}
    \Gamma
    \leftarrow
    \Gamma
    \setminus
    \{\gamma_i \mid n_i\in \mathcal{R}^{(r)}\}.
\end{equation}
All other DRC clean routes remain fixed. This prevents PROPEL from destroying stable geometry when the real problem is localized to a small number of difficult nets.

After the selected routes are removed, PROPEL rebuilds the dynamic obstacle map, the routed-geometry database, crossing records, and congestion map. The failed nets are then rerouted using the updated VFF guidance, updated history cost, current port-access reservations, and conflict-aware ordering. If a net succeeds, it is removed from the retry set. The next retry set contains only nets that still fail:
\begin{equation}
    \mathcal{R}^{(r+1)}
    =
    \{n_i\in\mathcal{R}^{(r)}
    \mid n_i \text{ still fails after rerouting}\}.
\end{equation}

If the same conflict appears repeatedly, PROPEL can selectively promote blocking nets into the retry set. For example, if a failed net $n_i$ is repeatedly blocked by an already committed route $n_j$, the relation
\begin{equation}
    n_j \rightarrow n_i
\end{equation}
indicates that $n_j$ affects the routability of $n_i$. In that case, $n_j$ may be added to the next retry set:
\begin{equation}
    \mathcal{R}^{(r)}
    \leftarrow
    \mathcal{R}^{(r)}
    \cup
    \mathcal{C}_{\mathrm{critical}}(n_i).
\end{equation}

\begin{figure}[!t]
    \centering
    \includegraphics[width=\linewidth]{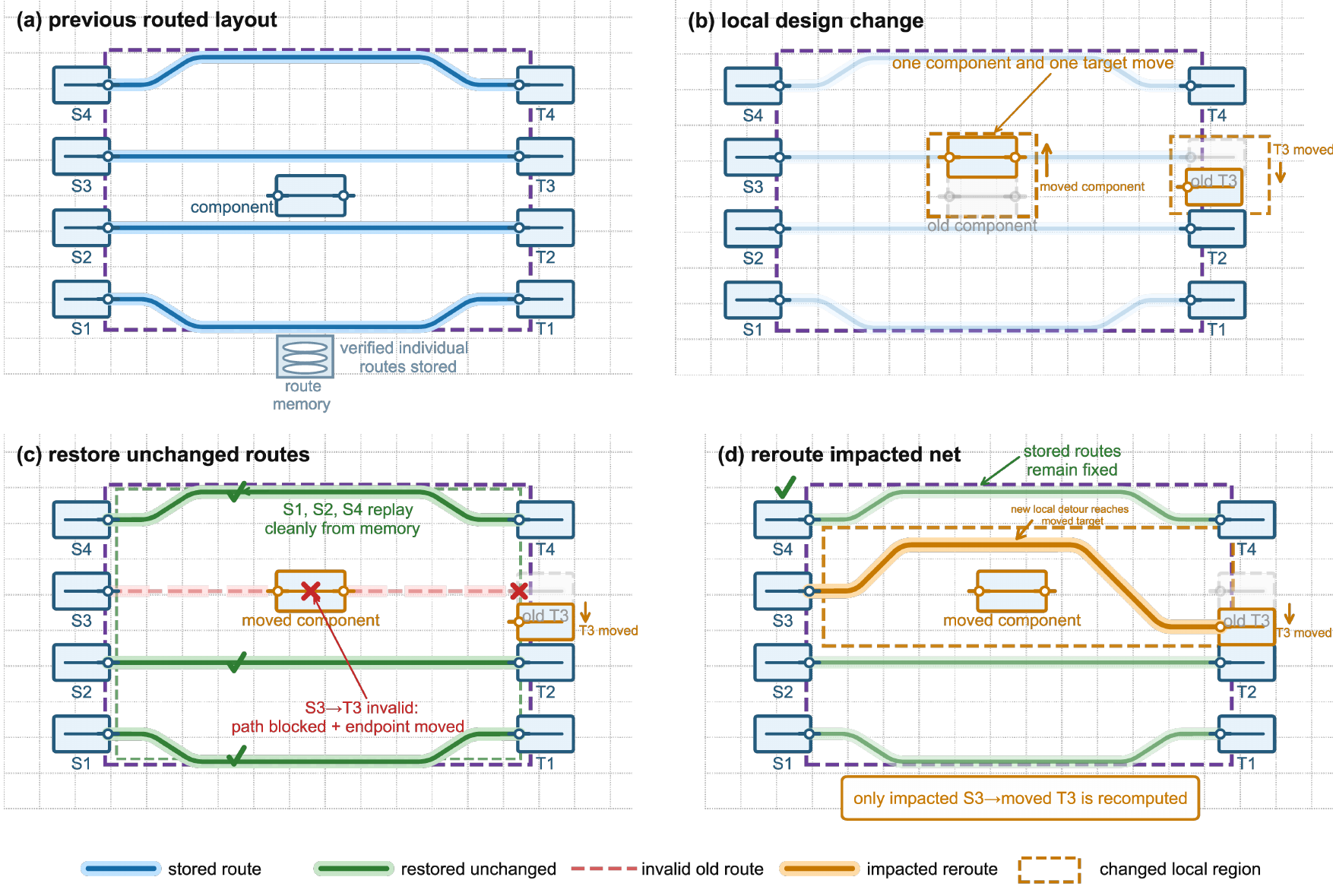}
    \caption{Incremental routing with route memory. PROPEL stores only verified routes after successful commitment. When a local design change occurs, unchanged routes are replayed through DRC before reuse, while moved, invalid, or conflict-related nets are rerouted. This allows local layout updates without discarding the entire previously routed solution.}
    \label{fig:incremental_route_memory}
\end{figure}

Route memory extends this idea across multiple design runs, as shown in Figure~\ref{fig:incremental_route_memory}. Large PIC layout is usually iterative: a designer may move a few components, change a local constraint, update a port location, or adjust a device placement. Rerouting the entire circuit from scratch after every small change is inefficient. PROPEL therefore stores verified route records after successful route commitment. When the design changes, unchanged routes are restored only after DRC replay, while moved, invalid, or conflict-related nets are placed into the rerouting queue. A saved route record contains the source and target port identifiers, saved port locations and orientations, routed centerline or geometry representation, crossing records, route length, domain metadata, and any information needed to restore the route during a later run. Failed routes, diagnostic relaxed routes, and rejected candidates are not stored as reusable memory. This keeps route memory from reintroducing invalid geometry into later layouts.

During a later routing run, PROPEL compares the current design against the saved route memory. A port is considered changed if its position moves beyond a configured tolerance or if its orientation changes:
\begin{equation}
    \left\|
    \mathbf{x}_p^{\mathrm{new}}
    -
    \mathbf{x}_p^{\mathrm{mem}}
    \right\|_2
    >
    \epsilon_{\mathrm{move}}
    \quad
    \text{or}
    \quad
    \theta_p^{\mathrm{new}}
    \neq
    \theta_p^{\mathrm{mem}} .
\end{equation}
Any net connected to a changed port is marked as impacted:
\begin{equation}
    \mathcal{N}_{\mathrm{impact}}
    =
    \{n_i \mid p_s(n_i) \text{ or } p_t(n_i) \text{ changed}\}.
\end{equation}

Unimpacted saved routes become candidates for restoration. A route can become invalid even if its own endpoints did not move. For example, another component may move into its corridor, a design rule may become stricter, or a newly inserted keep-out region may overlap the old path. If replay succeeds, the route is restored. If replay fails, the corresponding net is added to the rerouting queue.

The initial incremental rerouting queue is
\begin{equation}
    \mathcal{Q}_{\mathrm{reroute}}
    =
    \mathcal{N}_{\mathrm{impact}}
    \cup
    \mathcal{N}_{\mathrm{invalid}},
\end{equation}
where $\mathcal{N}_{\mathrm{invalid}}$ contains saved routes that failed restoration. 

The combined memory-aware rip-up and reroute process is summarized in Algorithm~\ref{alg:memory_aware_ripup}. The algorithm combines two levels of reuse. Across design runs, PROPEL restores valid routes from memory. Within a routing run, PROPEL selectively rips up failed or blocking nets instead of restarting the full design.

\begin{algorithm}[!t]
\caption{Memory-Aware Incremental Routing with Selective Rip-Up}
\label{alg:memory_aware_ripup}
\begin{algorithmic}[1]
\STATE \textbf{Input:} current design, saved route memory, movement tolerance $\epsilon_{\mathrm{move}}$, retry limit $R_{\max}$
\STATE Load saved route-memory records
\STATE Compare current ports with saved port locations and orientations
\STATE Mark moved ports and build impacted net set $\mathcal{N}_{\mathrm{impact}}$
\FOR{each saved net not in $\mathcal{N}_{\mathrm{impact}}$}
    \STATE Replay saved route through Python-side DRC
    \IF{replay succeeds}
        \STATE Restore route, crossing records, occupancy, and metrics
    \ELSE
        \STATE Add net to $\mathcal{N}_{\mathrm{invalid}}$
    \ENDIF
\ENDFOR
\STATE Initialize rerouting queue:
\[
    \mathcal{Q}_{\mathrm{reroute}}
    =
    \mathcal{N}_{\mathrm{impact}}
    \cup
    \mathcal{N}_{\mathrm{invalid}}
\]
\STATE Route nets in $\mathcal{Q}_{\mathrm{reroute}}$
\STATE Let $\mathcal{F}$ be the set of failed nets
\STATE $\mathcal{R}^{(0)} \leftarrow \mathcal{F}$
\FOR{$r=0$ to $R_{\max}-1$}
    \IF{$\mathcal{R}^{(r)}=\emptyset$}
        \STATE \textbf{break}
    \ENDIF
    \STATE Rip up only nets in $\mathcal{R}^{(r)}$
    \STATE Rebuild dynamic obstacles, occupancy, crossing records, and congestion maps
    \STATE Increase history cost in contested regions
    \STATE Reroute nets in $\mathcal{R}^{(r)}$ using strict DRC
    \STATE Use diagnostic conflicts to identify persistent blockers
    \STATE Promote only critical blockers into the retry set
    \STATE $\mathcal{R}^{(r+1)} \leftarrow$ nets that still fail plus promoted blockers
\ENDFOR
\STATE Save newly verified committed routes back to route memory
\end{algorithmic}
\end{algorithm}

\clearpage
\section{PROPEL Operating Modes}
\label{sec:PROPEL_operating_modes}

\subsection{Passive Optical Routing}
\label{subsec:passive_optical_mode}

Passive optical routing is the first operating mode of PROPEL. In this mode, the input is a fixed optical netlist, and every routed connection is implemented as an optical waveguide between two photonic ports.

\begin{figure}[!t]
    \centering
    \includegraphics[width=\linewidth]{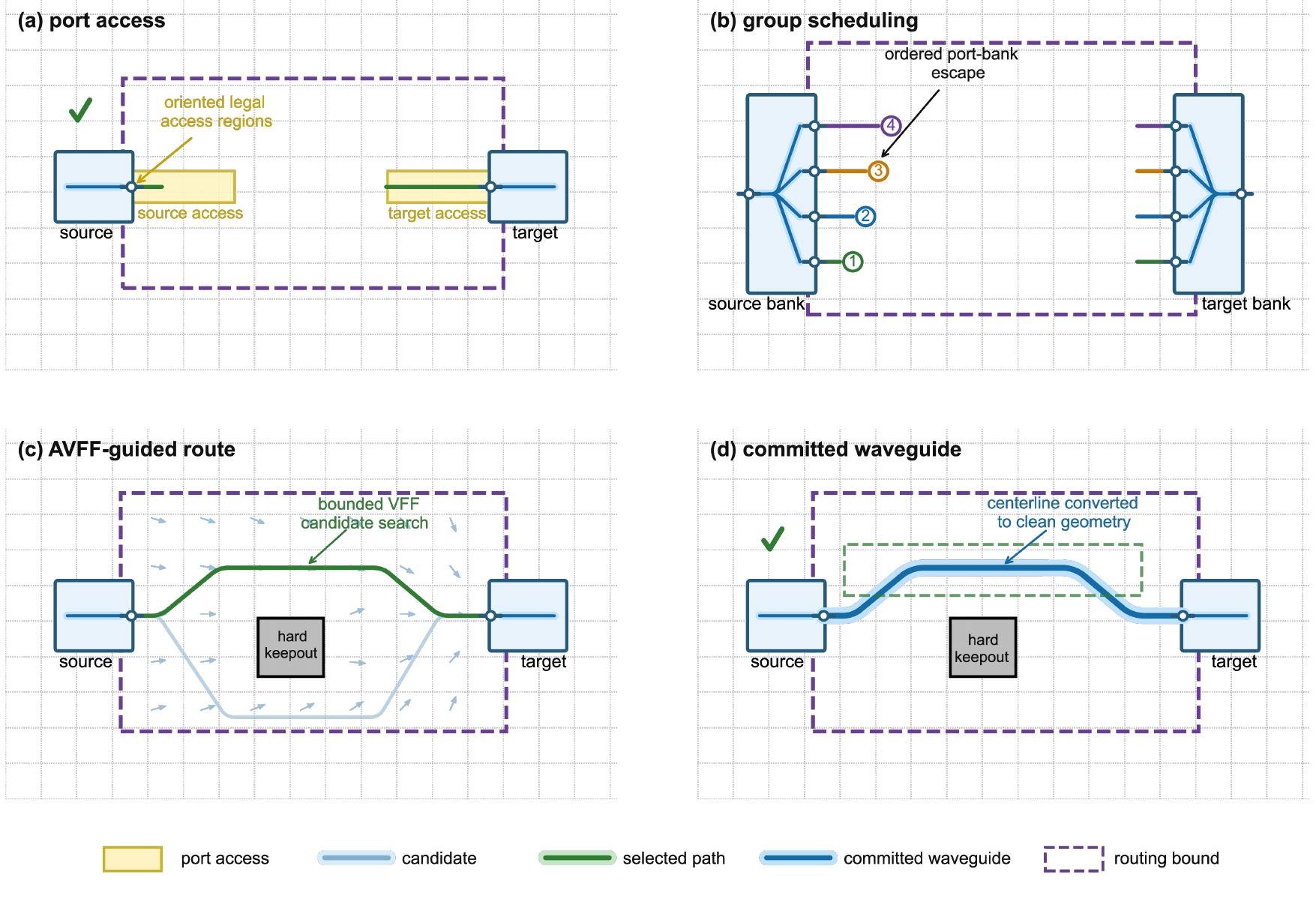}
    \caption{Passive optical routing flow in PROPEL. The router begins from oriented source and target ports, constructs legal access regions, and uses the bounded vector-flow field to guide a bend-aware route through the layout. Candidate routes are checked against hard obstacles, spacing rules, and terminal-access legality, and only the verified route is committed as a physical optical waveguide.}
    \label{fig:passive_optical_flow}
\end{figure}

The passive routing flow is shown in Figure~\ref{fig:passive_optical_flow}. PROPEL first extracts the oriented source and target ports, constructs legal port-access regions, and uses the bounded VFF field to guide the route through the available layout space. The selected route is then checked against hard obstacles, waveguide-spacing rules, bend legality, and port-entry constraints before being committed as a physical waveguide. The detailed VFF search, legality checking, and route commitment steps are the same kernel operations described in Section~\ref{sec:PROPEL_kernel}. This section explains how those kernel operations are specialized for passive optical waveguide routing.

Port access is the first passive-routing step. A photonic port is not just a point; it is an oriented optical interface with a layer, width, cross-section, and required approach direction. PROPEL therefore reserves an access region in front of each optical port before detailed routing begins. This access region gives the waveguide enough space to leave or enter the component without being immediately blocked by another route.

For an optical port
\begin{equation}
    p=(x_p,y_p,\theta_p,\ell_p,w_p,\tau_p),
\end{equation}
PROPEL defines the outward unit vector
\begin{equation}
    \hat{\mathbf{o}}_p =
    (\cos\theta_p,\sin\theta_p).
\end{equation}
The initial access region is constructed along this direction:
\begin{equation}
    \mathcal{A}(p)
    =
    \left\{
    \mathbf{x}_p+\lambda\hat{\mathbf{o}}_p
    \mid
    0\leq \lambda \leq L_{\mathrm{acc}}(p)
    \right\},
\end{equation}
where $L_{\mathrm{acc}}(p)$ is the reserved access length. This does not mean the final route must remain perfectly straight over the entire access length. Rather, the region protects the terminal escape direction so that the final route can connect to the port with a legal tangent. For ports located near or inside component bounding boxes, PROPEL propagates the usable routing point outward along the port orientation until the route is outside the component footprint and any required clearance region.

Dense multiport devices require additional access handling. Components such as MMIs, splitter arrays, grating-coupler banks, switch blocks, and interferometer stages can have many ports on the same side. If all nets are routed independently, an early route can occupy the natural escape corridor of a later route. PROPEL therefore groups ports by component instance and port orientation:
\begin{equation}
    G(m,\theta)
    =
    \{p\in\mathcal{P}(m)\mid \theta_p=\theta\}.
\end{equation}
Ports in the same group are sorted by their physical order along the device side. This gives PROPEL a local ordering model for port escape.

For a group with many ports, PROPEL can assign staggered or adaptive access lengths. Inner ports in a dense bank often need longer protected access because they are more likely to be blocked by neighboring routes. Outer ports may require shorter access because they have more available escape directions. A simple access policy is
\begin{equation}
    L_{\mathrm{acc}}(p_i)
    =
    \min
    \left(
    L_{\max},
    \max
    \left(
    L_{\min},
    L_{\mathrm{base}} + \Delta L_i
    \right)
    \right),
\end{equation}
where $L_{\min}$ and $L_{\max}$ clamp the reservation length, $L_{\mathrm{base}}$ is the nominal access distance, and $\Delta L_i$ is an adaptive offset based on port position, local density, bend radius, or previous failure history. This compact expression replaces a fixed access formula and allows the same mechanism to work for both sparse and compact layouts.

\begin{figure*}[!t]
    \centering
    \includegraphics[width=\textwidth]{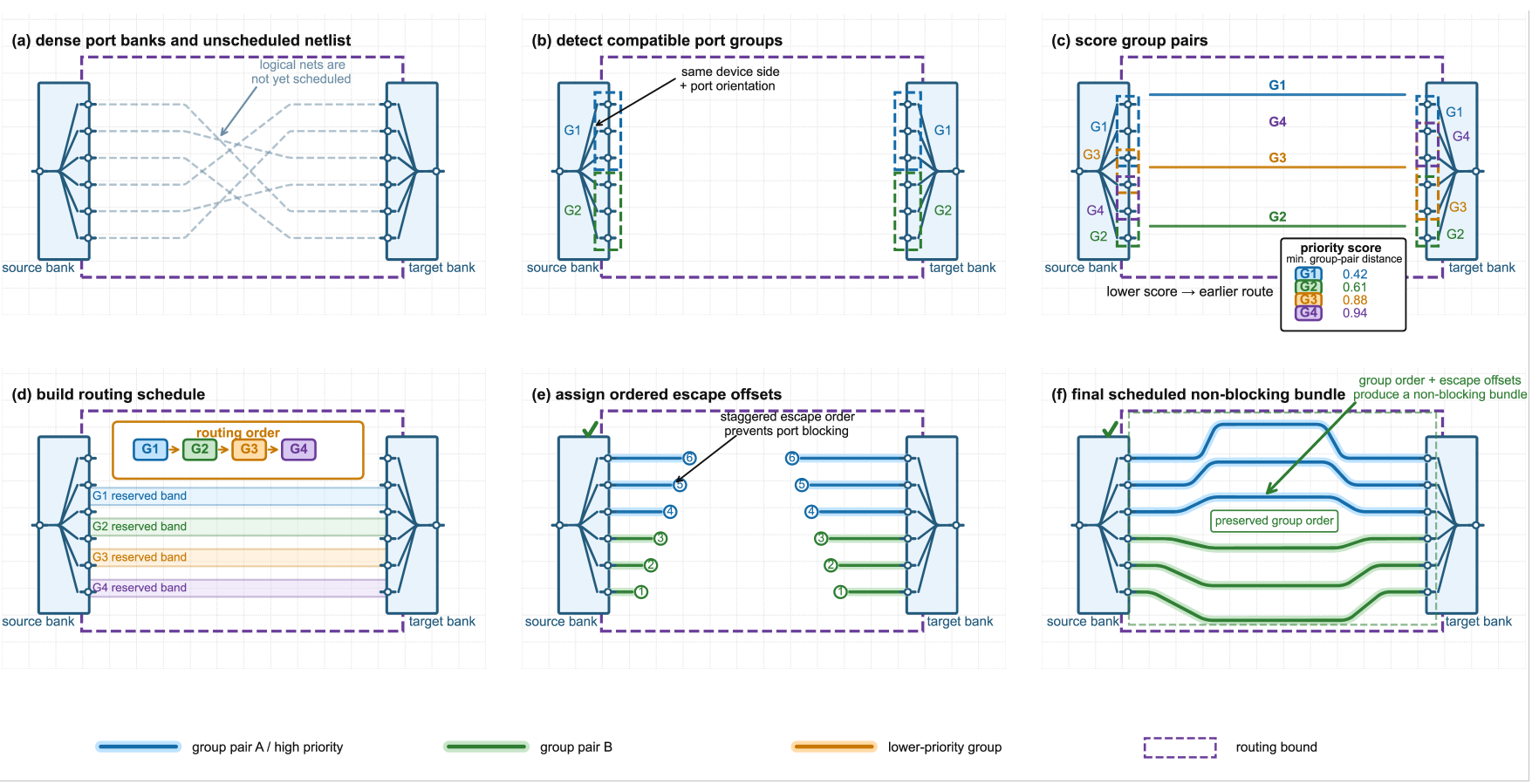}
    \caption{Group ordering and scheduling for dense passive optical port banks. PROPEL begins from an unscheduled logical netlist, detects compatible source--target port groups with matching device side and port orientation, scores candidate group pairs, and builds an ordered routing schedule. The ordered escape offsets preserve local port order and produce a final non-blocking bundle that remains compatible with downstream bend-aware waveguide routing.}
    \label{fig:passive_port_access_grouping}
\end{figure*}

Group information also affects routing order. As illustrated in Figure~\ref{fig:passive_port_access_grouping}, PROPEL first detects compatible source-target port groups from dense port banks, scores the group pairs using physical proximity and port compatibility, and then builds an ordered routing schedule before detailed routing begins. This prevents an unscheduled set of logical nets from immediately blocking the natural escape directions of nearby ports. PROPEL forms net groups from nets that share nearby source ports, nearby target ports, or common device-side access regions. A group priority can be estimated using the shortest source-to-target distance among nets in the group:
\begin{equation}
    S(g_k)
    =
    \min_{n_i\in g_k}
    \left\|
    \mathbf{x}_{p_s(n_i)}
    -
    \mathbf{x}_{p_t(n_i)}
    \right\|_2 .
\end{equation}
Groups with smaller values are usually routed earlier because short, local, and dense connections often have fewer feasible detours. Within a group, nets are ordered using spatial coherence so that adjacent ports escape in a staggered and ordered way rather than crossing or blocking each other immediately. The final scheduled bundle in Figure~\ref{fig:passive_port_access_grouping}(f) shows the intended behavior: the group order and escape offsets preserve the local port order while allowing the routes to separate into non-blocking waveguide corridors.

After port access and scheduling are initialized, each optical net is routed using the bounded VFF search described earlier. 

\subsection{Length, Optical-Path, Group-Delay, and Process-Map-Aware Geometry Generation}
\label{subsec:matching_process_refinement}

Some passive optical layouts require more than DRC-clean connectivity. Interferometers, coherent photonic paths, delay-sensitive circuits, programmable meshes, and phase-sensitive optical systems may require selected routes to satisfy matching constraints after routing. A route can be legal, short, and low-loss, but still be unacceptable if two nominally matched paths accumulate different geometric length, optical phase, or group delay. PROPEL therefore treats matching as a DRC-aware refinement step applied after the initial routed geometry is available.

PROPEL supports three related matching modes: geometric length matching, optical-path length (OPL) matching, and group-delay matching. Geometric matching uses the physical centerline length of the routed waveguide. For a routed net $n_i$, the geometric length is
\begin{equation}
    L_i =
    \int_{\gamma_i} ds .
\end{equation}
For a matching group $\mathcal{M}_k$, the target length is usually selected as the longest routed path in the group unless the user provides an explicit target. The remaining mismatch is then corrected by adding controlled delay geometry to shorter paths.

OPL matching is more physically meaningful when waveguides do not all have the same effective index. This can happen when routes use different waveguide widths, different layers, different local process conditions, or process-aware compensation. The optical path length is
\begin{equation}
    \mathrm{OPL}_i =
    \int_{\gamma_i} n_{\mathrm{eff}}(s)\,ds ,
\end{equation}
where $n_{\mathrm{eff}}(s)$ is the effective index along the routed path. If $n_{\mathrm{eff}}$ is constant, optical-path matching reduces to a scaled version of geometric length matching. If $n_{\mathrm{eff}}$ varies spatially, two routes with the same physical length can still accumulate different optical phase.

Group-delay matching is used when timing or broadband behavior is more important than phase at a single wavelength. The group delay of a routed net is
\begin{equation}
    \tau_i =
    \frac{1}{c}
    \int_{\gamma_i} n_g(s)\,ds ,
\end{equation}
where $n_g(s)$ is the group index and $c$ is the speed of light in vacuum. If all routes use the same constant group index, group-delay matching again reduces to a scaled length-matching problem. If group index varies with waveguide geometry, wavelength, or process condition, PROPEL evaluates the delay using the local group-index profile.

\begin{figure}[!t]
    \centering
    \includegraphics[width=\linewidth]{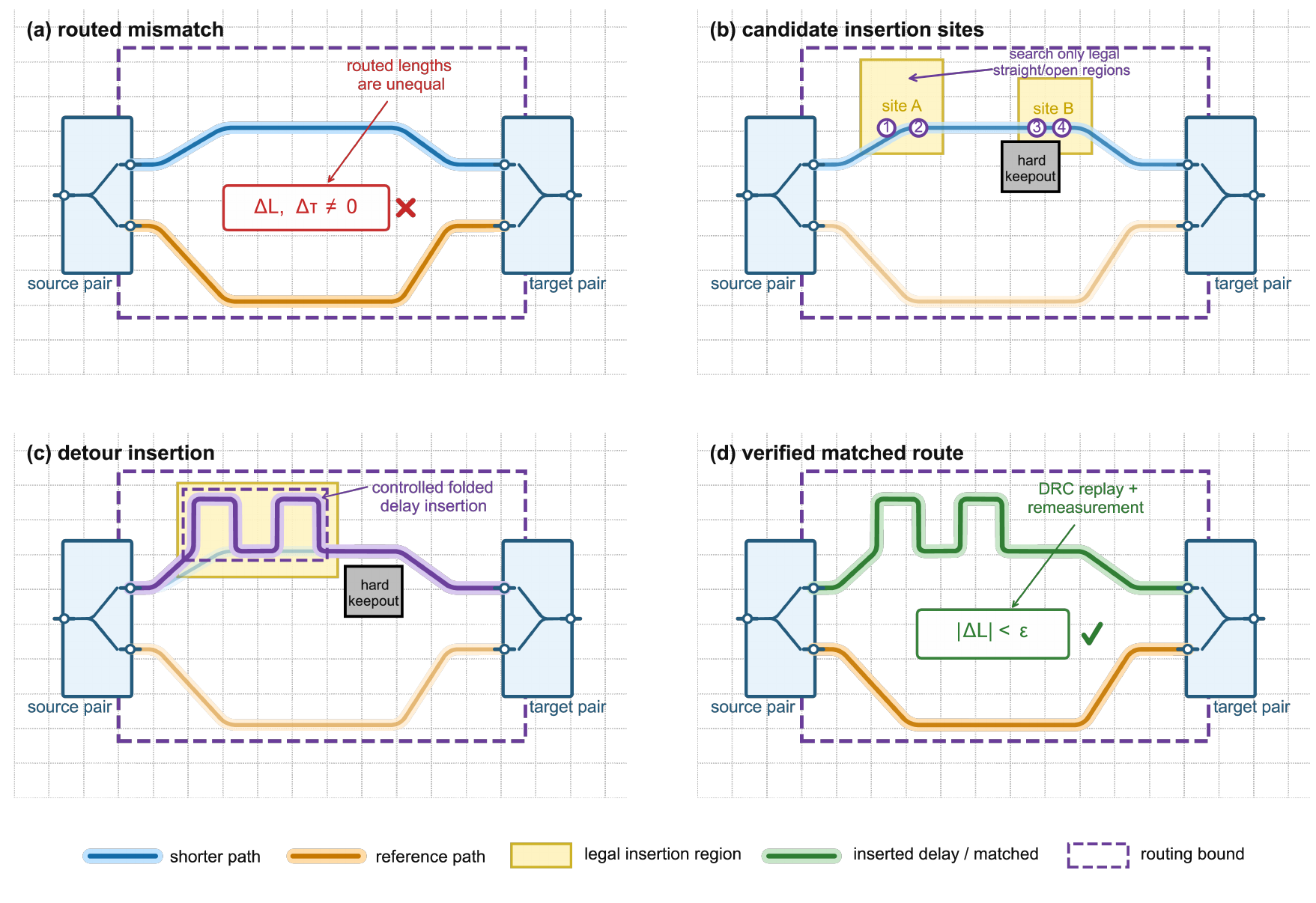}
    \caption{DRC-aware matching refinement in PROPEL. After initial routing, PROPEL measures the mismatch between matched paths, searches for legal insertion regions, and inserts controlled delay geometry only where bend radius, spacing, obstacle, and port-access rules remain valid. The corrected route is then replayed through DRC and remeasured before final commitment.}
    \label{fig:matching_refinement}
\end{figure}

The matching procedure is applied after the initial route commitment because the final geometry determines where tuning structures can be inserted. Figure~\ref{fig:matching_refinement} illustrates this refinement flow. Before routing, the available space for meanders, folded detours, compact spirals, or other delay structures is unknown. After routing, PROPEL inspects the committed layout, identifies legal insertion regions, evaluates how much length, optical path, or delay must be added, and checks whether candidate corrections remain DRC-clean before the corrected route is recommitted.

PROPEL first evaluates each matching group and computes the required correction for every route below the group target. The correction quantity depends on the selected matching mode. In geometric mode, PROPEL adds physical length. In optical-path mode, it adds 
OPL. In group-delay mode, it adds delay. These modes share the same geometric insertion machinery, but they evaluate the remaining error using different physical quantities.

Candidate delay structures are selected from legal regions of the committed route. Typical candidates include local meanders, side detours, folded delay sections, or compact spirals when more delay is required and sufficient open area exists. Each candidate is parameterized by its added length, footprint, bend count, spacing demand, and distance from protected features such as ports, crossings, bends, and dense routing corridors.

A candidate correction replaces a portion of the original route with a longer but still connected geometry. If $\gamma_{\mathrm{old}}$ is the original segment and $\gamma_{\mathrm{new}}$ is the corrected segment, the added geometric length is
\begin{equation}
    \Delta L_{\mathrm{cand}}
    =
    L(\gamma_{\mathrm{new}})
    -
    L(\gamma_{\mathrm{old}}).
\end{equation}
For optical-path or group-delay matching, PROPEL evaluates the corresponding added OPL or delay using the local effective-index or group-index profile. This lets the same geometry insertion stage support multiple physical matching objectives.

A candidate is accepted only if it remains design-rule-clean:
\begin{equation}
    \operatorname{Legal}(\gamma_{\mathrm{new}})=1,
\end{equation}
where $\operatorname{Legal}(\gamma)=1$ if and only if the generated physical geometry satisfies all applicable hard design rules.
PROPEL treats the added matching geometry as part of the routed layout and validates it using the same DRC model as the original route; it also uses the same rip-up method as the original kernel.

The matching objective balances accuracy and layout quality:
\begin{equation}
    C_{\mathrm{match}}
    =
    \lambda_m
    \left|
    \Delta_{\mathrm{target}}
    -
    \Delta_{\mathrm{cand}}
    \right|
    +
    \lambda_a A_{\mathrm{cand}}
    +
    \lambda_b B_{\mathrm{cand}}
    +
    \lambda_c C_{\mathrm{clearance}} .
\end{equation}
Here $\Delta_{\mathrm{target}}$ is the required correction in the selected matching mode, $\Delta_{\mathrm{cand}}$ is the correction supplied by the candidate, $A_{\mathrm{cand}}$ is the consumed area, $B_{\mathrm{cand}}$ is the bend contribution, and $C_{\mathrm{clearance}}$ penalizes candidates that are too close to obstacles or other routes. The first term improves matching accuracy, while the remaining terms discourage fragile or unnecessarily large corrections.

\begin{figure*}[!t]
    \centering
    \includegraphics[width=\textwidth]{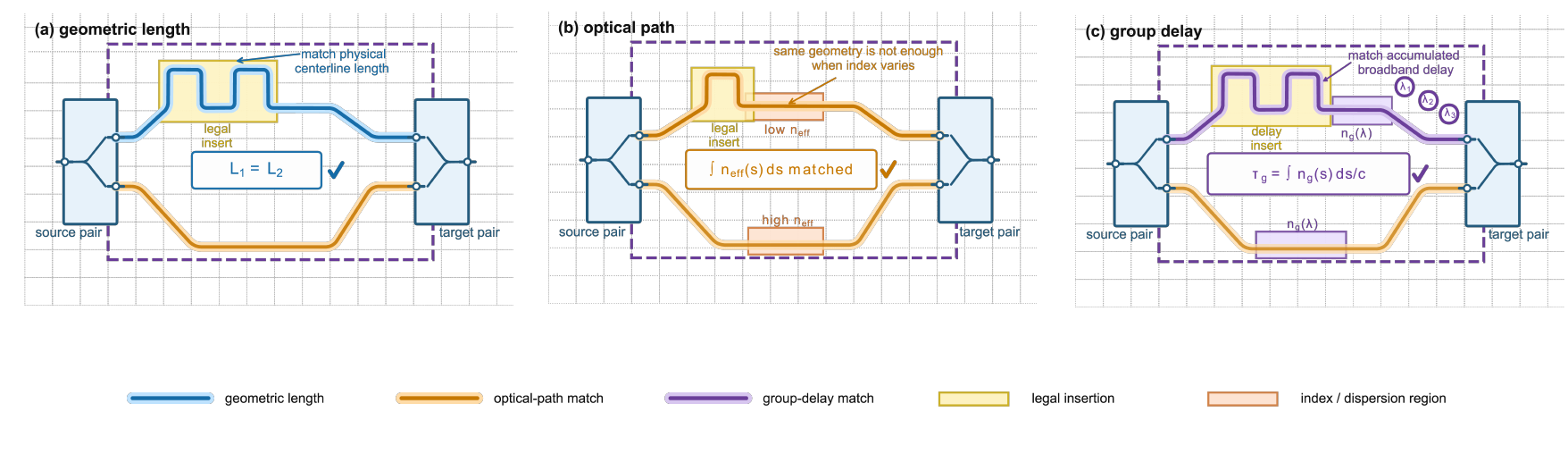}
    \caption{Matching modes supported by PROPEL. Geometric matching equalizes the physical centerline length of routed waveguides. Optical-path matching equalizes the accumulated effective-index-weighted path length, $\int n_{\mathrm{eff}}(s)\,ds$. Group-delay matching equalizes the accumulated group delay, $\int n_g(s)\,ds/c$, which is more appropriate for broadband or timing-sensitive photonic paths.}
    \label{fig:matching_modes}
\end{figure*}

After a correction is committed, PROPEL updates the routed geometry, dynamic obstacle map, route length, optical path length, group delay, and matching error. The route is then remeasured. This is important because matching structures consume area and may affect later matching operations in the same group. The refinement, therefore, proceeds as an iterative layout-aware correction rather than as an isolated formula applied to each route independently. Figure~\ref{fig:matching_modes} summarizes the three matching quantities supported by PROPEL: physical centerline length, accumulated optical path length, and accumulated group delay.

\begin{figure}[!t]
    \centering
    \includegraphics[width=\linewidth]{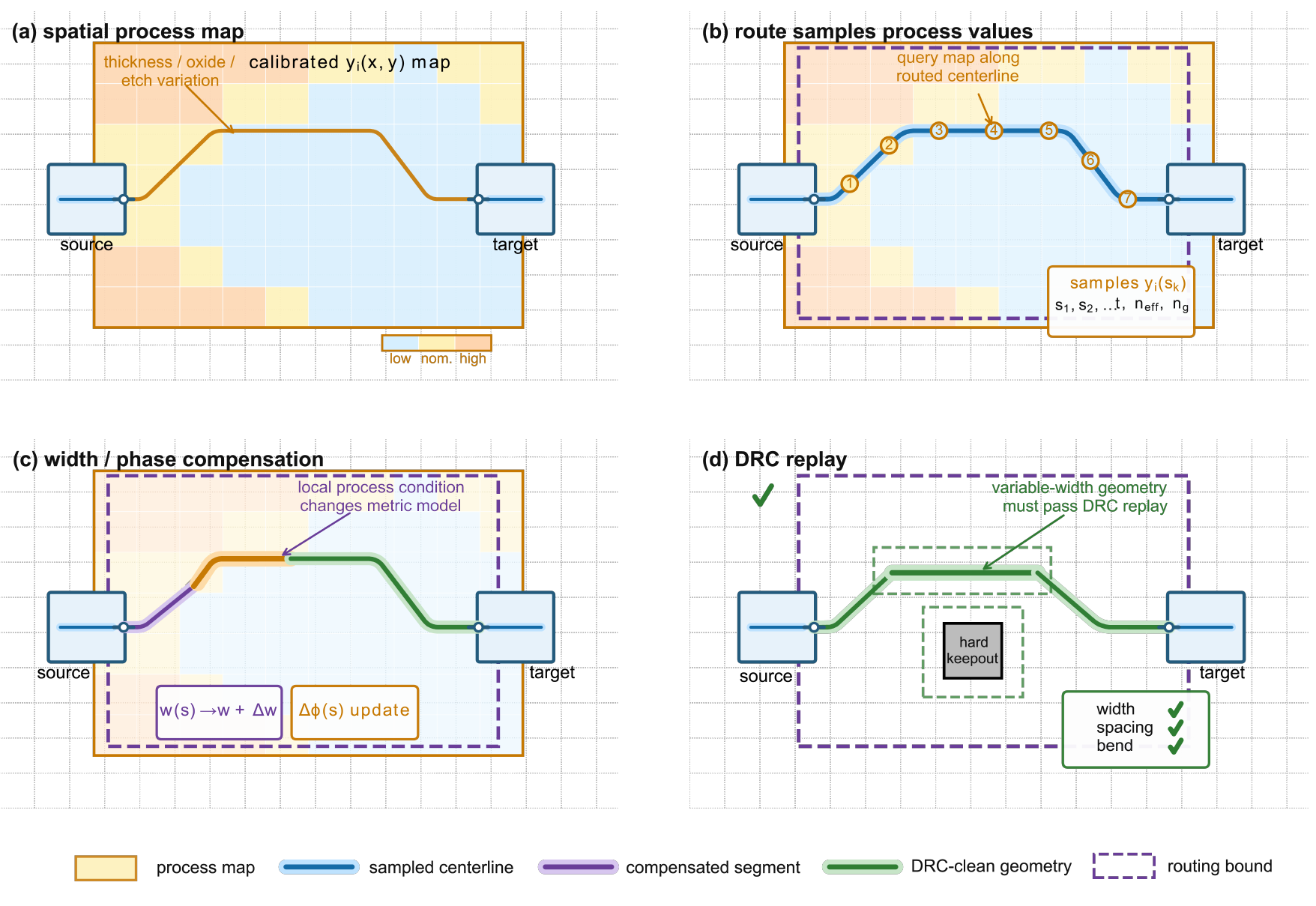}
    \caption{Process-aware routing hook in PROPEL. A calibrated spatial process map can encode local fabrication variation such as film thickness, oxide thickness, or etch-depth change. During replay, PROPEL samples the process map along the routed centerline, updates the local waveguide or phase model, applies geometry compensation when enabled, and then checks the compensated route against DRC before commitment.}
    \label{fig:process_aware_hook}
\end{figure}

Process-aware routing extends this refinement model by allowing routed waveguide geometry and optical metrics to depend on local process conditions. Figure~\ref{fig:process_aware_hook} shows how PROPEL can use a calibrated spatial process map during route generation and replay. In the default routing mode, a route is extruded with a fixed waveguide width and layer. This is appropriate for standard benchmarks and for designs where the process is assumed uniform. However, fabricated PICs can exhibit spatial variation in oxide thickness, film thickness, etch depth, sidewall profile, or other process-dependent quantities. These variations can change effective index, group index, propagation loss, and phase accumulation.

PROPEL represents such information through a spatial process map,
\begin{equation}
    t(x,y):\Omega\rightarrow\mathbb{R},
\end{equation}
where $t(x,y)$ is a process variable over the chip. Along a routed path, the local process value becomes
\begin{equation}
    t_i(s)=t(\gamma_i(s)).
\end{equation}
The final waveguide width can then be selected from a process-aware policy:
\begin{equation}
    w_i(s)=f_w(t_i(s)).
\end{equation}
This policy may be a binary rule, a continuous compensation function, or a user-provided lookup model. For example, a route in one process region may use the nominal width, while a route in another region may use a compensated width to reduce phase or delay error.

Because abrupt waveguide-width changes can create mode mismatch and reflection, Width changes are applied through the route builder, so transitions remain taper-compatible. The width profile can be smoothed or connected through taper-compatible transitions before final geometry construction. After the compensated route geometry is generated, PROPEL rechecks DRC using the actual waveguide footprint, because a variable-width route may consume more or less space than the original centerline route.

Process-aware routing also changes optical-path and group-delay evaluation. If width or local process condition varies along the route, then the effective index and group index become functions of position:
\begin{equation}
    n_{\mathrm{eff}}(s)
    =
    n_{\mathrm{eff}}
    \left(
    w_i(s),t_i(s),\lambda
    \right),
    \qquad
    n_g(s)
    =
    n_g
    \left(
    w_i(s),t_i(s),\lambda
    \right).
\end{equation}
The same OPL and group-delay equations above can then be evaluated using these spatially varying quantities. This is why matching and process-aware routing are described together: the physical metric being matched may depend on the process-aware geometry that PROPEL generates.

\subsection{Active Optical--Electrical Routing}
\label{subsec:active_optical_electrical_mode}

Active optical--electrical routing is the second operating mode of PROPEL. In this mode, the layout contains both optical waveguide nets and electrical metal nets. Optical nets connect photonic ports through waveguides, while electrical nets connect active-device terminals to pads, control lines, drivers, monitors, ground, bias sources, or readout circuitry. The goal is to produce a combined layout in which the optical routes remain DRC-clean and optically valid, while the electrical routes satisfy metal DRC and photonic-aware interaction constraints.

\begin{figure}[!t]
    \centering
    \includegraphics[width=\linewidth]{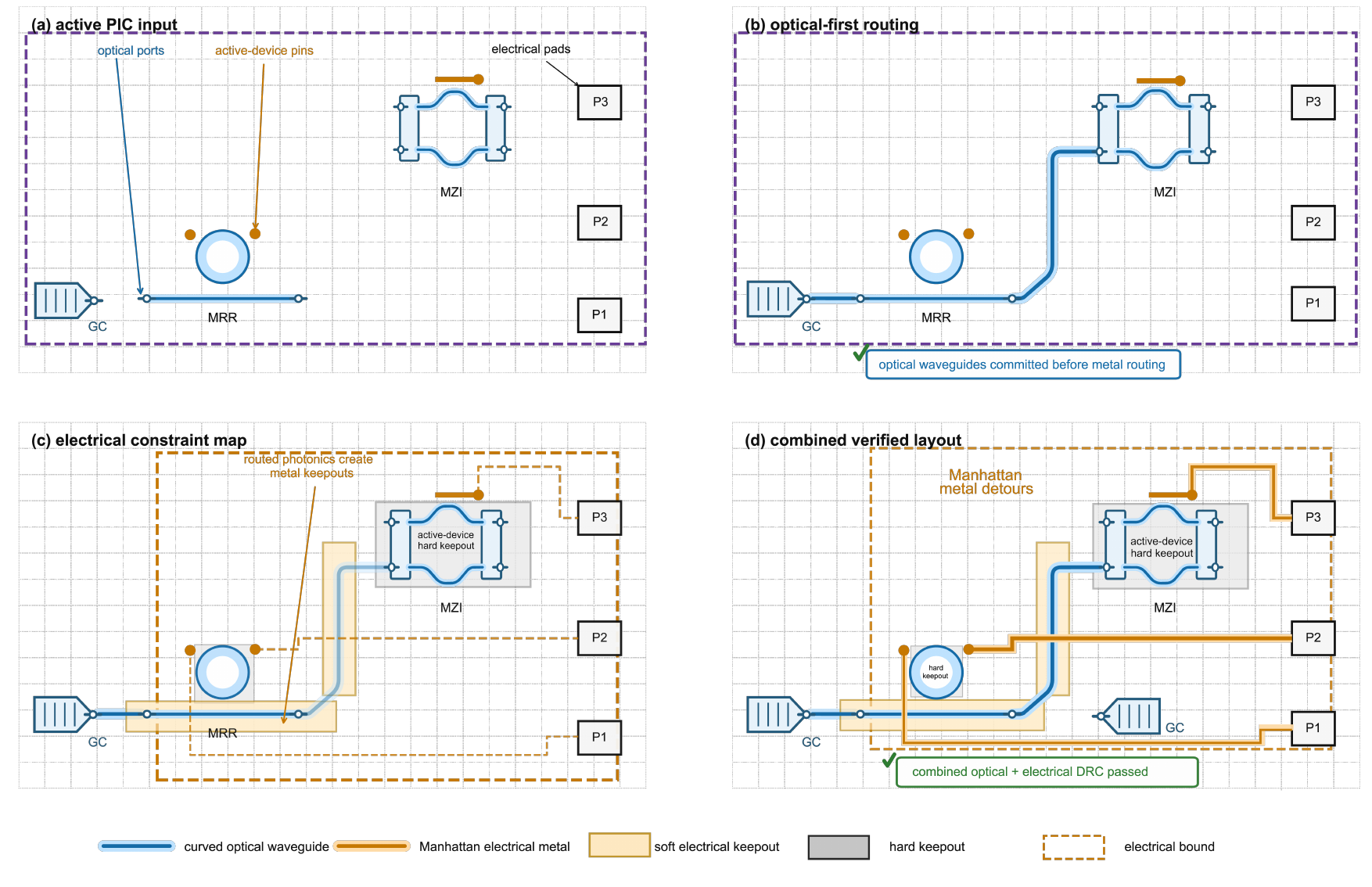}
    \caption{Active optical--electrical routing flow in PROPEL. The active PIC input contains optical ports, electrical pins, pads, and active photonic devices. PROPEL first routes and commits the optical waveguides, then builds an electrical constraint map from routed optical geometry, active-device hard keep-outs, soft electrical keep-outs, and pad-access requirements. The final layout is accepted only after the combined optical and electrical DRC checks pass.}
    \label{fig:active_optical_electrical_flow}
\end{figure}

PROPEL uses an optical-first flow for active PICs, as illustrated in Figure~\ref{fig:active_optical_electrical_flow}. The reason is that optical waveguides are usually more geometrically constrained than electrical metal routes. Waveguides require a valid bend radius, a smooth terminal approach, crossing-footprint clearance, and optical spacing. If the optical layer is distorted late in the design flow to accommodate electrical routing, the layout may lose optical validity or accumulate unnecessary loss. Therefore, PROPEL first routes and commits the optical waveguides using the passive optical routing flow. The committed optical layout is then converted into an electrical constraint map containing photonic device keep-outs, routed-waveguide keep-outs, pad-access regions, and user-defined hard or soft electrical--optical interaction regions.
Once committed, these optical routes become part of the active layout context for electrical routing.

\begin{figure*}[!t]
    \centering
    \includegraphics[width=\textwidth]{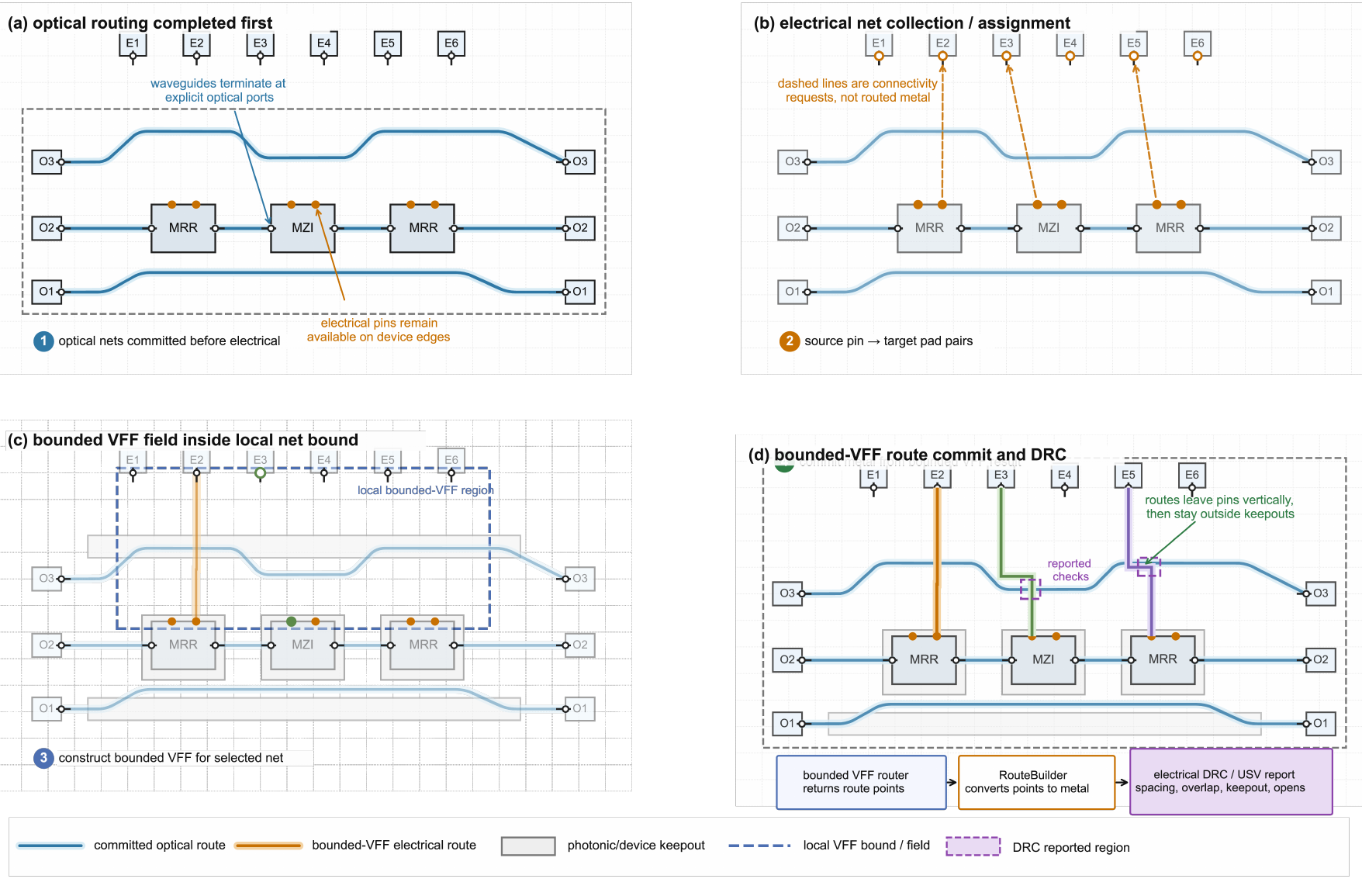}
    \caption{Bounded-VFF electrical routing after optical route commitment. PROPEL first commits the optical waveguides and keeps the active-device electrical pins available. Electrical source--target requests are then collected as pin-to-pad or pin-to-pin routing tasks. For each selected electrical net, PROPEL constructs a local bounded VFF field around the routed photonic layout, converts the returned route points into metal geometry, and performs electrical DRC and user-specified electrical--optical violation checks before committing the route.}
    \label{fig:active_bounded_vff_electrical}
\end{figure*}

After optical routing, PROPEL builds the electrical routing context. This process is detailed in Figure~\ref{fig:active_bounded_vff_electrical}. PROPEL collects the source--target electrical pin or pad pairs, constructs a local bounded VFF region for each selected electrical net, and routes metal while treating routed waveguides, active devices, pads, and photonic keep-outs as part of the electrical constraint map. The routed optical geometry is therefore not ignored during electrical routing. It can act as a hard obstacle, a soft penalty region, or an interaction region depending on the configured rule.
The electrical netlist is represented as
\begin{equation}
    n_i^{\mathrm{elec}}
    =
    (p_s,p_t,\eta_i^{\mathrm{elec}}),
\end{equation}
where $p_s$ and $p_t$ are electrical pins or pads, and $\eta_i^{\mathrm{elec}}$ stores the electrical rule class, metal width, spacing rule, allowed metal layers, via policy, pad-access requirement, and photonic-interaction constraints. Electrical nets may represent heater-control lines, modulator electrodes, photodetector readout lines, microring tuning lines, phase-shifter controls, ground connections, bias connections, or monitor connections.

PROPEL can classify electrical nets by role before routing:
\begin{equation}
    \mathcal{N}_{\mathrm{elec}}
    =
    \mathcal{N}_{\mathrm{ctrl}}
    \cup
    \mathcal{N}_{\mathrm{bias}}
    \cup
    \mathcal{N}_{\mathrm{gnd}}
    \cup
    \mathcal{N}_{\mathrm{readout}}
    \cup
    \mathcal{N}_{\mathrm{pad}} .
\end{equation}
This classification is useful because different electrical net types may require different widths, spacing values, layers, or via policies. For example, a heater line may use a different routing width and keep-out rule than a detector readout net or a ground connection. The classification also allows PROPEL to prioritize nets with difficult pin access, dense pad breakout, or repeated failure history.

Electrical pin and pad access are handled before detailed metal routing. Active-device pins may lie close to optical ports, waveguides, heaters, resonators, or sensitive regions. PROPEL therefore creates legal access regions for electrical pins so that metal routes connect to the correct terminal without shorting nearby pins or crossing forbidden device regions. Pad access is handled similarly. A pad route must enter through a legal pad-breakout region and must respect optical I/O keep-outs such as grating-coupler clearance or fiber-array access corridors.

The electrical route cost includes conventional metal-routing terms and photonic-aware interaction terms:
\begin{equation}
    C_{\mathrm{elec}}
    =
    C_{\mathrm{metal}}
    +
    C_{\mathrm{via}}
    +
    C_{\mathrm{cong}}
    +
    C_{\mathrm{eo}} .
\end{equation}
Here $C_{\mathrm{metal}}$ accounts for metal length and layer usage, $C_{\mathrm{via}}$ penalizes or restricts via insertion, $C_{\mathrm{cong}}$ captures metal congestion, and $C_{\mathrm{eo}}$ captures electrical--optical interaction. The interaction term discourages metal from passing too close to sensitive photonic structures unless the rule allows it.

PROPEL represents photonic-aware electrical constraints using hard and soft regions. A hard keep-out is a forbidden region whereas a soft keep-out is a penalty region.

The electrical--optical interaction penalty is written as
\begin{equation}
    C_{\mathrm{eo}}
    =
    \int_{\gamma^{\mathrm{elec}}}
    \rho_{\mathrm{eo}}(x,y,\ell)\,ds ,
\end{equation}
where $\rho_{\mathrm{eo}}(x,y,\ell)$ is a layer-dependent penalty density. The penalty is increased near photonic devices. Lower metal layers can be assigned larger penalties if they are more likely to interact with optical modes or thermal tuning structures. If a region is strictly forbidden, the penalty is treated as a hard rejection rather than as a soft cost.

Metal--waveguide co-propagation is treated as a separate interaction check. Even when a metal route satisfies minimum spacing, a long parallel segment near an optical waveguide may be undesirable. PROPEL therefore measures or estimates the length over which an electrical route runs close and nearly parallel to a routed waveguide. If the co-propagation exceeds a configured limit, the route can be rejected or penalized depending on whether the rule is strict or soft.

Electrical legality is checked before route commitment. The checks include metal width, metal spacing, legal layer usage, via rules, pin access, pad access, shorts, opens, optical hard keep-outs, optical I/O clearance, device keep-outs, and co-propagation limits. These checks are applied during search and again after the final metal geometry is generated. A metal route is committed only if the final geometry satisfies the electrical and photonic-aware rule set.

Table~\ref{tab:active_electrical_checks} summarizes the active electrical checks and reported interaction metrics.

\begin{table}[!t]
\centering
\caption{Electrical routing checks and interaction metrics in active PROPEL routing.}
\label{tab:active_electrical_checks}
\small
\begin{tabularx}{\columnwidth}{@{}lX@{}}
\toprule
\textbf{Check or metric} & \textbf{Purpose} \\
\midrule
Metal width and spacing &
Enforces conventional electrical DRC for routed metal polygons \\

Layer legality &
Restricts each electrical net to its allowed metal-layer set \\

Via policy &
Allows, penalizes, or forbids layer transitions depending on the net class \\

Pin access &
Ensures metal connects to the correct active-device terminal without illegal device overlap \\

Pad access &
Ensures metal enters assigned pads through legal breakout regions \\

Optical hard keep-out &
Rejects metal routes that enter forbidden photonic regions \\

Optical soft keep-out &
Penalizes routes that pass near sensitive photonic structures but remain allowed \\

Metal--waveguide co-propagation &
Measures long nearby parallel metal--waveguide interaction \\

$\mathrm{DRV}_{\mathrm{elec}}$ &
Reports hard electrical design-rule violations \\

$\mathrm{USV}_{\mathrm{eo}}$ &
Reports strict user-specified electrical--optical violations \\

\bottomrule
\end{tabularx}
\end{table}

If some electrical--optical rules are configured as soft preferences, they are reported through $C_{\mathrm{eo}}$ instead of being counted as hard violations. This allows PROPEL to distinguish between manufacturing errors and design-quality penalties.

\subsection{Netlist-Free Topological Routing}
\label{subsec:netlist_free_topological_mode}

\begin{figure*}[!t]
    \centering
    \includegraphics[width=\textwidth]{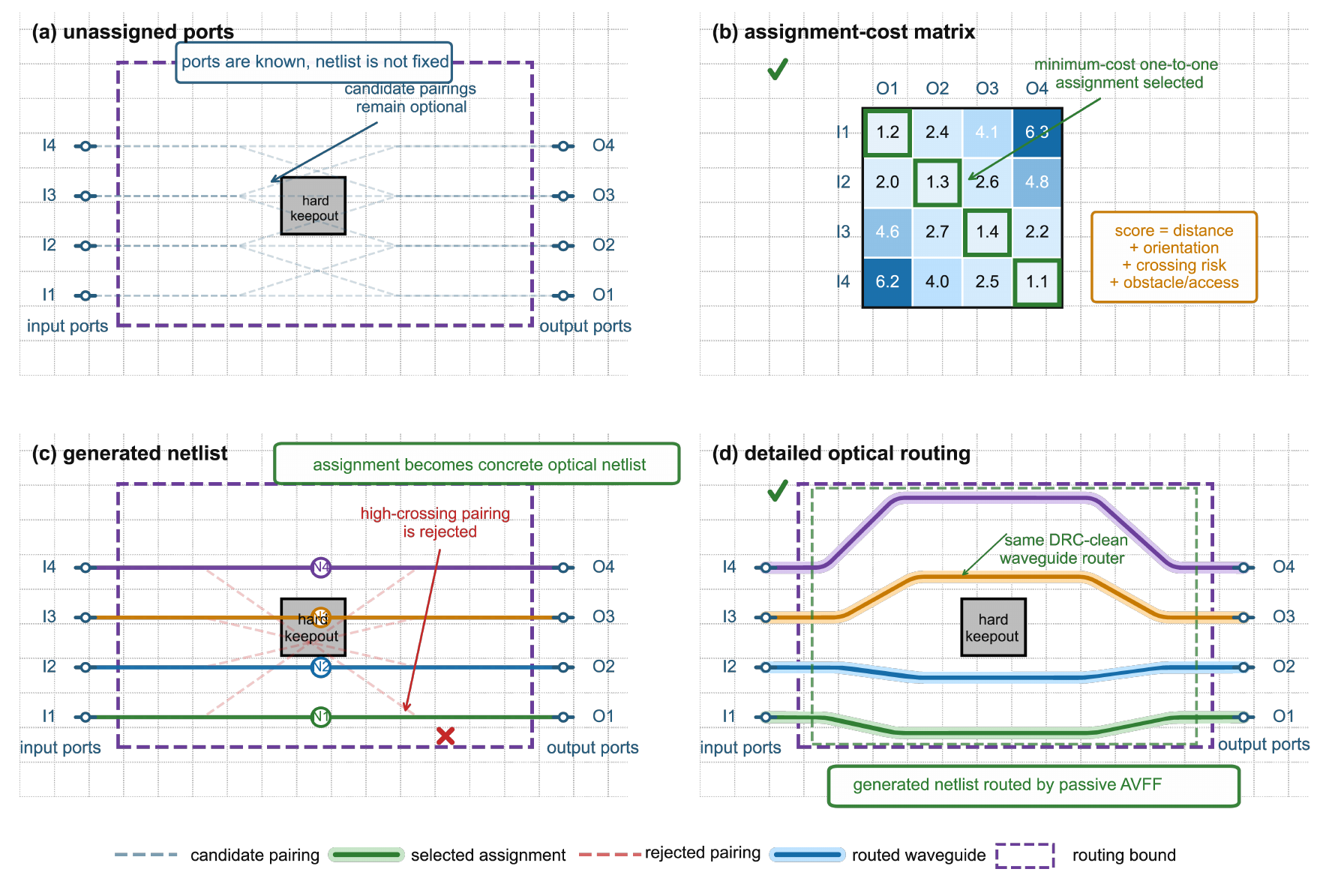}
    \caption{Netlist-free topological routing in PROPEL. When the physical input and output ports are known but the netlist is not fixed, PROPEL first evaluates candidate input--output pairings using distance, orientation, crossing-risk, obstacle, and access costs. The selected minimum-cost one-to-one assignment becomes a concrete optical netlist, which is then routed by the same DRC-aware passive PROPEL optical router.}
    \label{fig:netlist_free_topological_routing}
\end{figure*}

Netlist-free topological routing is the third operating mode of PROPEL. In ordinary detailed routing, the netlist is already fixed: each source port is assigned to a specific target port before routing begins. In netlist-free topological routing, the physical input and output ports are known, but the one-to-one connectivity is not fixed. As shown in Figure~\ref{fig:netlist_free_topological_routing}, the router must therefore solve two problems: first decide which input should connect to which output using a routing-aware assignment cost, and then route the generated connections as optical waveguides using the passive PROPEL routing kernel.

This mode is useful for early-stage PIC floorplanning, synthetic benchmark generation, optical switch fabrics, generic interconnect fabrics, and layouts where the designer knows that one set of ports must connect to another set but does not yet require a specific pairing. In such cases, using an arbitrary index-based assignment can create unnecessary crossings, long detours, port-access conflicts, and bend infeasibility. PROPEL instead generates the netlist using physical routability costs before detailed routing begins.

Let the input and output port sets be
\begin{equation}
    \mathcal{P}_{\mathrm{in}}
    =
    \{p_1^{\mathrm{in}},p_2^{\mathrm{in}},\ldots,p_n^{\mathrm{in}}\},
    \qquad
    \mathcal{P}_{\mathrm{out}}
    =
    \{p_1^{\mathrm{out}},p_2^{\mathrm{out}},\ldots,p_n^{\mathrm{out}}\}.
\end{equation}
The goal is to find an assignment
\begin{equation}
    A:\mathcal{P}_{\mathrm{in}}\rightarrow\mathcal{P}_{\mathrm{out}},
\end{equation}
where each input port is paired with exactly one output port. Once the assignment is selected, PROPEL converts it into a generated optical netlist:
\begin{equation}
    \mathcal{N}_{\mathrm{gen}}
    =
    \left\{
    \left(
    p_i^{\mathrm{in}},
    A(p_i^{\mathrm{in}})
    \right)
    \mid
    p_i^{\mathrm{in}}\in\mathcal{P}_{\mathrm{in}}
    \right\}.
\end{equation}
The generated netlist is then routed using the same passive optical PROPEL router described earlier.

The assignment is selected using a routing-aware cost. For a candidate pairing between input port $p_i$ and output port $p_j$, PROPEL evaluates
\begin{equation}
    C_{\mathrm{assign}}(p_i,p_j)
    =
    \lambda_d C_{\mathrm{dist}}
    +
    \lambda_o C_{\mathrm{orient}}
    +
    \lambda_s C_{\mathrm{side}}
    +
    \lambda_x C_{\mathrm{cross}}
    +
    \lambda_b C_{\mathrm{block}}
    +
    \lambda_a C_{\mathrm{access}} .
\end{equation}
Here $C_{\mathrm{dist}}$ penalizes long input--output connections, $C_{\mathrm{orient}}$ penalizes incompatible port directions, $C_{\mathrm{side}}$ penalizes poor side compatibility, $C_{\mathrm{cross}}$ estimates crossing pressure, $C_{\mathrm{block}}$ estimates obstacle interaction, and $C_{\mathrm{access}}$ penalizes assignments that are likely to block dense port-access regions.

The full assignment is chosen by minimizing total assignment cost:
\begin{equation}
    A^\star
    =
    \arg\min_A
    \sum_{p_i\in\mathcal{P}_{\mathrm{in}}}
    C_{\mathrm{assign}}
    \left(
    p_i,
    A(p_i)
    \right).
\end{equation}
This formulation does not guarantee that the final detailed route will be DRC-clean by itself. It only gives a detailed router a physically better netlist than an arbitrary pairing would. Final validity is still determined by the passive optical routing stage.
\begin{algorithm}[!t]
\small
\caption{Netlist-Free Topological Assignment}
\label{alg:netlist_free_topological_assignment}
\begin{algorithmic}[1]
\STATE \textbf{Input:} input ports $\mathcal{P}_{\mathrm{in}}$, output ports $\mathcal{P}_{\mathrm{out}}$, layout geometry, obstacles, port orientations
\STATE Initialize assignment-cost matrix $M$
\FOR{each input port $p_i\in\mathcal{P}_{\mathrm{in}}$}
    \FOR{each output port $p_j\in\mathcal{P}_{\mathrm{out}}$}
        \STATE Estimate distance cost $C_{\mathrm{dist}}(p_i,p_j)$
        \STATE Estimate orientation cost $C_{\mathrm{orient}}(p_i,p_j)$
        \STATE Estimate side-compatibility cost $C_{\mathrm{side}}(p_i,p_j)$
        \STATE Estimate crossing-pressure cost $C_{\mathrm{cross}}(p_i,p_j)$
        \STATE Estimate obstacle/blockage cost $C_{\mathrm{block}}(p_i,p_j)$
        \STATE Estimate port-access cost $C_{\mathrm{access}}(p_i,p_j)$
        \STATE Store weighted assignment cost:
        \[
        M_{ij}
        =
        \lambda_d C_{\mathrm{dist}}
        +
        \lambda_o C_{\mathrm{orient}}
        +
        \lambda_s C_{\mathrm{side}}
        +
        \lambda_x C_{\mathrm{cross}}
        +
        \lambda_b C_{\mathrm{block}}
        +
        \lambda_a C_{\mathrm{access}}
        \]
    \ENDFOR
\ENDFOR
\STATE Solve the one-to-one assignment problem using $M$
\STATE Optionally refine the assignment using crossing-aware swaps or local reordering
\STATE Generate optical netlist:
\[
\mathcal{N}_{\mathrm{gen}}
=
\{(p_i,A(p_i))\mid p_i\in\mathcal{P}_{\mathrm{in}}\}
\]
\STATE Route $\mathcal{N}_{\mathrm{gen}}$ using the passive PROPEL optical router
\STATE \textbf{return} generated netlist, routed layout, and assignment metrics
\end{algorithmic}
\end{algorithm}

The assignment cost uses approximate physical information rather than full detailed routing. Distance and side compatibility are fast geometric terms. Orientation cost estimates whether the input and output ports naturally face each other or require immediate detours. Crossing cost estimates whether a proposed assignment is likely to intersect other proposed connections. Obstacle cost estimates whether the coarse connection corridor passes through dense device regions or blocked areas. Access cost penalizes assignments that force many routes through the same dense port bank or narrow escape corridor.

After the first assignment is generated, PROPEL can optionally refine the pairing before detailed routing. A simple refinement step swaps two output assignments if the swap reduces estimated crossing pressure, improves port-side ordering, or reduces obstacle interaction. This is useful when the initial assignment has low total distance but produces a poor physical ordering. The refinement is lightweight and runs before detailed routing.

This mode generates a netlist from available port sets; it is invoked only when no fixed netlist is supplied, and it never modifies a netlist the designer provides. Netlist-free topological routing is used only when the input design explicitly allows the router to choose the pairing.

The generated netlist is treated as an ordinary optical netlist after assignment. Each generated connection receives the same port-access handling, bounded VFF guidance, bend-aware candidate expansion, crossing-aware legality checking, DRC replay, and route commitment as passive optical routing. Therefore, topological routing does not introduce a separate detailed router. It only adds a routability-aware assignment stage before the existing PROPEL optical routing flow.


\section{Experimental Evaluation}
\label{sec:evaluation}

\subsection{Benchmark Suite}
\label{subsec:benchmark_suite}

\subsubsection{Passive Optical Benchmarks}
\label{subsec:passive_optical_benchmarks}

The passive benchmark group evaluates PROPEL as a detailed optical router. Each benchmark provides a fixed optical netlist and a placed photonic layout, and the router must generate DRC-clean waveguide geometry between predefined optical ports. The selected benchmarks cover both photonic-computing layouts and optical-switching layouts. Clements, ADEPT, and BENES circuits follow the photonic tensor-core benchmark families used in recent detailed PIC routing evaluation~\cite{zhou2025lidar2}. Clements meshes provide regular low-crossing MZI-array baselines, while ADEPT and BENES introduce denser multiport access, fanout, and local congestion. Quantum graph photonic circuit extends this group with dense port banks and repeated photonic-computing motifs.

The switch-fabric benchmarks are included because large-scale optical switches are a central application of automated PIC routing. $\lambda$-router, standard-crossbar, Light, GWOR, and Benes benchmarks stress long-range connectivity, distributed I/O, crossing pressure, and routing-order sensitivity. These designs are motivated by earlier ONoC and WRONoC physical-design studies~\cite{boos2013proton,beuningen2016protonplus,beuningen2016platon,chuang2018planaronoc,zheng2021topro}, by silicon photonic switch topology studies~\cite{cheng2020siliconswitch,soref2018switching}, and by large MZI-based electro-optic switch demonstrations~\cite{qiao2017switch}. Tables~\ref{tab:bench_spacious_revised} and~\ref{tab:bench_compact_revised} summarize the spacious and compact passive benchmark statistics before detailed routing.

\begin{table*}[t]
\centering
\caption{Spacious passive PIC benchmark statistics in the PROPEL benchmark suite.}
\label{tab:bench_spacious_revised}
\scriptsize
\setlength{\tabcolsep}{3pt}
\begin{tabular}{llrrrr}
\hline
Benchmark & Source & Comp. & Nets & Topo. CR & Die Size ($\mu$m$^2$) \\
\hline
Clements\_8$\times$8 & \cite{zhou2025lidar2} & 52 & 79 & 0 & $4800 \times 1600$ \\
Clements\_16$\times$16 & \cite{zhou2025lidar2} & 168 & 287 & 0 & $8000 \times 3200$ \\
ADEPT\_8$\times$8 & \cite{zhou2025lidar2} & 82 & 111 & 18 & $4520 \times 1600$ \\
ADEPT\_16$\times$16 & \cite{zhou2025lidar2} & 162 & 223 & 21 & $6910 \times 3200$ \\
ADEPT\_32$\times$32 & \cite{zhou2025lidar2} & 318 & 447 & 73 & $13000 \times 6400$ \\
$\lambda$Router\_8$\times$8 & \cite{boos2013proton,beuningen2016platon} & 44 & 52 & 58 & $8000 \times 8000$ \\
StandardCrossbar\_8$\times$8 & \cite{boos2013proton,chuang2018planaronoc} & 80 & 100 & 12 & $9000 \times 9000$ \\
Light\_a & \cite{zhou2025lidar2} & 32 & 24 & 34 & $9000 \times 9000$ \\
Light\_b & \cite{zhou2025lidar2} & 24 & 12 & 5 & $9000 \times 9000$ \\
Light\_c & \cite{zhou2025lidar2} & 32 & 24 & 21 & $9000 \times 9000$ \\
Light\_d & \cite{zhou2025lidar2} & 32 & 24 & 39 & $9000 \times 9000$ \\
\hline
\end{tabular}
\end{table*}

\begin{table*}[t]
\centering
\caption{Compact passive PIC benchmark statistics in the PROPEL benchmark suite.}
\label{tab:bench_compact_revised}
\scriptsize
\setlength{\tabcolsep}{3pt}
\begin{tabular}{llrrrr}
\hline
Benchmark & Source & Comp. & Nets & Topo. CR & Die Size ($\mu$m$^2$) \\
\hline
Clements\_8$\times$8\_C & \cite{zhou2025lidar2} & 60 & 87 & 0 & $3500 \times 1000$ \\
Clements\_16$\times$16\_C & \cite{zhou2025lidar2} & 184 & 303 & 0 & $6800 \times 2000$ \\
ADEPT\_8$\times$8\_C & \cite{zhou2025lidar2} & 104 & 127 & 30 & $4400 \times 1000$ \\
ADEPT\_16$\times$16\_C & \cite{zhou2025lidar2} & 227 & 319 & 80 & $6200 \times 2000$ \\
ADEPT\_32$\times$32\_C & \cite{zhou2025lidar2} & 529 & 767 & 160 & $7800 \times 4000$ \\
GWOR\_8$\times$8\_C & \cite{tan2011gwor,zhou2025lidar2} & 7 & 16 & - & $6000 \times 6000$ \\
GWOR\_16$\times$16\_C & \cite{tan2011gwor,zhou2025lidar2} & 17 & 32 & - & $4000 \times 4000$ \\
GWOR\_32$\times$32\_C & \cite{tan2011gwor,zhou2025lidar2} & 33 & 64 & - & $6000 \times 6000$ \\
Benes\_16$\times$16\_C & \cite{qiao2017switch,zhou2025lidar2} & 224 & 416 & 88 & $2200 \times 1300$ \\
Benes\_32$\times$32\_C & \cite{qiao2017switch,zhou2025lidar2} & 576 & 1024 & 416 & $3500 \times 2500$ \\
\hline
\end{tabular}
\end{table*}

\subsubsection{Length-Matching and Process-Aware Benchmarks}
\label{subsec:length_process_benchmarks}

The length-matching benchmark group evaluates whether PROPEL can satisfy path-balancing constraints after ordinary optical routing. These benchmarks are included because interferometers, coherent paths, true-time-delay structures, and programmable photonic systems can require matched physical length, optical path length, or group delay in addition to DRC-clean connectivity. The benchmark set includes MZI-style matched paths, multi-arm interferometric paths, optical ring-resonator routing cases, optical true-time-delay cases, and process-aware OPL cases. The evaluation is motivated by recent work on automatic PIC routing under delay-matching constraints, which shows that phase- and delay-matching constraints require routing objectives beyond ordinary shortest-path routing~\cite{wu2025delaymatching}.

The process-aware benchmarks use representative passive and active layouts with and without process-aware geometry generation. These cases are not separate circuit topologies; instead, they evaluate the same routed circuits under a measured wafer-map hook. This isolates the effect of process-aware width assignment from route-search behavior: the route topology remains fixed, while the generated waveguide width follows the sampled process map.

\begin{table*}[t]
\centering
\caption{Length-matching and process-aware benchmark groups.}
\label{tab:bench_length_process}
\scriptsize
\setlength{\tabcolsep}{3pt}
\begin{tabular}{llrrrl}
\hline
Benchmark Group & Source & Comp. & Nets & Matched Nets & Purpose \\
\hline
lm\_mzi\_2x2 / lm\_mzi\_4x4 & PROPEL & 4--16 & 2--8 & 2--8 & MZI path-length closure \\
lm\_multiarm\_4 / lm\_multiarm\_8 & PROPEL & 16--32 & 12--24 & 4--8 & Multi-arm interferometric matching \\
lm\_orr\_4 / lm\_orr\_8 & PROPEL & 12--24 & 8--16 & 8--16 & Ring-resonator path balancing \\
lm\_ottd\_4 / lm\_ottd\_8 & PROPEL & 12--24 & 8--16 & 8--16 & True-time-delay matching \\
lm\_process\_mzi / lm\_process\_multiarm & PROPEL & 16--32 & 8--24 & 8 & Process-aware OPL matching \\
Process-aware passive cases & PROPEL & -- & -- & -- & Wafer-map-based width assignment \\
Process-aware active cases & PROPEL & -- & -- & -- & Active layout with process-aware waveguides \\
\hline
\end{tabular}
\end{table*}

\subsubsection{Active, Topological, and quantum-graph-inspired Benchmarks}
\label{subsec:active_quantum_topological_placement_benchmarks}

The expanded benchmark group evaluates PROPEL beyond passive fixed-netlist routing. Active PIC benchmarks contain both optical waveguide nets and electrical control, bias, heater, readout, or pad-access nets. They are included because active photonic systems require metal routing around sensitive photonic devices, routed waveguides, optical I/O, and pad-access regions. The larger active Clements and ADEPT cases follow the active optical--electrical routing setting used in prior active PIC routing evaluation~\cite{zhou2026lidar3}, while the smaller active Clements and heater cases provide compact GDS-level examples for visual inspection.

The topology-driven benchmarks evaluate netlist-free routing. In these cases, the physical input and output port sets are known, but the one-to-one optical connectivity is not fixed. PROPEL first generates a routing-aware assignment between the port sets and then routes the generated optical netlist using the passive routing kernel. These benchmarks are included because switch fabrics and WRONoC-style layouts can be strongly affected by port pairing, port-side ordering, and crossing pressure before detailed routing begins~\cite{tseng2019wronoc,tan2011gwor,zheng2021topro}.

The quantum-graph-inspired benchmark group is included as a heterogeneous photonic stress case. The original quantum-graph-inspired photonic layout integrates hundreds of photonic components, hundreds of transmission lines, many waveguide crossings, and hundreds of length-matching delay lines, and both optical and electronic I/O~\cite{bao2023boya}. It therefore provides a useful reference for evaluating whether a routing framework can coordinate passive routing, active electrical access, crossing-rich optical layouts, and delay-sensitive connectivity in one physical-design flow.

\begin{table*}[t]
\centering
\caption{Expanded benchmark cases used to evaluate active routing, topology-driven routing, and quantum-photonic layout stress.}
\label{tab:bench_active_topo_quantum}
\scriptsize
\setlength{\tabcolsep}{3pt}
\begin{tabular}{llrrrrl}
\hline
Benchmark & Source & Mode & Comp. & Opt. Nets & Elec. Nets & Die Size ($\mu$m$^2$) \\
\hline
Topological\_26IO\ & PROPEL & Topology & 26 & generated & -- & $5000 \times 5000$ \\
Topological\_100IO\_10$\times$10 & PROPEL & Topology & 200 & generated & -- & $9000 \times 9000$ \\
QuantumGraph\_32$\times$32 & \cite{bao2023boya} & Heterogeneous & 268 & 423 & -- & $12000 \times 15000$ \\
\hline
\end{tabular}
\end{table*}
\subsection{Experimental Setup and Evaluation Metrics}
\label{sec:setup}

Unless otherwise stated, silicon-photonic-style benchmarks use a routing grid resolution of \SI{2}{\micro\meter}, a minimum bend radius of \SI{5}{\micro\meter}, 45-degree neighbor expansion, bounded VFF search, group-aware routing order, adaptive port-access reservation, nested failed-net rip-up/reroute, and strict DRC before committing any route. Relaxed DRC is used only as a diagnostic mode to identify blocking nets and conflict regions after strict routing fails; relaxed routes are not directly committed to the final layout.

\begin{table}[t]
\centering
\caption{Default PROPEL routing configuration used for the main benchmark evaluation.}
\label{tab:router_config}
\scriptsize
\setlength{\tabcolsep}{4pt}
\begin{tabular}{@{}ll@{}}
\toprule
\textbf{Parameter} & \textbf{Default setting} \\
\midrule
Routing mode & Bounded PROPEL guidance \\
Grid resolution & \SI{2}{\micro\meter} \\
Minimum bend radius & \SI{5}{\micro\meter} \\
Neighbor expansion & 45-degree enabled \\
Port access & Adaptive reservation enabled \\
Routing groups & Group-aware ordering enabled \\
Fluid VFF bounds & Enabled for selected dense/topological cases \\
Crossing awareness & Crossing legality and crossing-space checks enabled \\
Rip-up/reroute & Nested failed-net rerouting enabled \\
DRC mode & Strict for committed routes \\
Relaxed DRC & Diagnostic only, not committed \\
Geometry generation & GDSFactory-based final route construction \\
Acceleration & C++ numerical calculation backend \\
Correctness authority & Python DRC replay and GDS validation \\
\bottomrule
\end{tabular}
\end{table}

For optical evaluation, PROPEL reports routed waveguide length, crossing count, design-rule violations, runtime, and maximum insertion loss. 

\begin{table}[t]
\centering
\caption{Default optical loss parameters used for baseline insertion-loss evaluation.}
\label{tab:loss_params}
\scriptsize
\setlength{\tabcolsep}{4pt}
\begin{tabular}{@{}lll@{}}
\toprule
\textbf{Loss term} & \textbf{Symbol} & \textbf{Default value} \\
\midrule
Propagation loss & $\alpha_w$ & \SI{1.5}{dB/cm} \\
Bend loss & $\alpha_b$ & \SI{0.005}{dB/bend} \\
Crossing loss & $\alpha_c$ & \SI{0.52}{dB/crossing} \\
Y-branch loss & -- & \SI{0.3}{dB} \\
MZI loss & -- & \SI{1.2}{dB} \\
MMI loss & -- & \SI{0.1}{dB} \\
\bottomrule
\end{tabular}
\end{table}

\begin{table}[t]
\centering
\caption{Experimental platform used for runtime measurement.}
\label{tab:experimental_platform}
\scriptsize
\setlength{\tabcolsep}{8pt}
\begin{tabular}{ll}
\hline
Item & Value \\
\hline
CPU & Apple M2, 8-core CPU \\
Memory & 8 GB unified memory \\
Operating system & macOS on Apple Silicon \\
Python version & Python 3.11 \\
GDSFactory version & 8.x \cite{gdsfactory} \\
C++ compiler & Apple Clang / LLVM \\
Native binding & pybind11 \\
\hline
\end{tabular}
\end{table}

\subsection{Routing Results}
\label{sec:results}
PROPEL runtimes were measured on the platform in Table~\ref{tab:experimental_platform}, whereas LiDAR~2.0 runtimes were reproduced from the prior publication. These cross-platform runtimes are reported for reference and are not treated as hardware-normalized speedups.
This section evaluates PROPEL across passive optical routing, length matching, process-aware routing, active optical--electrical routing, topology-driven routing, and quantum-graph-inspired heterogeneous routing. The passive optical results provide the main benchmark-level comparison with prior detailed PIC routers. The length-matching results evaluate whether matched optical groups can be equalized after routing. The process-aware results evaluate how measured SiN wafer-map variation changes the routed waveguide width while keeping the geometric route unchanged. The active-routing results evaluate whether PROPEL can close electrical control and pad-access nets while preserving photonic keep-outs. The topology-driven results are reported separately because they evaluate capabilities beyond fixed passive netlist routing.

\subsubsection{Cross-Paper Routing Scope}
\label{subsec:cross_paper_routing_comparison}

The comparisons below are separated into direct benchmark-level comparisons and scope-level comparisons. Direct comparisons are used when prior work reports the same benchmark family and compatible metrics. Scope-level comparisons are used only to explain functional coverage across optical routing, electrical routing, topology generation, and GDS-level closure. All PROPEL routing-quality metrics are computed after final geometry generation and DRC replay.
Table~\ref{tab:router_paper_scope_comparison} summarizes the scope of the routing-related photonic physical-design works considered in this paper. The purpose of this table is not to rank all methods directly, because these works target different design stages, benchmark families, and physical constraints. Instead, it clarifies which comparisons are direct benchmark-level comparisons and which comparisons are scope-level references.

\begin{table*}[t]
\centering
\caption{Scope-level comparison of routing-related photonic physical-design papers.}
\label{tab:router_paper_scope_comparison}
\scriptsize
\setlength{\tabcolsep}{3pt}
\begin{tabular}{lccccp{0.28\textwidth}}
\toprule
Work & Optical & Electrical & Topology & GDS & Main use \\
\midrule
PROTON~\cite{boos2013proton} & Yes & No & Yes & Partial & ONoC/PIC routing baseline \\
PROTON+~\cite{beuningen2016protonplus} & Yes & No & Yes & Partial & WRONoC/topology baseline \\
PLATON~\cite{beuningen2016platon} & Yes & No & Yes & Partial & High-level physical design \\
PlanarONoC~\cite{chuang2018planaronoc} & Yes & No & Yes & Partial & Crossing-aware ONoC routing \\
ToPro~\cite{zheng2021topro} & Yes & No & Yes & Partial & Topology-driven comparison \\
LiDAR~1.0~\cite{zhou2025lidar} & Yes & No & No & Yes & Passive GDS-level routing \\
LiDAR~2.0~\cite{zhou2025lidar2} & Yes & No & Partial & Yes & Main passive-routing comparison \\
LiDAR~3.0~\cite{zhou2026lidar3} & Yes & Yes & No & Yes & Active-routing comparison \\
PROPEL & Yes & Yes & Yes & Yes & This work \\
\bottomrule
\end{tabular}
\end{table*}

\subsubsection{Passive Optical Routing Results}
\label{subsec:passive_optical_results}

Passive optical routing is the primary detailed waveguide-routing evaluation. Tables~\ref{tab:passive_noncompact_comparison} and~\ref{tab:passive_compact_comparison} preserve the original row-wise result format: each benchmark is followed by one row per router, and each row reports the critical-path crossing count, critical-path waveguide length, maximum insertion loss, total design-rule violations, and runtime. The comparison tables include the benchmark families reported by prior passive PIC routing works.

The passive benchmarks cover different routing regimes. Clements circuits are regular MZI meshes with low topological crossing pressure. ADEPT circuits include dense multiport access and higher congestion. Benes compact circuits stress hierarchy, repeated routing patterns, and crossing-space preservation. GWOR and Light switch-style circuits evaluate larger-radius routing and the trade-off between crossing avoidance and long propagation detours. The large-scale quantum graph circuit is used as a stress test because of its chip-scale size, length matching routing requirements, and overall routing complexity.

\begin{table*}[t]
\centering
\caption{Passive optical routing comparison on non-compact benchmarks.}
\label{tab:passive_noncompact_comparison}
\scriptsize
\setlength{\tabcolsep}{3pt}
\begin{tabular}{llrrrrr}
\hline
Benchmark & Router & CR & WL (mm) & IL$_{\max}$ (dB) & DRV & Time (s) \\
\hline
Clements\_8x8 & PROTON & 0 & 3.39 & 16.99 & 0 & 112 \\
 & LiDAR 1.0 & 0 & 2.94 & 16.38 & 0 & 29 \\
 & LiDAR 2.0 & 0 & 2.89 & 15.98 & 0 & 7 \\
 & PROPEL & \textbf{0} & \textbf{3.08} & \textbf{16.40} & \textbf{0} & \textbf{5.42} \\
\hline
Clements\_16x16 & PROTON & 5 & 5.06 & 29.31 & 12 & 527 \\
 & LiDAR 1.0 & 0 & 4.38 & 26.74 & 0 & 144 \\
 & LiDAR 2.0 & 0 & 4.07 & 26.03 & 0 & 61 \\
 & PROPEL & \textbf{0} & \textbf{4.94} & \textbf{26.77} & \textbf{0} & \textbf{18.68} \\
\hline
ADEPT\_8x8 & PROTON & 16 & 4.70 & 17.12 & 26 & 194 \\
 & LiDAR 1.0 & 18 & 4.10 & 18.00 & 0 & 71 \\
 & LiDAR 2.0 & 18 & 3.99 & 17.63 & 0 & 65 \\
 & PROPEL & \textbf{18} & \textbf{4.308} & \textbf{17.736} & \textbf{0} & \textbf{13.19} \\
\hline
ADEPT\_16x16 & PROTON & 28 & 7.84 & 24.07 & 98 & 1395 \\
 & LiDAR 1.0 & 16 & 7.38 & 17.80 & 0 & 243 \\
 & LiDAR 2.0 & 16 & 6.95 & 17.20 & 0 & 243 \\
 & PROPEL & \textbf{30} & \textbf{7.75} & \textbf{24.50} & \textbf{0} & \textbf{43.90} \\
\hline
ADEPT\_32x32 & PROTON & 66 & 16.13 & 44.57 & 355 & 10894 \\
 & LiDAR 1.0 & 50 & 15.04 & 36.34 & 0 & 1348 \\
 & LiDAR 2.0 & 50 & 14.70 & 36.06 & 0 & 1038 \\
 & PROPEL & \textbf{50} & \textbf{16.47} & \textbf{36.46} & \textbf{0} & \textbf{397.08} \\
\hline
Light\_a & PROTON & 12 & 32.98 & 11.09 & 0 & 39 \\
 & LiDAR 1.0 & 0 & 31.11 & 7.78 & 0 & 101 \\
 & LiDAR 2.0 & 0 & 29.99 & 7.61 & 0 & 103 \\
 & PROPEL & \textbf{0} & \textbf{31.04} & \textbf{7.67} & \textbf{0} & \textbf{67.42} \\
\hline
Light\_b & PROTON & 6 & 18.71 & 5.89 & 0 & 8 \\
 & LiDAR 1.0 & 0 & 21.55 & 6.31 & 0 & 44 \\
 & LiDAR 2.0 & 0 & 21.12 & 6.24 & 0 & 47 \\
 & PROPEL & \textbf{0} & \textbf{24.43} & \textbf{6.89} & \textbf{0} & \textbf{21.1} \\
\hline
Light\_c & PROTON & 14 & 20.81 & 10.23 & 1 & 52 \\
 & LiDAR 1.0 & 0 & 35.29 & 8.40 & 0 & 72 \\
 & LiDAR 2.0 & 0 & 35.92 & 8.57 & 0 & 74 \\
 & PROPEL & \textbf{0} & \textbf{34.88} & \textbf{8.41} & \textbf{0} & \textbf{49.36} \\
\hline
Light\_d & PROTON & 13 & 28.49 & 10.94 & 1 & 53 \\
 & LiDAR 1.0 & 0 & 33.52 & 8.14 & 0 & 80 \\
 & LiDAR 2.0 & 0 & 32.82 & 8.09 & 0 & 83 \\
 & PROPEL & \textbf{0} & \textbf{32.41} & \textbf{8.03} & \textbf{0} & \textbf{53.24} \\
\hline
StandardCrossbar\_8x8 & PROTON & 36 & 15.26 & 8.50 & \NA & 606.9 \\
 & PLATON & 50 & 26.48 & 12.40 & \NA & 35.2 \\
 & PlanarONoC & 13 & 32.13 & 7.35 & \NA & 0.2 \\
 & PROPEL & \textbf{9} & \textbf{32.82} & \textbf{15.49} & \textbf{0} & \textbf{119.97} \\
\hline
LambdaRouter\_8x8 & PROTON & 41 & 10.41 & 8.60 & \NA & 95.8 \\
 & PROPEL & \textbf{37} & \textbf{8.22} & \textbf{27.92} & \textbf{0} & \textbf{63.32} \\

\hline
\end{tabular}
\end{table*}

Table~\ref{tab:passive_noncompact_comparison} shows that PROPEL closes all reported non-compact passive benchmarks with zero final DRV while reducing runtime in the largest shared cases. For Clements\_16x16, runtime drops from 61~s with LiDAR~2.0 to 18.68~s with PROPEL, giving a 3.3$\times$ speedup. For ADEPT\_32x32, runtime drops from 1038~s to 397.08~s, giving a 2.6$\times$ speedup. The main trade-off is optical quality in some ADEPT cases: PROPEL is faster and DRC-clean, but it accepts slightly longer WL and higher IL$_{\max}$ than LiDAR~2.0.

\begin{figure*}[t]
\centering
\includegraphics[width=0.95\textwidth]{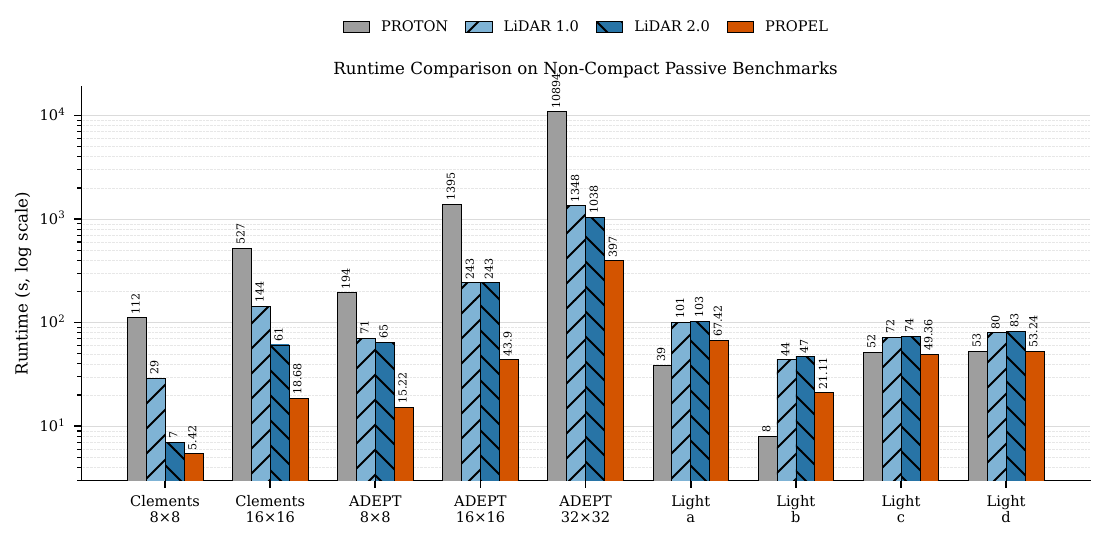}
\caption{Runtime comparison on non-compact passive optical routing benchmarks using a logarithmic scale.}
\label{fig:time_comparison_noncompact_log}
\end{figure*}
\FloatBarrier

\paragraph{Compact-Circuit Comparison.}

Figure~\ref{fig:time_comparison_noncompact_log} makes the runtime gap easier to see across both small and large benchmarks. The largest improvements appear in routing-heavy cases such as ADEPT\_16x16 and ADEPT\_32x32, where PROPEL reduces repeated Python-side exploration through bounded VFF guidance and native acceleration.

Compact layouts reduce available routing area and expose local congestion, port-access blocking, bend-feasibility limits, and crossing-space conflicts more aggressively than the large-circuit cases.

\begin{table*}[t]
\centering
\caption{Passive optical routing comparison on compact benchmarks.}
\label{tab:passive_compact_comparison}
\scriptsize
\setlength{\tabcolsep}{3pt}
\begin{tabular}{llrrrrr}
\hline
Benchmark & Router & CR & WL (mm) & IL$_{\max}$ (dB) & DRV & Time (s) \\
\hline
Clements\_8x8\_C & PROTON & 0 & 2.12 & 16.28 & 2 & 11 \\
 & LiDAR 1.0 & 0 & 1.91 & 16.23 & 2 & 11 \\
 & LiDAR 2.0 & 0 & 1.91 & 16.23 & 0 & 9 \\
 & PROPEL & \textbf{0} & \textbf{2.17} & \textbf{16.29} & \textbf{0} & \textbf{2.81} \\
\hline
Clements\_16x16\_C & PROTON & 1 & 3.91 & 27.13 & 5 & 93 \\
 & LiDAR 1.0 & 0 & 3.56 & 26.56 & 5 & 101 \\
 & LiDAR 2.0 & 0 & 3.56 & 26.56 & 0 & 89 \\
 & PROPEL & \textbf{0} & \textbf{4.02} & \textbf{26.63} & \textbf{0} & \textbf{35.58} \\
\hline
ADEPT\_8x8\_C & PROTON & 14 & 4.53 & 16.26 & 11 & 96 \\
 & LiDAR 1.0 & 14 & 3.71 & 16.12 & 2 & 82 \\
 & LiDAR 2.0 & 14 & 3.71 & 16.12 & 0 & 68 \\
 & PROPEL & \textbf{18} & \textbf{4.31} & \textbf{18.04} & \textbf{0} & \textbf{16.04} \\
\hline
ADEPT\_16x16\_C & PROTON & 34 & 8.24 & 27.60 & 42 & 515 \\
 & LiDAR 1.0 & 21 & 8.63 & 21.09 & 7 & 304 \\
 & LiDAR 2.0 & 21 & 8.35 & 21.02 & 0 & 236 \\
 & PROPEL & \textbf{30} & \textbf{7.22} & \textbf{24.75} & \textbf{0} & \textbf{92.16} \\
\hline
ADEPT\_32x32\_C & PROTON & 49 & 10.78 & 36.20 & 171 & 1496 \\
 & LiDAR 1.0 & 31 & 10.58 & 27.07 & 30 & 735 \\
 & LiDAR 2.0 & 30 & 9.83 & 26.48 & 3 & 568 \\
 & PROPEL & \textbf{50} & \textbf{13.42} & \textbf{36.03} & \textbf{0} & \textbf{112.32} \\
\hline
GWOR\_16x16\_C & PROTON & 30 & 3.09 & 16.72 & 35 & 19 \\
 & LiDAR 1.0 & 27 & 2.78 & 15.02 & 26 & 15 \\
 & LiDAR 2.0 & 21 & 2.80 & 12.03 & 0 & 7 \\
 & PROPEL & \textbf{0} & \textbf{3.84} & \textbf{12.64} & \textbf{0} & \textbf{1.18} \\
\hline
GWOR\_32x32\_C & PROTON & 62 & 7.22 & 35.26 & 126 & 60 \\
 & LiDAR 1.0 & 56 & 6.81 & 31.26 & 44 & 45 \\
 & LiDAR 2.0 & 44 & 7.82 & 25.40 & 0 & 11 \\
 & PROPEL & \textbf{0} & \textbf{5.92} & \textbf{25.95} & \textbf{0} & \textbf{3.00} \\
\hline
Benes\_16x16\_C & PROTON & 33 & 4.42 & 22.69 & 76 & 721 \\
 & LiDAR 1.0 & 26 & 4.28 & 18.84 & 3 & 175 \\
 & LiDAR 2.0 & 23 & 3.57 & 17.15 & 0 & 33 \\
 & PROPEL & \textbf{26} & \textbf{3.77} & \textbf{17.02} & \textbf{0} & \textbf{32.75} \\
\hline
Benes\_32x32\_C & PROTON & 56 & 9.24 & 36.76 & 365 & 5117 \\
 & LiDAR 1.0 & 68 & 9.08 & 42.39 & 78 & 1690 \\
 & LiDAR 2.0 & 58 & 6.51 & 36.56 & 0 & 251 \\
 & PROPEL & \textbf{58} & \textbf{6.968} & \textbf{36.15} & \textbf{0} & \textbf{258} \\
\hline
\end{tabular}
\end{table*}

\begin{figure*}[t]
\centering
\includegraphics[width=0.95\textwidth]{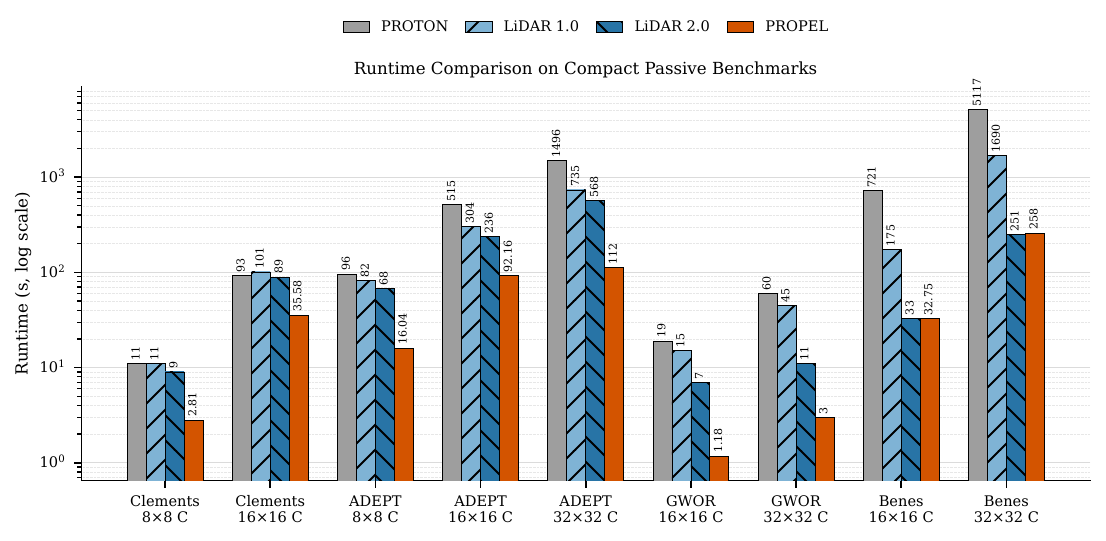}
\caption{Runtime comparison on compact passive optical routing benchmarks using a logarithmic scale.}
\label{fig:time_comparison_compact_log}
\end{figure*}
Table~\ref{tab:passive_compact_comparison} shows that PROPEL is especially useful when compact routing creates residual DRC failures in earlier baselines. In ADEPT\_32x32\_C, PROPEL removes the three residual DRV reported by LiDAR~2.0 and reduces runtime from 568~s to 112.32~s, a $5.1\times$ speedup, although the critical-path crossing count increases from 30 to 50. GWOR\_32x32\_C exhibits a different trade-off: PROPEL reduces the critical-path crossing count from 44 to 0 and runtime from 11~s to 3.00~s, but IL$_{\max}$ increases slightly from 25.40~dB to 25.95~dB. This case should therefore be interpreted as a crossing-count and runtime improvement, not as an insertion-loss improvement.

\begin{figure*}[t]
\centering

\begin{minipage}[t]{0.48\textwidth}
\centering
\includegraphics[width=\linewidth]{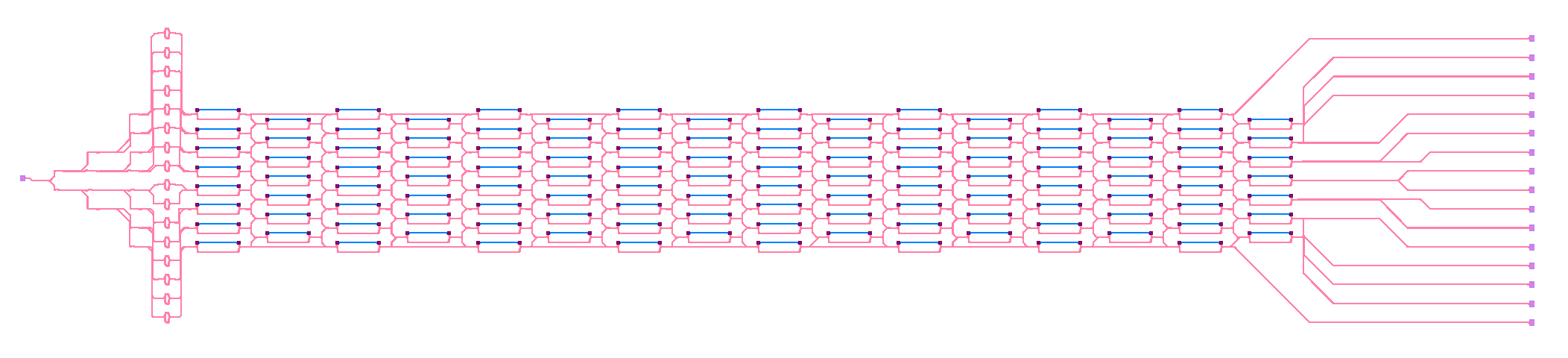}
\vspace{0.2em}

\footnotesize (a) Routed Clements $16\times16$ mesh
\end{minipage}
\hfill
\begin{minipage}[t]{0.48\textwidth}
\centering
\includegraphics[width=\linewidth]{gds/mmi32.png}
\vspace{0.2em}

\footnotesize (b) Routed multiport MMI $32\times32$ benchmark
\end{minipage}

\vspace{0.8em}

\begin{minipage}[t]{0.32\textwidth}
\centering
\includegraphics[width=\linewidth]{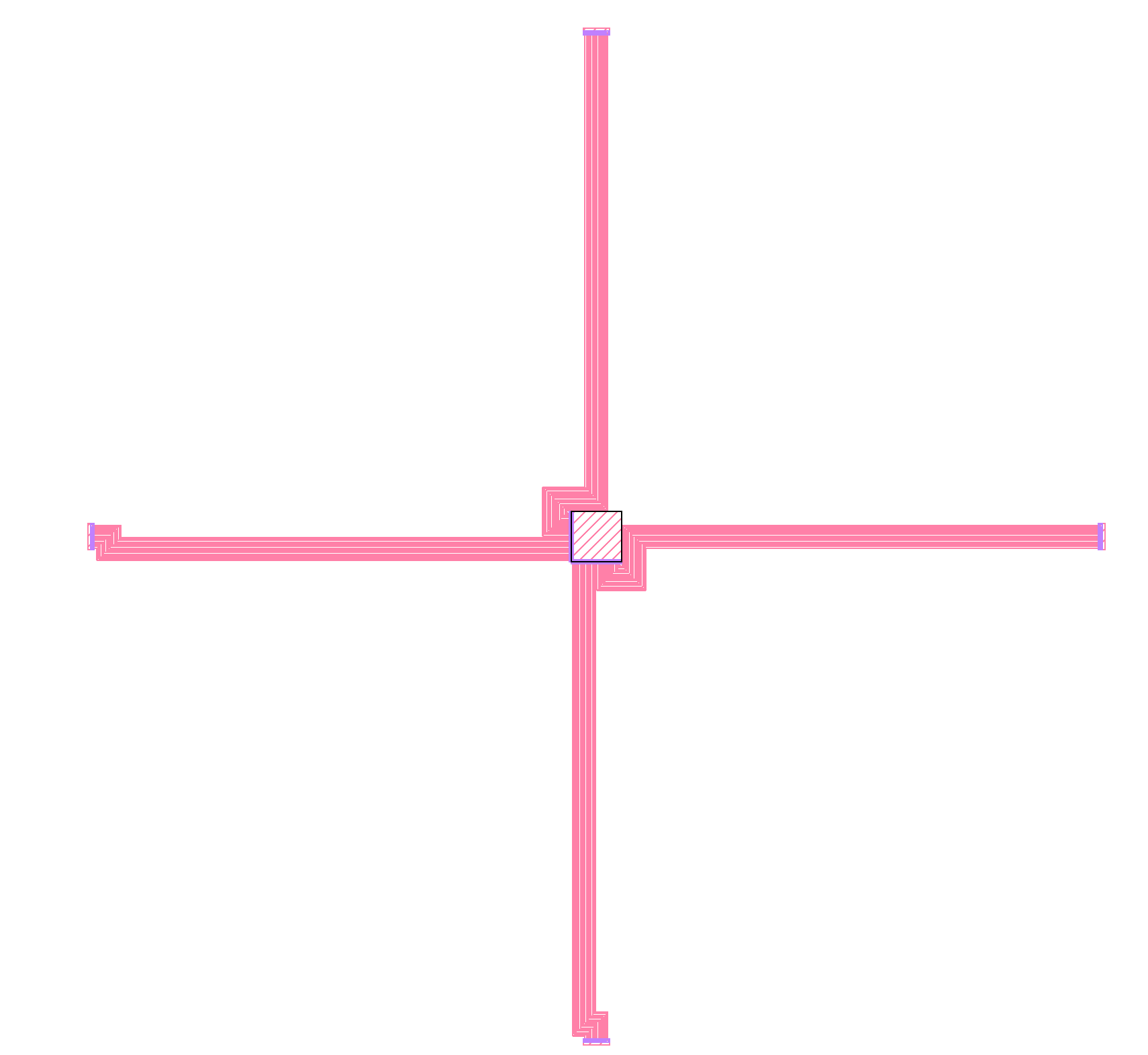}
\vspace{0.2em}

\footnotesize (c) Routed GWOR $32\times32$ benchmark
\end{minipage}
\hspace{0.08\textwidth}
\begin{minipage}[t]{0.32\textwidth}
\centering
\includegraphics[width=\linewidth]{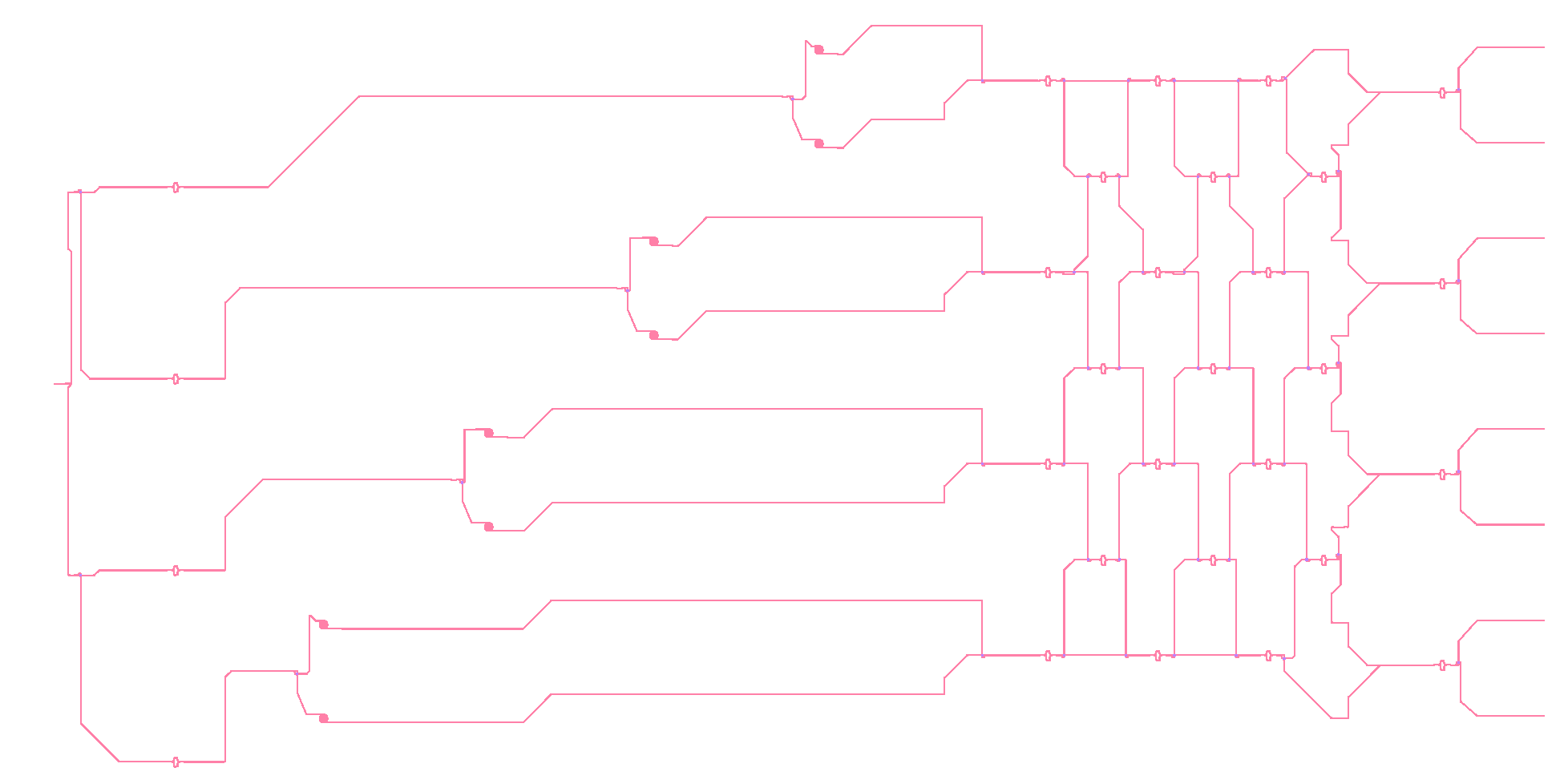}
\vspace{0.2em}

\footnotesize (d) Quantum-graph-inspired $8\times8$ photonic layout
\end{minipage}

\begin{minipage}[t]{0.32\textwidth}
\centering
\includegraphics[width=\linewidth]{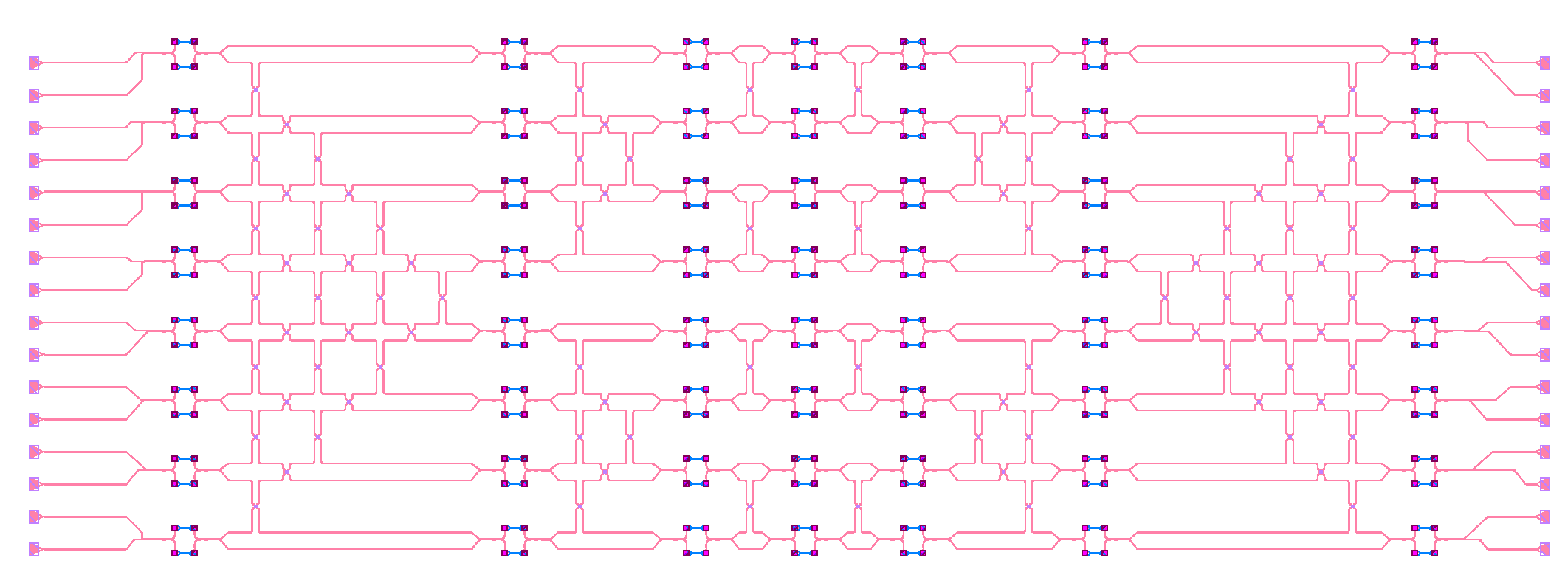}
\vspace{0.2em}

\footnotesize (e) Routed  $16\times16$ C photonic layout
\end{minipage}

\caption{Representative passive optical layouts generated by PROPEL. The examples include regular MZI-array routing, multiport MMI routing, multi-side switch-style routing, and quantum-graph-inspired heterogeneous photonic routing.}
\label{fig:passive_routing_gds_examples}
\end{figure*}

The passive results should be interpreted through both routing quality and physical validity. A lower maximum insertion loss is useful only when the final geometry is legal, and the ports are accessed correctly. For this reason, the DRV column is retained in the comparison tables. On regular Clements-style meshes, PROPEL preserves DRC-clean routing with competitive runtime. On denser ADEPT, GWOR, and Benes-style layouts, the results show the importance of port-access reservation, routing order, and crossing-space preservation. The compact cases are especially useful because they expose whether the router can close difficult layouts without relying on manual post-routing edits.

\begin{table*}[t]
\centering
\caption{Implementation-path comparison between the Python VFF search and the C++-accelerated PROPEL search on representative passive benchmarks. CR is critical-path crossing count, WL is critical-path waveguide length in millimeters, IL$_{\max}$ is maximum critical-path insertion loss in dB, DRV is final design-rule violation count, and Time is runtime in seconds. Because the two implementations can produce different search orders and routing outcomes, this comparison does not isolate programming language alone.}
\label{tab:propel_ablation}
\scriptsize
\setlength{\tabcolsep}{4pt}
\begin{tabular}{llrrrrr}
\toprule
Benchmark & Configuration & CR & WL (mm) & IL$_{\max}$ (dB) & DRV & Time (s) \\
\midrule
Clements\_16x16
    & Python VFF Search & 0 & 4.510 & 26.527 & 0 & 77.74 \\
    & C++ PROPEL Search & 0 & 4.94 & 26.77 & 0 & 18.68 \\
\midrule
ADEPT\_32x32
    & Python VFF Search & 49 & 14.999 & 35.700 & 9 & 1567.33 \\
    & C++ PROPEL Search & 50 & 16.47 & 36.46 & 0 & 397.08 \\
\midrule
Benes\_16x16
    & Python VFF Search & 25 & 5.988 & 15.368 & 5 & 1300.23 \\
    & C++ PROPEL Search & 21 & 6.334 & 13.450 & 0 & 51.92 \\
\midrule
QuantumGraph\_8x8
    & Python VFF Search & 4 & 15.596 & 16.809 & 0 & 49.53 \\
    & C++ PROPEL Search & 4 & 16.294 & 16.964 & 0 & 38.87 \\
\bottomrule
\end{tabular}
\end{table*}

Table~\ref{tab:propel_ablation} shows that the C++-accelerated PROPEL search substantially reduces runtime compared with the Python VFF search. Clements\_16x16 improves from 77.74~s to 18.68~s, a $4.2\times$ reduction, while ADEPT\_32x32 improves from 1567.33~s to 397.08~s, a $3.9\times$ reduction, and its final DRV decreases from 9 to 0. For Benes\_16x16, the C++ configuration reduces CR from 25 to 21, reduces IL$_{\max}$ from 15.368~dB to 13.450~dB, and reduces DRV from 5 to 0. Because the Python and C++ configurations produce different CR, WL, IL$_{\max}$, and DRV values, this table is an implementation-path comparison rather than a runtime-only language ablation.

\paragraph{Incremental Route-Memory Reuse.}

To evaluate the memory-driven capability of PROPEL, we apply small local design changes after a complete DRC-clean routing run. Verified routes are loaded from route memory and replayed through final DRC. Routes whose ports, nearby obstacles, or committed geometries are unchanged are restored directly, while only impacted or invalidated nets are returned to the routing queue. The incremental runtime includes loading the saved route records, replaying restored routes through DRC, identifying impacted nets, rerouting those nets, and performing final legality validation.

\begin{table*}[t]
\centering
\caption{Incremental routing using verified route-memory reuse after small local design changes. Moved Comp. is the number of relocated components. Restored Routes is the percentage of previously committed routes restored after DRC replay. Rerouted Nets includes routes directly connected to moved components and any additional routes invalidated by the changed local geometry. Incremental Time includes memory loading, DRC replay, invalidation analysis, selective rerouting, and final validation.}
\label{tab:route_memory_reuse}
\scriptsize
\setlength{\tabcolsep}{4pt}
\begin{tabular}{lrrrrrr}
\toprule
Benchmark &
Moved Comp. &
Restored Routes (\%) &
Rerouted Nets &
Incremental Time (s) &
From-Scratch Time (s) &
DRV \\
\midrule
Clements\_8x8
    & 0
    & 100.0
    & 0
    & 0.06
    & 5.42
    & \textbf{0} \\

Clements\_16x16
    & 2
    & 97.6
    & 7
    & 0.82
    & 18.68
    & \textbf{0} \\

ADEPT\_16x16
    & 3
    & 96.9
    & 7
    & 1.36
    & 43.90
    & \textbf{0} \\

ADEPT\_32x32
    & 5
    & 97.1
    & 13
    & 4.76
    & 397.08
    & \textbf{0} \\

Benes\_16x16\_C
    & 3
    & 98.6
    & 6
    & 1.11
    & 32.75
    & \textbf{0} \\

QuantumGraph\_16x16
    & 4
    & 95.5
    & 7
    & 3.47
    & 221.56
    & \textbf{0} \\
\bottomrule
\end{tabular}
\end{table*}

Table~\ref{tab:route_memory_reuse} demonstrates the intended behavior of PROPEL's route-memory mechanism. When the design is unchanged, all previously verified routes are restored and no detailed rerouting is required. After small local placement changes, between 95.5\% and 98.6\% of the previously committed routes are restored successfully, and only 6--13 nets require selective rerouting in the evaluated modified layouts. The incremental runs remain DRC-clean and complete in a few seconds, compared with from-scratch runtimes ranging from 18.68~s to 397.08~s for the modified benchmark cases. These results show that route memory is not only a route-storage mechanism: it allows PROPEL to preserve unaffected physical-design work and restrict detailed routing to the portion of the layout changed by a local design modification.

\begin{table}[t]
\centering
\caption{Runtime speedup summary on shared passive benchmarks against LiDAR~2.0.}
\label{tab:speedup_summary}
\small
\begin{tabular}{lr}
\toprule
Metric & Value \\
\midrule
Shared passive benchmarks & 18 \\
Cases faster than LiDAR~2.0 & 17 \\
Median speedup & 2.6$\times$ \\
Geometric-mean speedup & 2.5$\times$ \\
Final PROPEL DRV & 0 \\
\bottomrule
\end{tabular}
\end{table}

\paragraph{Length Matching Results.}

In addition to ordinary shortest-path optical routing, PROPEL supports length-sensitive routing constraints. These cases are reported inside the passive-routing subsection because they still use optical waveguide routing, but they evaluate a different objective from insertion-loss minimization. The standard length-matching benchmarks use the match-length net attribute to group routes that should be equalized after routing. Table~\ref{tab:length_matching_benchmarks} summarizes the length-matching benchmark families generated from the project source. 

\begin{table*}[t]
\centering
\caption{Length-matching optical-routing benchmarks. ``Mismatch'' reports the final maximum residual mismatch among length-constrained nets after PROPEL applies length, delay, or optical-path-length matching, in micrometers. Lower mismatch is better; DRV reports final design-rule violations.}
\label{tab:length_matching_benchmarks}
\scriptsize
\setlength{\tabcolsep}{3pt}
\begin{tabular}{lrrrlllrr}
\hline
Benchmark & Components & Nets & Matched Nets & Die & Compact Die & Target & Mismatch & DRV \\
\hline
lm\_mzi\_2x2 & 4 & 2 & 2 & 2200x1200 & 1600x900 & Length & \textbf{0.00} & \textbf{0} \\
lm\_mzi\_4x4 & 16 & 8 & 8 & 3600x1800 & 2600x1300 & Length & \textbf{0.00} & \textbf{0} \\
lm\_multiarm\_4 & 16 & 12 & 4 & 4200x1800 & 3000x1300 & Length & \textbf{0.00} & \textbf{0} \\
lm\_multiarm\_8 & 32 & 24 & 8 & 6400x2400 & 4600x1700 & Length & \textbf{0.00} & \textbf{0} \\
lm\_orr\_4 & 12 & 8 & 8 & 3200x1600 & 2400x1200 & Length & \textbf{0.00} & \textbf{0} \\
lm\_orr\_4\_C & 12 & 8 & 8 & 3200x1600 & 2400x1200 & Length & \textbf{0.00} & \textbf{0} \\
lm\_orr\_8\_C & 24 & 16 & 16 & 5200x2200 & 3800x1600 & Length & \textbf{0.00} & \textbf{0} \\
lm\_ottd\_4 & 12 & 8 & 8 & 5000x1800 & 3600x1300 & Delay & \textbf{0.00} & \textbf{0} \\
lm\_ottd\_8 & 24 & 16 & 16 & 8000x2400 & 5600x1800 & Delay & \textbf{0.00} & \textbf{0} \\
lm\_process\_mzi\_4 & 16 & 8 & 8 & 4200x1800 & 3000x1300 & OPL & \textbf{0.00} & \textbf{0} \\
lm\_process\_multiarm\_8 & 32 & 24 & 8 & 7000x2400 & 5000x1700 & OPL & \textbf{0.00} & \textbf{0} \\
\hline
\end{tabular}
\end{table*}
Table~\ref{tab:length_matching_benchmarks} shows that PROPEL closes all 11 reported matching benchmarks with 0.00~$\mu$m final mismatch and zero final DRV. This means the matching stage is not only numerically equalizing path lengths, but also inserting compensation geometry that remains legal after DRC replay. The result is important for MZI, multi-arm, ORR, OTTD, and process-aware OPL cases, where ordinary shortest-path routing would not be sufficient.
\begin{figure*}[t]
\centering

\begin{minipage}[t]{0.48\textwidth}
\centering
\includegraphics[width=\linewidth]{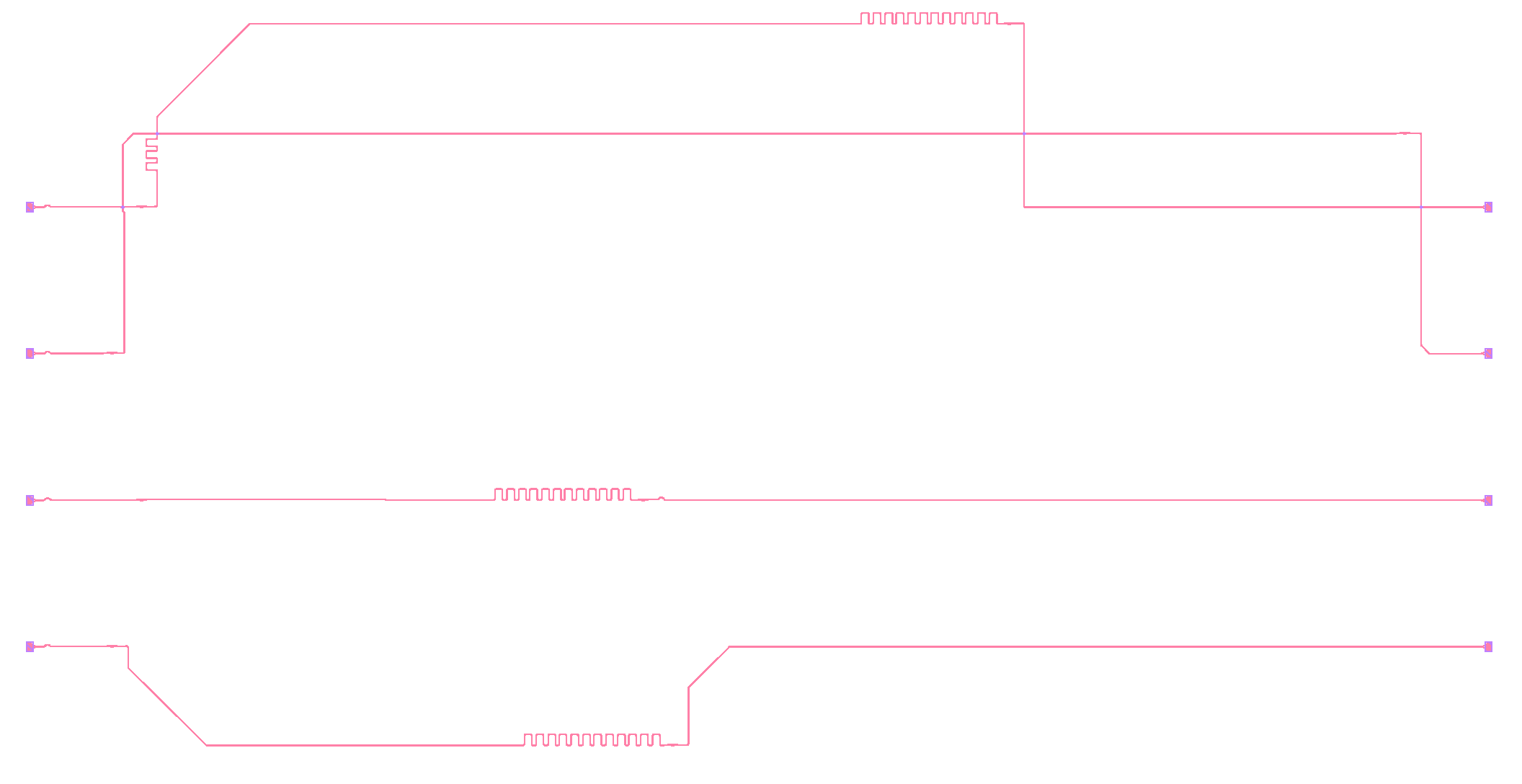}
\vspace{0.2em}

\footnotesize (a) Multi-arm matched routing
\end{minipage}
\hfill
\begin{minipage}[t]{0.48\textwidth}
\centering
\includegraphics[width=\linewidth]{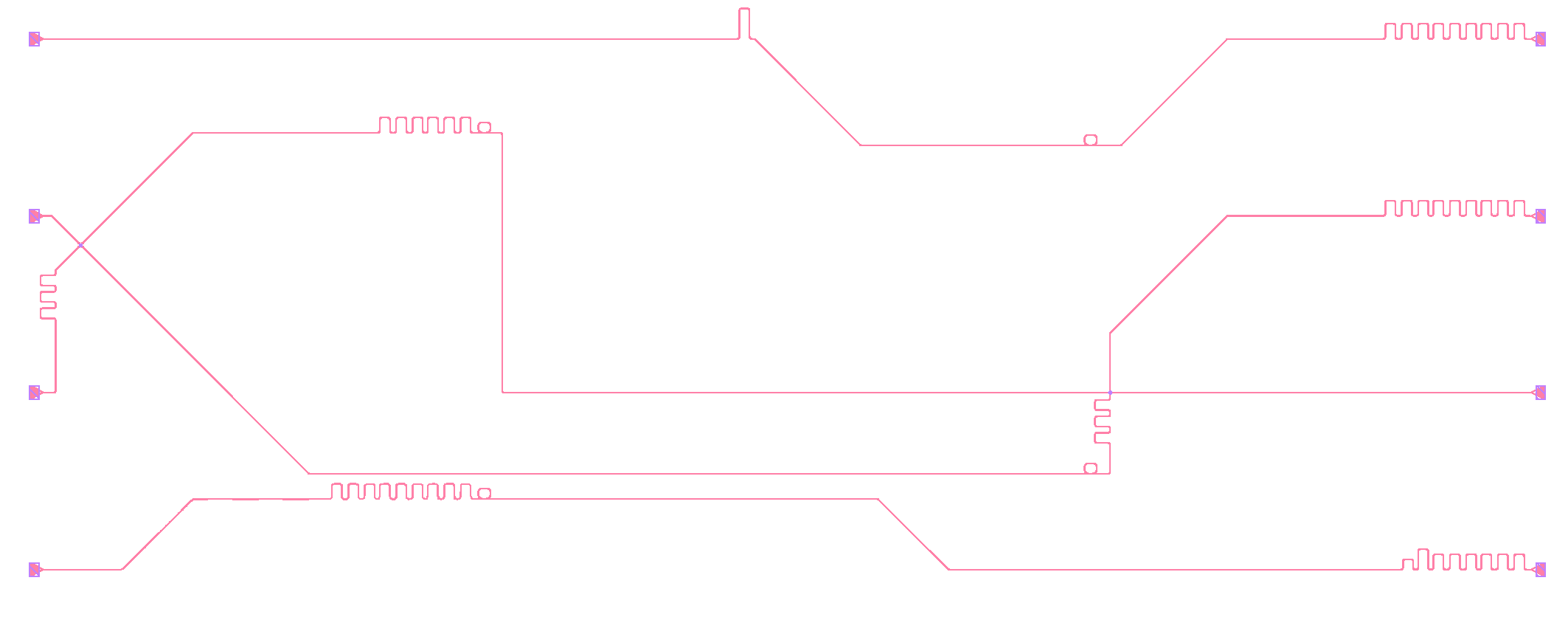}
\vspace{0.2em}

\footnotesize (b) Optical ring-resonator matched routing
\end{minipage}

\vspace{0.8em}

\begin{minipage}[t]{0.48\textwidth}
\centering
\includegraphics[width=\linewidth]{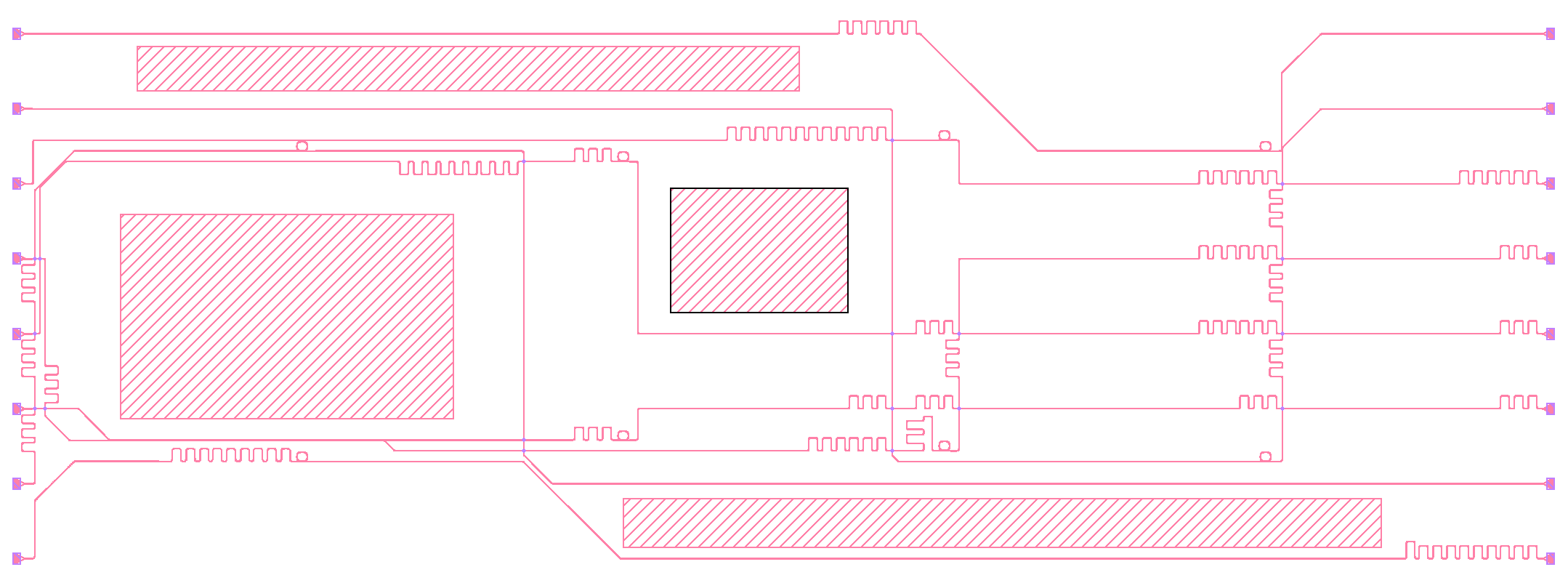}
\vspace{0.2em}

\footnotesize (c) Larger optical ring-resonator matching case
\end{minipage}
\hfill
\begin{minipage}[t]{0.48\textwidth}
\centering
\includegraphics[width=\linewidth]{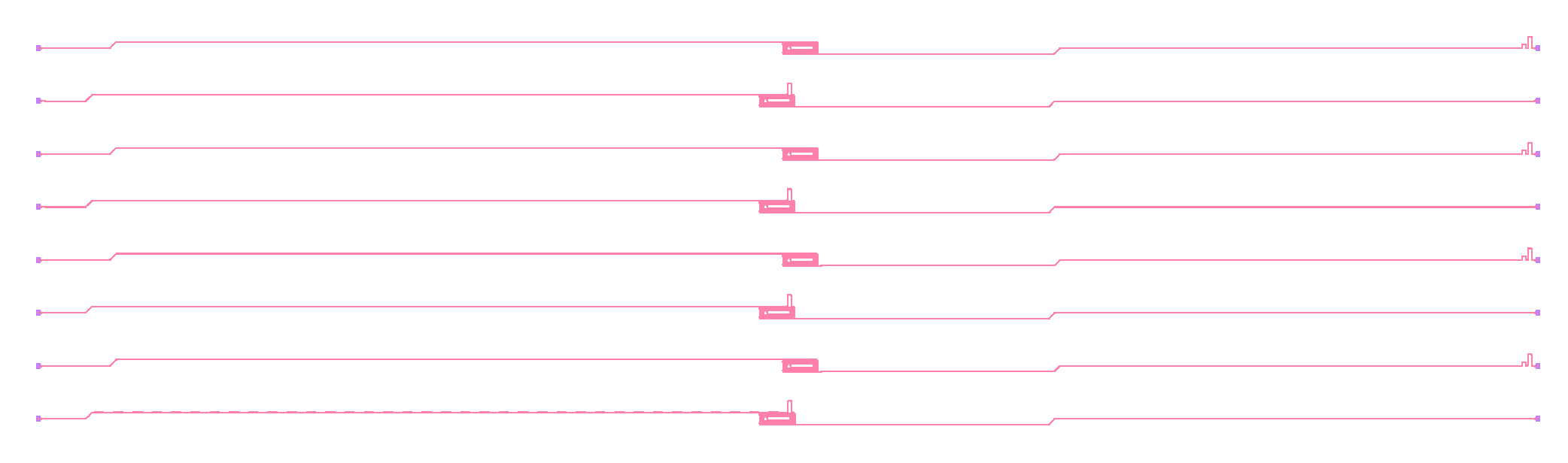}
\vspace{0.2em}

\footnotesize (d) Spiral delay insertion for path-length compensation
\end{minipage}

\caption{Representative length-matching layouts. The examples show multi-arm matching, optical ring-resonator matching, larger matched-ring routing, and spiral delay insertion for path-length equalization.}
\label{fig:length_matching_gds_examples}
\end{figure*}

The length-matching results show that PROPEL can apply matching constraints on top of ordinary detailed optical routing. Across the evaluated length, delay, and optical-path-length matching benchmarks, the final reported mismatch is zero within the measurement precision, while all cases remain DRC-clean.

\paragraph{Process-Aware Routing Results.}

\begin{figure*}[t]
\centering

\begin{minipage}[t]{0.95\textwidth}
\centering
\includegraphics[width=\linewidth]{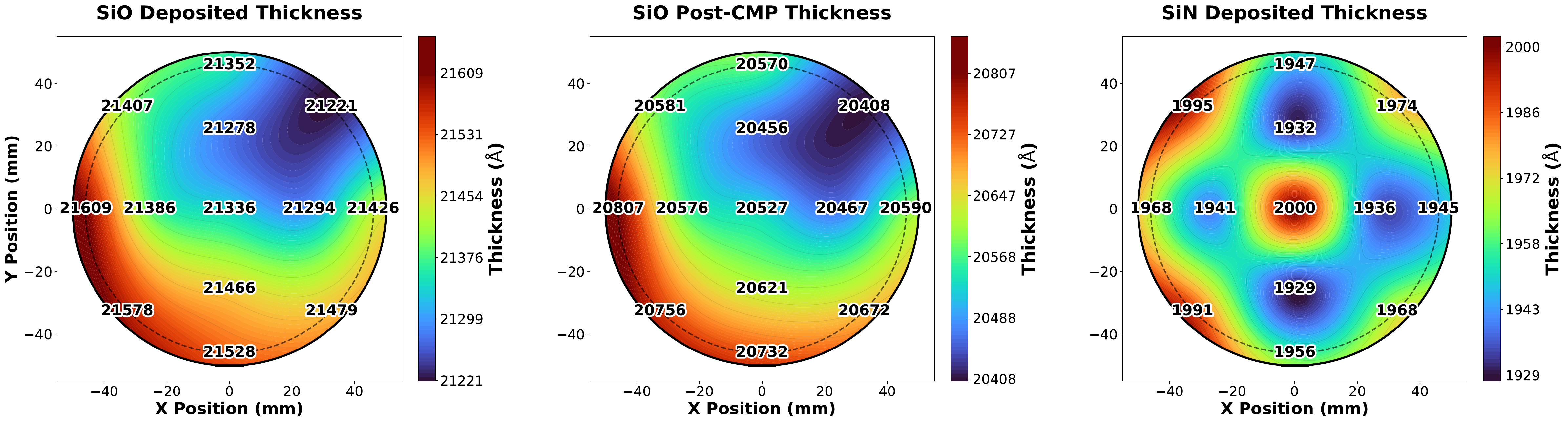}
\vspace{0.2em}

\footnotesize (a) Wafer map thickness distribution of Oxide deposition, Post CMP Oxide and deposited SiN.
\end{minipage}

\vspace{0.8em}

\begin{minipage}[t]{0.95\textwidth}
\centering
\includegraphics[width=\linewidth]{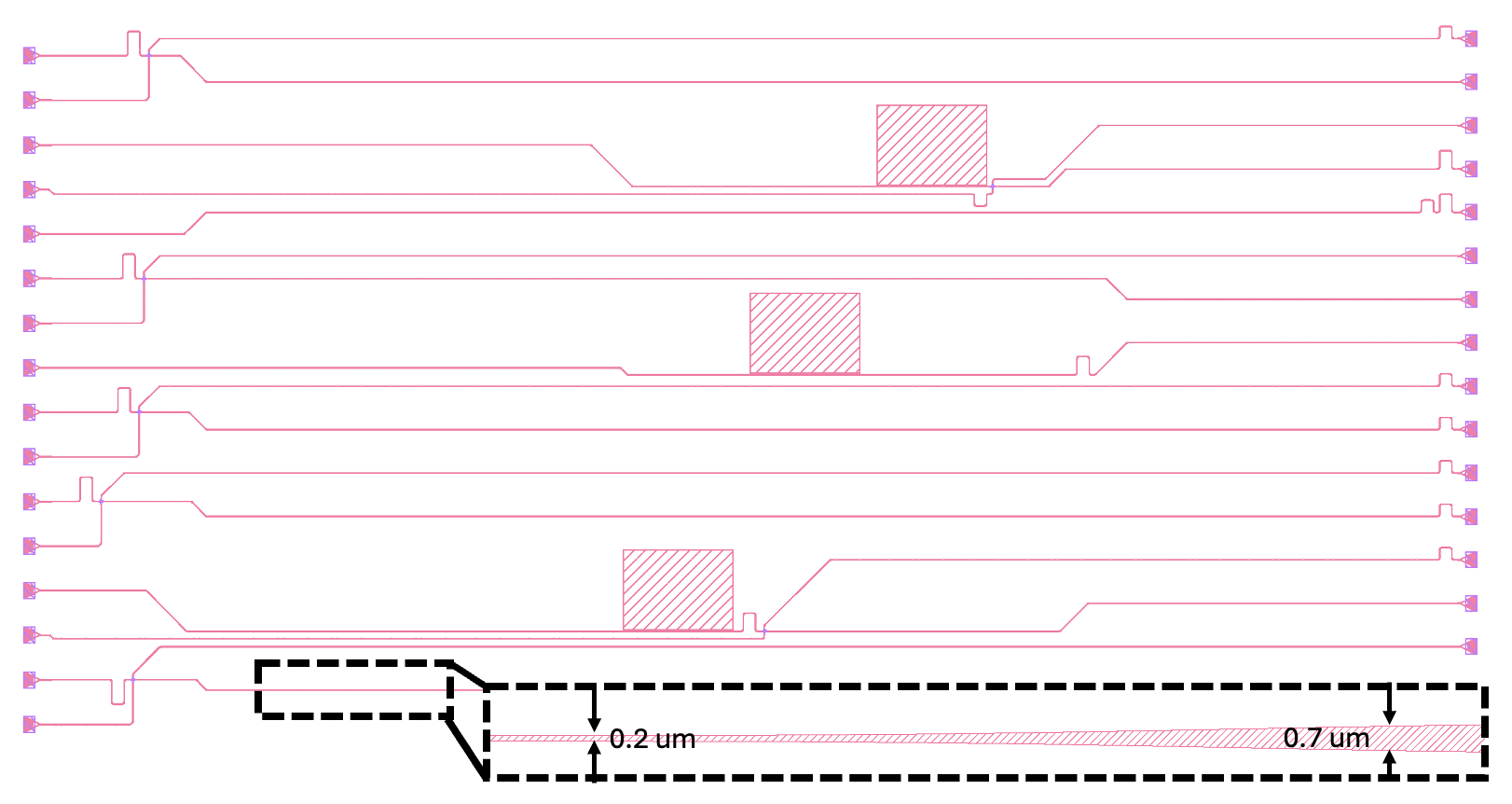}
\vspace{0.2em}

\footnotesize (b) Process-aware routed layout over the selected die window
\end{minipage}

\caption{SiN wafer-map characterization and process-aware routing setup. The first three panels show the measured process maps, while the bottom panel shows the routed GDS response over the selected process-map window.}
\label{fig:sin_wafer_map_process_window}
\end{figure*}

Process-aware routing is evaluated as an extension of normal passive and active routing. In this experiment, process awareness does not change the topological path, routing order, crossing count, runtime, or DRC result. Instead, it changes the generated waveguide width along the already selected route according to the wafer-map value sampled at each route segment. Therefore, the process-aware results are reported with identical geometric routing metrics for nominal and process-aware modes, while the process map is shown separately in Fig.~\ref{fig:sin_wafer_map_process_window}. This separates routing quality from process-aware width assignment.

\begin{table*}[t]
\centering
\caption{Process-map settings used for process-aware width assignment. The router samples the process map along each waveguide and changes width accordingly, but does not alter the selected route.}
\label{tab:process_map_summary}
\scriptsize
\setlength{\tabcolsep}{4pt}
\begin{tabular}{lrrrrrrr}
\hline
Map & Center X (mm) & Center Y (mm) & Inner step ($\mu$m) & Max radius (mm) & Values (\si{\angstrom}) & Threshold (\si{\angstrom}) & Width range ($\mu$m) \\
\hline
SiO as deposited & 0 & 0 & 100 & 50 & 21221--21609 & 21412 & 0.20--0.7 \\
SiO after CMP    & 0 & 0 & 100 & 50 & 20408--20807 & 20597 & 0.20--0.7 \\
SiN as deposited & 0 & 0 & 100 & 50 & 1929--2000   & 1960  & 0.20--0.7 \\
\hline
\end{tabular}
\end{table*}
\begin{table*}[t]
\centering
\caption{Process-aware routing comparison on selected passive and active circuits. Nominal and process-aware modes have the same geometric routing result because process awareness only changes segment width based on the sampled wafer map.}
\label{tab:process_aware_comparison}
\scriptsize
\setlength{\tabcolsep}{4pt}
\begin{tabular}{lllrrr}
\hline
Benchmark & Type & Mode & WL (mm) & IL$_{\max}$ (dB) & Time (s) \\
\hline
Clements\_16x16\_C & Passive & Nominal & 4.02 & 26.63 & 35.58 \\
 & Passive & Process-aware & \textbf{4.02} & \textbf{26.63} & \textbf{35.58} \\
\hline
ADEPT\_16x16\_C & Passive & Nominal & 7.22 & 24.75 & 92.16 \\
 & Passive & Process-aware & \textbf{7.22} & \textbf{24.75} & \textbf{92.16} \\
\hline
GWOR\_16x16 & Passive & Nominal & 4.84 & 0.79 & 1.45 \\
 & Passive & Process-aware & \textbf{4.84} & \textbf{0.79} & \textbf{1.45} \\
\hline
GWOR\_16x16\_C & Passive & Nominal & 3.84 & 0.64 & 1.18 \\
 & Passive & Process-aware & \textbf{3.84} & \textbf{0.64} & \textbf{1.18} \\
\hline
\end{tabular}
\end{table*}
Table~\ref{tab:process_aware_comparison} confirms that the current process-aware mode changes geometry generation rather than route search. Nominal and process-aware rows have identical WL, IL$_{\max}$, and runtime for all six tested passive and active cases; for example, Clements\_16x16\_C remains at 4.02~mm, 26.63~dB, and 35.58~s in both modes. This is therefore a fabrication-aware width-assignment result, not yet a process-cost-aware routing result.
Because the current process-aware implementation is a width-assignment step rather than a route-search perturbation, it does not change the selected route, waveguide length, insertion-loss estimate, or runtime. The process-aware value is still useful because it produces a layout where the waveguide width follows the local wafer-map condition. In the next version, this map can also be pushed into the routing cost so that the path itself avoids high-variation regions.

\subsubsection{Active Optical--Electrical Routing Results}
\label{subsec:active_results}

Table~\ref{tab:active_prior_results} reports the active optical--electrical routing comparison. Unlike passive optical routing, active PIC routing must complete electrical control and pad-access nets while preserving the already routed photonic layout. The comparison therefore focuses on via count, metal-layer usage, electrical wirelength, and runtime, which directly reflect active-routing complexity and layout overhead.

\begin{table*}[t]
\centering
\caption{Active PIC electrical-routing comparison. Via reports electrical layer transitions, Layer reports the maximum number of metal layers used, WL reports electrical wirelength, and Time reports runtime in seconds.}
\label{tab:active_prior_results}
\scriptsize
\setlength{\tabcolsep}{4pt}
\begin{tabular}{llrrrr}
\hline
Benchmark & Router & Via & Layer & WL (mm) & Time (s) \\
\hline
Clements\_8x8 & Anaroute & 1 & 2 & 42.99 & 45 \\
 & Anaroute* & 2 & 3 & 45.86 & 114 \\
 & LiDAR 3.0 & 0 & 1 & 46.94 & 33 \\
 & PROPEL & \textbf{0} & \textbf{1} & \textbf{48.20} & \textbf{19} \\
\hline
Clements\_16x16 & Anaroute & 69 & 3 & 340.97 & 2383 \\
 & Anaroute* & 117 & 3 & 348.69 & 2324 \\
 & LiDAR 3.0 & 0 & 1 & 360.51 & 207 \\
 & PROPEL & \textbf{0} & \textbf{1} & \textbf{372.40} & \textbf{430} \\
\hline
Clements\_32x32 & Anaroute & 159 & 3 & 2084.42 & 13235 \\
 & Anaroute* & 205 & 3 & 2095.78 & 20093 \\
 & LiDAR 3.0 & 0 & 2 & 2124.23 & 930 \\
 & PROPEL & \textbf{0} & \textbf{2} & \textbf{2256.80} & \textbf{1212} \\
\hline
ADEPT\_8x8 & Anaroute & 8 & 3 & 95.58 & 212 \\
 & Anaroute* & 16 & 3 & 97.20 & 481 \\
 & LiDAR 3.0 & 0 & 1 & 100.70 & 65 \\
 & PROPEL & \textbf{0} & \textbf{1} & \textbf{108.50} & \textbf{74} \\
\hline
ADEPT\_16x16 & Anaroute & 54 & 3 & 422.90 & 1558 \\
 & Anaroute* & 83 & 3 & 425.98 & 3184 \\
 & LiDAR 3.0 & 0 & 1 & 440.56 & 245 \\
 & PROPEL & \textbf{0} & \textbf{1} & \textbf{474.30} & \textbf{296} \\
\hline
\end{tabular}
\end{table*}

Table~\ref{tab:active_prior_results} compares active PIC electrical routing using via count, metal-layer usage, electrical wirelength, and runtime. PROPEL completes all evaluated active-routing cases without vias, using one metal layer for the 8x8 and 16x16 cases and two metal layers for the 32x32 cases. On the smaller Clements benchmarks, PROPEL is faster than LiDAR~3.0 \cite{zhou2026lidar3}, reducing runtime from 33~s to 19~s for Clements\_8x8. On the larger Clements and ADEPT cases, PROPEL remains via-free but is slower and has higher WL, indicating that the current active electrical router would benefit from stronger global planning, pad assignment, and metal-corridor selection before detailed VFF routing. Because the prior-work and PROPEL results were measured on different hardware platforms and use different electrical-quality definitions, the runtime and wirelength values should not be interpreted as hardware-normalized or one-to-one quality comparisons. The results nevertheless show that PROPEL can produce via-free active-routing solutions, while also indicating the need for stronger global planning, pad assignment, and metal-corridor selection.

\begin{figure*}[t]
\centering

\begin{minipage}[t]{0.48\textwidth}
\centering
\includegraphics[width=\linewidth]{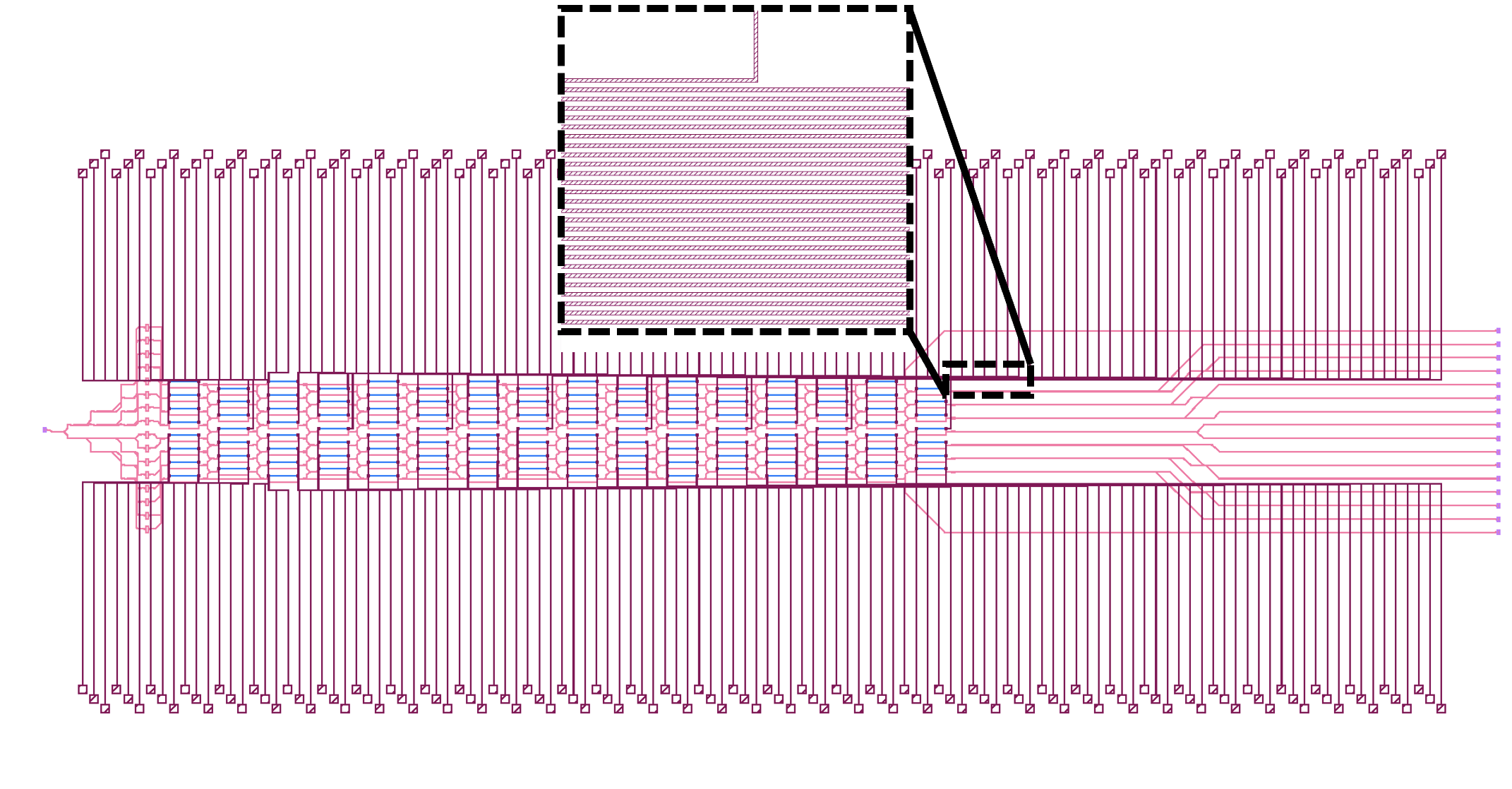}
\vspace{0.2em}

\footnotesize (a) Active 16x16 Clements mesh with electrical routing
\end{minipage}
\hfill
\begin{minipage}[t]{0.48\textwidth}
\centering
\includegraphics[width=\linewidth]{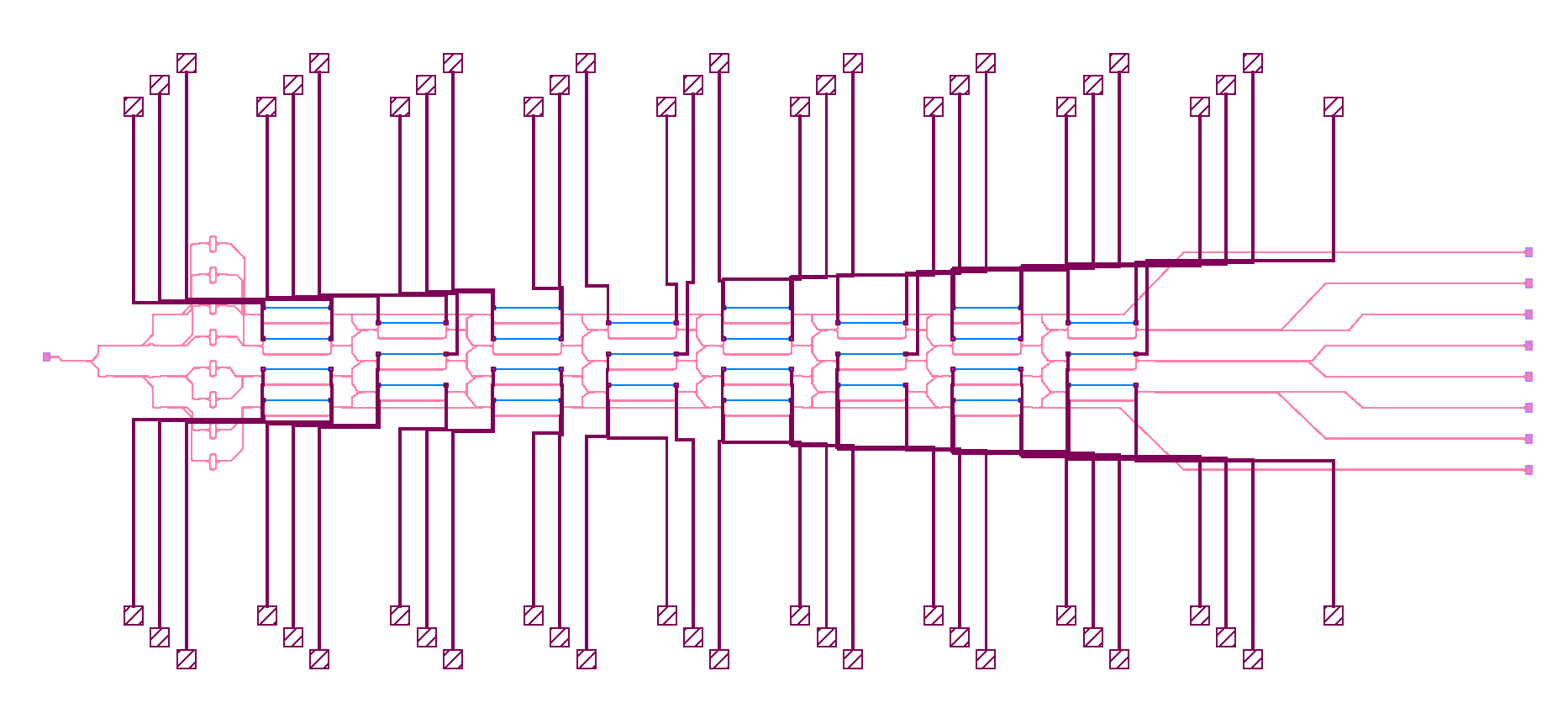}
\vspace{0.2em}

\footnotesize (b) Active 8x8 Clements mesh with electrical routing
\end{minipage}

\caption{Active PIC layouts generated by PROPEL. The examples show electrical control routing with photonic layout constraints.}
\label{fig:active_routing_gds_examples}
\end{figure*}

For active PICs, lower via count and fewer user-specified violations are usually more important than small changes in metal wirelength. A slightly longer metal route can be acceptable if it reduces via usage, avoids sensitive photonic regions, or preserves routing planarity.

\subsubsection{Topology-Driven Routing Results}
\label{subsec:topological_results}

Table~\ref{tab:topological_results_final} reports the topology-driven routing result format. These benchmarks evaluate whether PROPEL can generate a physically meaningful optical netlist before detailed routing begins. The assignment stage and the detailed routing stage are reported separately because a poor generated netlist can increase crossing pressure, path length, port blocking, and detailed routing failure.

\begin{table*}[t]
\centering
\caption{Topology-driven routing results.}
\label{tab:topological_results_final}
\scriptsize
\setlength{\tabcolsep}{5pt}
\begin{tabular}{lrrrrrr}
\hline
Benchmark & Input IO & Output IO & Nets & Routing Layers & Length Matching & DRV \\
\hline
Topological\_26IO & 13 & 13 & 26 & 1 & No & \textbf{0} \\
Topological\_100IO\_10x10 & 100 & 100 & 100 & 1 & No & \textbf{0} \\
\hline
\end{tabular}
\end{table*}
\begin{figure*}[t]
\centering

\begin{minipage}[t]{0.48\textwidth}
\centering
\includegraphics[width=\linewidth]{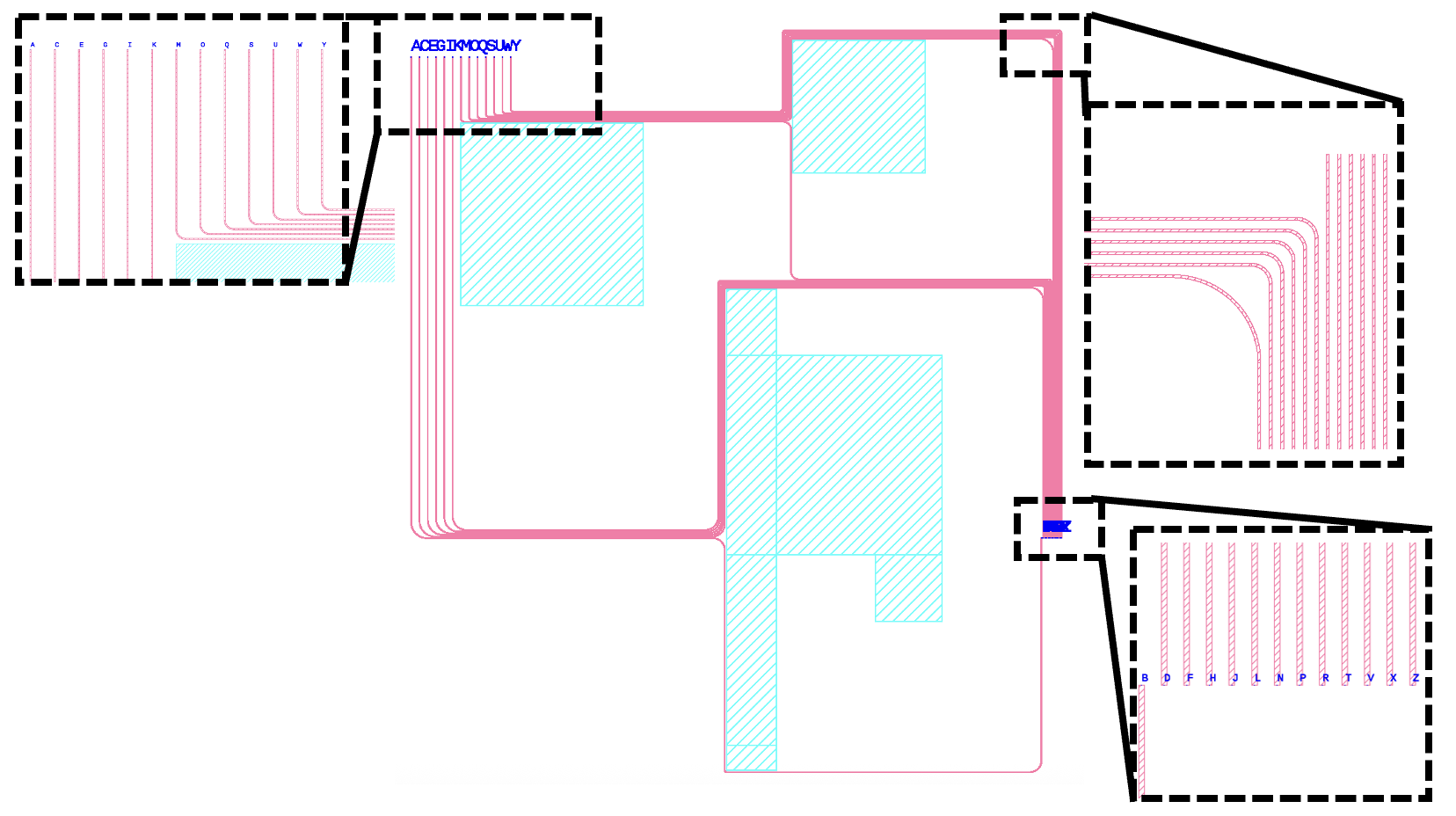}
\vspace{0.2em}

\footnotesize (a) Routing-aware topology generation
\end{minipage}
\hfill
\begin{minipage}[t]{0.48\textwidth}
\centering
\includegraphics[width=\linewidth]{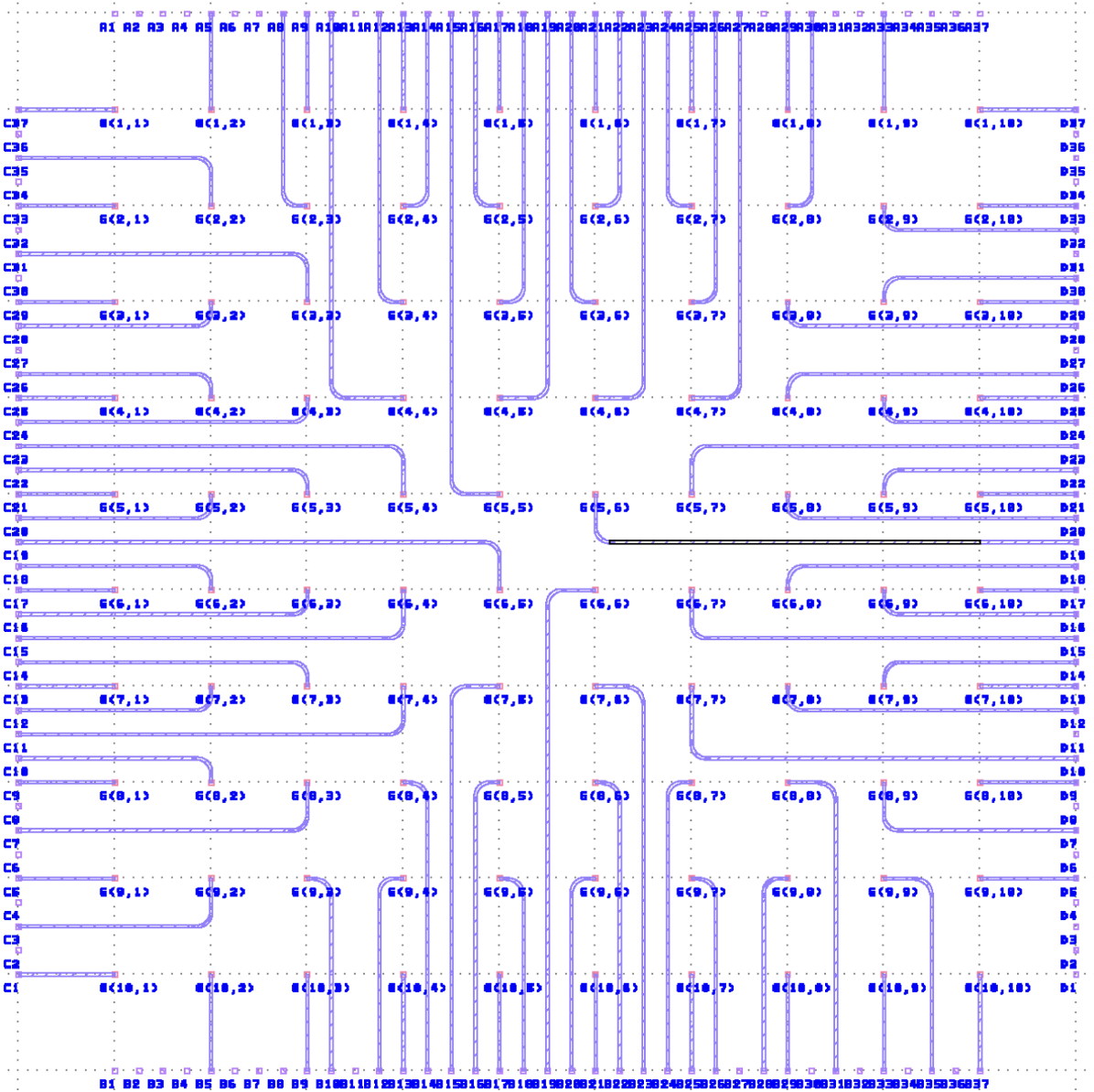}
\vspace{0.2em}

\footnotesize (b) Topology-driven routed layout
\end{minipage}

\caption{Topology-driven routing visualization. The panels show the generated routing-aware topology and the corresponding routed optical interconnect structure.}
\label{fig:topology_driven_gds_examples}
\end{figure*}

The topology-driven metrics answer two questions. First, the assignment metrics show whether routing-aware pairing reduces the estimated difficulty of the generated netlist. Second, the detailed routing metrics show whether the generated netlist can be converted into a strict DRC-clean waveguide geometry.

\paragraph{Quantum Graph Photonic Stress Test with Bend-Radius Variation.}
\label{subsec:quantum_graph_results}

Table~\ref{tab:quantum_graph_results} reports the quantum-graph-inspired photonic stress test. This benchmark is kept under passive routing because the reported result focuses on optical route closure, but the layout also includes heterogeneous stress features such as dense optical I/O, crossing-rich structure, electrical access, and delay-sensitive connectivity. The compact cases are evaluated with a more restrictive bend-radius setting, which reduces legal detour space and makes route closure more difficult. The largest case, QuantumGraph\_32x32, uses a 20~$\mu$m bend radius and routes 423 optical nets with 64 matched nets in 890.00~s. The 16x16 and 8x8 cases use a tighter 10~$\mu$m bend radius, which reduces the bend footprint but also requires the router to preserve legal curvature and port access in denser local regions. Comparing the non-compact and compact variants under the same 10~$\mu$m bend-radius setting shows the cost of layout-space restriction: QuantumGraph\_16x16\_C takes 420.80~s compared with 221.56~s for QuantumGraph\_16x16, and QuantumGraph\_8x8\_C takes 92.40~s compared with 38.87~s for QuantumGraph\_8x8. These results show that PROPEL can scale beyond regular mesh and switch-fabric layouts to heterogeneous quantum-graph-inspired photonic layouts while handling matched-net routing and bend-radius-constrained geometry.

\begin{table*}[t]
\centering
\caption{Quantum-graph-inspired photonic stress test with bend-radius variation.}
\label{tab:quantum_graph_results}
\scriptsize
\setlength{\tabcolsep}{5pt}
\begin{tabular}{lrrrrr}
\hline
Metrics &
QuantumGraph\_32x32 &
QuantumGraph\_16x16 &
QuantumGraph\_16x16\_C &
QuantumGraph\_8x8 &
QuantumGraph\_8x8\_C \\
\hline
Bend radius & 20 & 10 & 10 & 10 & 10 \\
Comp. & 268 & 138 & 138 & 61 & 61 \\
Nets & 423 & 155 & 155 & 85 & 85 \\
Matched Nets & 64 & 16 & 16 & 8 & 8 \\
Routed Opt. & \textbf{423} & \textbf{155} & \textbf{155} & \textbf{85} & \textbf{85} \\
Time (s) & \textbf{890.00} & \textbf{221.56} & \textbf{420.80} & \textbf{38.87} & \textbf{92.40} \\
\hline
\end{tabular}
\end{table*}
\begin{figure*}[t]
\centering
\includegraphics[width=0.95\textwidth]{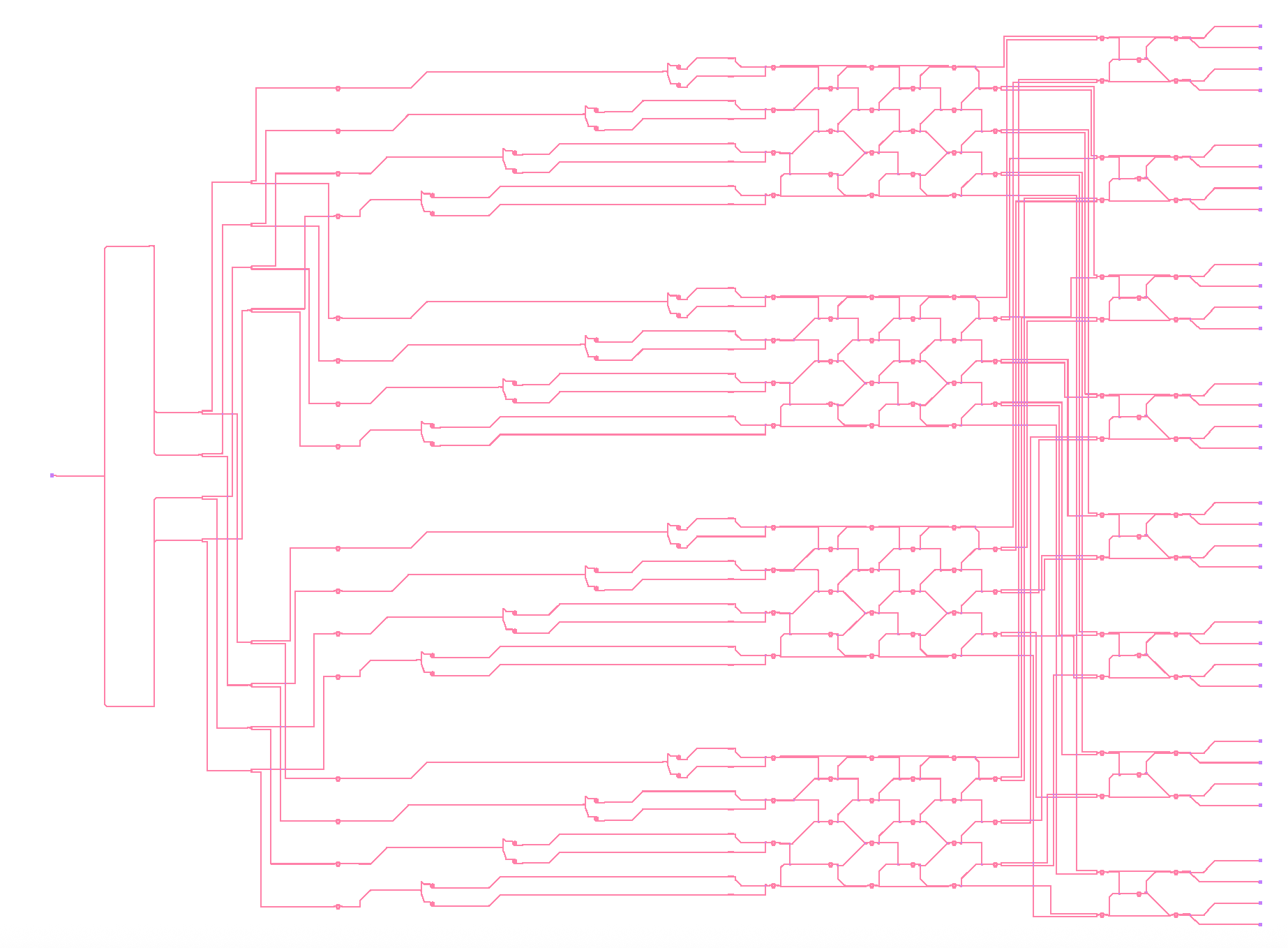}
\caption{Representative quantum-graph-32x32 photonic layout generated by PROPEL.}
\label{fig:quantum_graph_layout}
\end{figure*}

The quantum-graph-inspired benchmark evaluates whether PROPEL can route heterogeneous photonic layouts beyond standard mesh and switch-fabric cases. The non-compact 8x8 and 16x16 cases close without optical DRV, while the compact bend-radius-constrained variants expose the expected difficulty of routing dense layouts under reduced spacing and detour freedom. This stress test is useful because it combines optical route closure, crossing-rich structure, electrical-access awareness, and delay-sensitive layout constraints in one benchmark family.

\section{Discussion}
\label{sec:discussion}

The same kernel closes passive, matched, active, and topology-driven cases, which is the framework's main result. The same VFF-guided and DRC-aware kernel is used for passive optical routing, length matching, process-aware layout generation, active optical--electrical routing, and topology-driven net generation. This unified structure is useful because modern PIC layouts often combine these constraints in the same design, especially in large switch fabrics, photonic-computing meshes, programmable photonic circuits, and heterogeneous quantum-photonic layouts~\cite{boos2013proton,beuningen2016protonplus,beuningen2016platon,chuang2018planaronoc,zhou2025lidar,zhou2025lidar2,zhou2026lidar3}.

In the passive-routing benchmarks, PROPEL demonstrates competitive runtime and DRC-clean route commitment across regular Clements meshes, ADEPT circuits, Light and GWOR switch fabrics, Benes-style layouts, crossbar cases, and tensor-core-style examples. The main strength of the current implementation is its ability to produce legal GDS-level routing across different circuit families using the same routing kernel in the shortest amount of time. This is important because a route is only useful if it can be converted into manufacturable geometry with valid port access, bend radius, crossing structures, and spacing.

The comparison also shows a minor limitation: in some benchmarks, PROPEL produces more crossings than LiDAR~2.0. This is undesirable because optical crossings increase insertion loss, crosstalk risk, and layout footprint. The higher crossing count is mainly a consequence of the bounded-VFF search strategy. By limiting the search to a compact routing region, PROPEL gains substantial runtime improvement and better scalability, but the router may miss a wider detour that would reduce crossings. Expanding the VFF window can expose those alternatives, but it also increases field construction cost, candidate expansion, DRC checking, and rerouting time. Therefore, the present results reflect a trade-off between fast bounded routing and crossing-optimal exploration.

This trade-off is most visible in compact benchmarks, where routing resources are limited, and early decisions can strongly affect later nets. A locally legal route may block a port-access corridor, consume a crossing-compatible region, or force later nets into longer and more lossy paths. PROPEL addresses this through group-aware ordering, adaptive port-access reservation, route memory, and selective rip-up/reroute. These mechanisms improve convergence, but global visibility remains limited. A global routing-aware bounding box or a polygon-shaped bounding box could be used in the future to avoid a higher amount of crossing in those particular benchmark circuits.

The length-matching results show that PROPEL can apply matching constraints after ordinary detailed routing. This is important for interferometers, true-time-delay structures, coherent paths, and programmable photonic systems where DRC-clean connectivity alone is insufficient. The reported zero mismatch indicates that PROPEL can insert legal compensation geometry for geometric, OPL, and delay matching circuits while preserving final DRC cleanliness. This confirms that matching should be treated as a layout-aware refinement step rather than as a numerical length adjustment disconnected from physical routing.

The process-aware results should be interpreted as a geometry-generation capability rather than as a route-search improvement. In the current implementation, process awareness does not change the selected route, crossing count, waveguide length, insertion-loss estimate, or runtime. Instead, PROPEL samples the wafer map along the routed path and modifies the generated waveguide width according to the local process condition. This provides a first step toward fabrication-aware layout generation. In the future, we want this feature to include process maps directly in the routing cost so that the router can avoid high-risk regions or optimize optical path length under spatially varying effective-index and group-index conditions.

The active-routing results demonstrate the need for integrated optical and electrical layout closure. Active PICs require electrical routes for heaters, phase shifters, modulators, photodetectors, microring tuning electrodes, bias lines, and pads, while the optical layer must remain valid. PROPEL handles this using an optical-first flow in which the routed waveguides become part of the electrical routing context. However, the electrical router is still less mature than the passive optical router. LiDAR~3.0 uses a more specialized two-stage strategy with global planning followed by detailed electrical routing~\cite{zhou2026lidar3}. PROPEL currently relies more directly on VFF guidance and local legality feedback, which can lead to unnecessary detours, wandering routes, and additional rip-up/reroute cycles. Future work should add stronger electrical global planning, pad assignment, metal-corridor selection, and better coupling between optical keep-outs and electrical search.

The topology-driven results show that routing quality can be improved before detailed routing begins. In WRONoCs, GWOR-style routers, switch fabrics, and graph-structured photonic systems, the physical input--output assignment can strongly affect crossing pressure and routability~\cite{tseng2019wronoc,tan2011gwor,zheng2021topro}. PROPEL uses routing-aware assignment and crossing-aware refinement to generate a more physically meaningful netlist before invoking detailed routing. This reduces avoidable routing difficulty and helps distinguish between failures caused by poor connectivity choices and failures caused by detailed routing limitations.

Overall, the results identify both the strengths and the current limitations of PROPEL. The framework provides fast DRC-aware routing across several PIC design modes, but crossing reduction and active electrical routing still require improvement. Future work will focus on adaptive VFF visibility, crossing-aware global detour discovery, stronger electrical global planning, process-aware routing costs, and tighter feedback between topology assignment and detailed routing. AI-assisted routing and placement may also be useful as a supervisory layer that predicts difficult nets, selects routing parameters, recommends VFF-bound expansion, and proposes placement changes before detailed routing. Such methods should guide the deterministic router rather than replace it, because final layout commitment must remain governed by geometry construction and DRC replay.

\paragraph{Threats to validity.}
The reported comparisons combine direct benchmark-level results and scope-level positioning. Direct runtime and routing-quality comparisons are made only where prior work reports the same benchmark family and compatible metrics. The active-routing comparison is not a one-to-one electrical-quality comparison because PROPEL does not use the same USV model or wirelength-reporting definition as the prior active-routing baselines. Similarly, the current process-aware mode modifies generated waveguide geometry after route selection rather than optimizing the search path using process cost. The raw ablation table should also be interpreted separately from the final benchmark tables when a configuration has nonzero residual DRV. These limitations do not invalidate the routing framework, but they define the boundary of the present evidence.

\section{Conclusion}
\label{sec:conclusion}

This paper presented PROPEL, a memory-driven, native-accelerated, adaptive vector-flow-field routing framework with process-aware waveguide generation for large-scale PIC physical design. The central contribution is a common geometry-aware routing kernel that supports passive optical routing, length matching, process-aware geometry generation, active optical--electrical routing, and netlist-free topology-driven routing within one DRC-aware physical-design framework.

Across the shared passive benchmark set, PROPEL is faster than LiDAR~2.0 in 17 of 18 cases, with a median speedup of $2.6\times$ and a geometric-mean speedup of $2.5\times$, while maintaining zero final DRC violations. The results also show that PROPEL can close length-, delay-, and optical-path-length-matching benchmarks with zero residual mismatch, generate process-aware waveguide geometry from wafer-map data, demonstrate active electrical routing in the context of routed optical geometry, and route topology-driven and quantum-graph-inspired photonic layouts.

The results also identify limitations. Bounded VFF search improves runtime but can miss longer detours that reduce crossings or insertion loss. The current active electrical router is less mature than the passive optical router and needs stronger global planning, pad assignment, and metal-corridor generation. Process awareness is currently implemented as post-route waveguide-width assignment rather than process-cost-aware route search. Future work will address these limitations by adding adaptive VFF visibility, stronger electrical global planning, process-aware routing costs, and feedback from detailed-routing failures into topology assignment.

Overall, PROPEL shows that fast directional search, strict DRC replay, route memory, selective rerouting, matching refinement, active routing, and topology-driven assignment can be integrated into a single extensible routing-centered framework for heterogeneous PIC layout closure.
\section*{Funding}
This work was supported by the Defense Advanced Research Projects Agency (DARPA) under the Heterogeneous Adaptively Produced Photonic Interfaces (HAPPI) program, Award/Contract No. HR0011-25-C-0044.

\section*{Disclaimer}
The views, opinions and/or findings expressed are those of the author and should not be interpreted as representing the official views or policies of DARPA or the U.S. Government. Approved for public release, distribution unlimited.

\section*{Conflict of Interest}
The authors declare that they have no conflicts of interest related to this work.

\section*{Data and Code Availability}
The benchmark definitions, configuration files, and routing scripts used in this work are available from the corresponding author upon reasonable request.

\bibliographystyle{ACM-Reference-Format}
\bibliography{references}
\clearpage

\end{document}